\documentclass[twocolumn]{aastex631}

\accepted{27 July, 2026}

\submitjournal{ApJS}

\shorttitle{The \emph{s}-process in Magellanic Cloud Planetary Nebulae}
\shortauthors{Sterling \& Stephenson et al.}

\begin{document}

\title{The \emph{s}-process at Subsolar Metallicity: Insights from High-Resolution Infrared Spectroscopy of Magellanic Cloud Planetary Nebulae}

\author[0000-0002-9604-1434]{N.\ C.\ Sterling}
\affiliation{University of West Georgia, Carrollton, GA 30118, USA}
\correspondingauthor{N.\ C.\ Sterling}
\email{nsterlin@westga.edu}

\author[0000-0003-4717-0376]{M.\ G.\ Stephenson}
\affiliation{University of West Georgia, Carrollton, GA 30118, USA}
\affiliation{Department of Astronomy, University of Texas at Austin, Austin, TX, 78712, USA}

\author[0000-0002-4017-5572]{Harriet L.\ Dinerstein}
\affiliation{Department of Astronomy, University of Texas at Austin, Austin, TX, 78712, USA}

\author[0000-0002-7294-9288]{A.\ Yag\"ue L\'opez}
\affiliation{Los Alamos National Laboratory, Los Alamos, NM 87545, USA}

\author[0000-0001-6909-3856]{Kyle F.\ Kaplan}
\affiliation{Department of Astronomy, University of Texas at Austin, Austin, TX, 78712, USA}

\author{J.\ Beaumont}
\affiliation{University of West Georgia, Carrollton, GA 30118, USA} 

\author[0000-0002-6972-3958]{Maria Lugaro}
\affiliation{Konkoly Observatory, HUN-REN Research Centre for Astronomy and Earth Sciences, Konkoly Thege Mikl\'os \'ut 15-17, H-1121 Budapest, Hungary}
\affiliation{ELTE E\"otv\"os Lor\'and University, Institute of Physics and Astronomy, Budapest 1117, P\'azm\'any P\'eter s\'et\'any 1/A, Hungary}
\affiliation{School of Physics and Astronomy, Monash University, Clayton VIC 3800, Australia}

\author[0000-0002-3625-6951]{Amanda I.\ Karakas}
\affiliation{School of Physics and Astronomy, Monash University, Clayton VIC 3800, Australia}

\author[0000-0002-4619-8492]{Umberto Battino}
\affiliation{Department of Physics, University of Naples Federico II, Via Cintia, Napoli, 80126, NA, Italy}
\affiliation{INAF – Osservatorio Astronomico d’Abruzzo, Via M. Maggini, 64100 Teramo, Italy}
\affiliation{The NuGrid Collaboration, http:// www.nugridstars.org}

\author[0000-0002-6138-1869]{J.\ Garc\'ia-Rojas}
\affiliation{Instituto de Astrofísica de Canarias (IAC), E-38205 La Laguna, Spain}
\affiliation{Departamento de Astrofísica, Universidad de La Laguna, E-38206 La Laguna, Spain}

\author{Samuel M.\ H.\ Erben}
\affiliation{Georgia Institute of Technology, Atlanta, GA, 30332, USA}

\begin{abstract}

We present observations of 12 Magellanic Cloud planetary nebulae (PNe) obtained with the high-resolution near-infrared spectrometer IGRINS on Gemini South. In 10 targets we detect multiple neutron(\emph{n})-capture elements that can be synthesized by the \emph{s}-process during the preceding asymptotic giant branch (AGB) phase, including the widely-observed species Se and Kr, and first detections of Rb, Cd, and Te in extragalactic PNe. The derived abundances significantly expand the inventory of trans-iron element abundance determinations from PNe in subsolar metallicity stellar populations. Seven targets exhibit \emph{s}-process enrichments relative to the metallicity references O and Ar, with abundances elevated by factors of up to $\sim$40--50 for Te and Kr. We compare these results with AGB nucleosynthesis predictions, using a machine-learning algorithm to find the best-fitting Monash, FRUITY, and NuGrid models. For five of the PNe, Monash and FRUITY models with $2-4$~M$_{\odot}$ and metallicities from 1/3 solar to near-solar match the observations well, although the Monash models more successfully fit the largest Kr enhancements. NuGrid models predict smaller \emph{s}-process enrichments than observed, but the limited number of available models makes it difficult to assess their ability to reproduce the observed abundances.  We were unable to find models that provide good fits to three other enriched PNe, including the most metal-poor objects in our sample. These comparisons address uncertainties in mass loss, convection, and other mixing mechanisms during the late evolution of low- and intermediate-mass stars, and improve the accuracy of \emph{s}-process yields of AGB stars, which are key parameters for modeling galactic chemical evolution.

\end{abstract}


\vspace{.1in}

\section{Introduction}\label{sec:intro}

Low- and intermediate-mass stars (1--8~M$_{\odot}$) are major sources of C, N, and trans-iron nuclides produced by slow neutron(\emph{n})-capture nucleosynthesis (the ``\emph{s}-process") in galaxies \citep[e.g.,][]{Travaglio_Etal_2004, Kobayashi_Etal_2020, Lian_Etal_2023, Arcones_Thielemann_2023}. These species are produced during the asymptotic giant branch (AGB) evolutionary stage, and are ejected into the interstellar medium via stellar winds and planetary nebulae (PNe). The abundances of PNe represent the final stellar envelope composition at the end of the AGB, and therefore nebular spectroscopy provides a valuable probe of AGB nucleosynthesis and elemental yields \citep[e.g.,][]{Sterling_Dinerstein_2008, Henry_Etal_2015, Ventura_Etal_2017, Kwitter_Henry_2022}. This information is complementary to the extensive body of work on AGB stars \citep[e.g.,][]{Busso_Etal_1999}. An advantage of studying PNe is that their compositions are not subject to further alteration by nucleosynthesis, in contrast to AGB stars. Additionally, PNe provide access to elements that are not observable in the cool photospheres of AGB stars, such as noble gases.

Recent work has demonstrated that despite their low cosmic abundances, \emph{n}-capture elements (atomic number $Z>30$) can be detected in numerous Galactic PNe, providing a new tool to study \emph{s}-process nucleosynthesis in low- and intermediate-mass stars. Emission lines of trans-iron elements were first recognized in the optical spectrum of the bright PN NGC~7027 \citep{Pequignot_Baluteau_1994, Baluteau_Etal_1995}, and subsequent high-resolution investigations confirmed several of these identifications \citep{Sharpee_Etal_2007, Garcia-Rojas_Etal_2015, Otsuka_Hyung_J900}. \citet{Sterling_Dinerstein_2008} studied K~band features of [\ion{Kr}{3}] at 2.1986 and [\ion{Se}{4}] at 2.2864~$\mu$m\footnote{Vacuum wavelengths are used throughout in this paper.} first identified by \citet{Dinerstein_2001} in a survey of 120 Galactic PNe. They detected Se ($Z=34$) and/or Kr (36) in more than two-thirds of their targets. The development of sensitive, high-resolution near-infrared (NIR) spectrometers such as the Immersion GRating INfrared Spectrometer \citep[IGRINS;][]{Park_Etal_2014}, the Habitable-zone Planet Finder \citep[HPF;][]{Mahadevan_Etal_2014_HPF}, and iSHELL \citep{Rayner_Etal_2022} have enabled weak NIR features of Br ($Z=35$), Rb (37), Cd (48), Te (52), and Xe (54) to be detected in PNe for the first time \citep[][Dinerstein et al., in prep.]{Sterling_Etal_2016, Madonna_Etal_2018, Sterling_2020, Dinerstein_Etal_2021_AAS, Dinerstein_Etal_2022_AAS}.

Most previous nebular studies of \emph{n}-capture elements have targeted Milky Way PNe due to the greater ease of detecting faint emission lines in brighter, more proximate sources. However, most \emph{s}-process enriched Galactic PNe have solar or near-solar metallicity \citep[e.g.,][]{Sterling_Dinerstein_2008, Manea_Etal_2022}, and thus do not probe predicted metallicity effects on \emph{s}-process enrichments. New empirical constraints on \emph{s}-process enhancements at subsolar metallicities in particular are keenly needed (\S\ref{sec:constraints}). The Large and Small Magellanic Clouds (LMC and SMC, respectively) hold the nearest (and hence relatively bright) and largest population of subsolar metallicity PNe that span a wide range of initial masses. 

We present high-resolution spectroscopic data of Magellanic Cloud PNe obtained with IGRINS on the 8.1-m Gemini South Telescope. Our observations yield a high detection rate, with at least two \emph{n}-capture elements observed in 10 of the 12 targets, and include the first detections of Rb, Cd, and Te in extragalactic PNe. In many cases, the line strengths are enhanced by elevated abundances that result from \emph{s}-process  nucleosynthesis. These results expand the inventory of nebular \emph{n}-capture element abundance determinations in these galaxies, especially the SMC.  We compare the empirical abundances with theoretical enrichments predicted by different models of AGB evolution and nucleosynthesis, which aids in identifying the most realistic treatments of currently poorly-understood processes such as mass loss, convection, and other internal mixing mechanisms.

This paper is organized as follows. In \S\ref{sec:context} we discuss AGB evolution and nucleosynthesis, its sensitivity to metallicity and initial mass, and the need for new empirical information on \emph{s}-process enrichments. In \S\ref{sec:sample} we describe our survey and the target selection. The observations, data reduction, and line identifications are detailed in \S\ref{sec:obs_data}, and the abundance analysis in \S\ref{sec:abunds}. The \emph{s}-process enhancements of \emph{n}-capture elements are compared to theoretical predictions from AGB evolutionary codes in \S\ref{sec:models}. We discuss and interpret results in \S\ref{sec:discuss}, and provide concluding remarks in \S\ref{sec:Conclusions}.

\section{AGB Stars as Sources of Neutron-Capture Elements}\label{sec:context}

\subsection{\emph{s}-Process Nucleosynthesis During the AGB} \label{sec:agb}

The most dramatic nuclear processing that ultimately determines elemental abundances in PNe largely occurs during the late, ``thermally-pulsing" AGB phase. In this evolutionary stage, ashes from the H-burning shell build up on the quiescent He shell until He fusion ignites, triggering runaway triple-alpha fusion called a ``He-shell flash." This phenomenon is also referred to as a thermal pulse, as it injects a large amount of energy deep in the star's interior, propelling the outer layers to larger radii. The H shell cools, temporarily pausing H fusion, and an internal convective zone is generated that dredges up C- and \emph{s}-process rich material (``third dredge-up") from the intershell region between the H- and He-burning shells into the envelope \citep[see reviews of][]{Iben_Renzini_1983, Busso_Etal_1999, Herwig_2005, Karakas_Lattanzio_2014, Ventura_2022}. He shell fusion quickly dies out, the star contracts, and the H-shell resumes energy production.  This cycle of thermal pulses can recur several times for stars above a metallicity-dependent minimum mass of $\sim1.0$--1.5~M$_{\odot}$ \citep{Stancliffe_Etal_2005, Karakas_Lugaro_2016, Shetye_Etal_2019}.

The episodic thermal pulses set the stage for \emph{s}-process production of \emph{n}-capture elements. Protons from the H-rich envelope are mixed into the C-rich, H-poor intershell region, where they can be captured by $^{12}$C nuclei to form a ``$^{13}$C pocket" or ``partial mixing zone" (PMZ; see \S\ref{sec:Monash}). During the $\sim$10$^3$--10$^4$ year interval between thermal pulses, $^{13}$C($\alpha$,\emph{n})$^{16}$O reactions release free neutrons, with typical densities of 10$^7$--10$^8$~cm$^{-3}$ \citep{Busso_Etal_1999, Karakas_Lattanzio_2014, Lugaro_Etal_2023}. Iron-peak nuclei experience a series of \emph{n}-captures that are \textit{slow} relative to the $\beta$-decay timescales of the participating isotopes, producing nuclei near the valley of $\beta$ stability up to Pb ($Z=82$). Isotopes with closed neutron shells (``neutron-magic" nuclei) have small \emph{n}-capture cross sections, and thus act as bottlenecks in the \emph{s}-process path, leading to large enrichments of these nuclides. Consequently, the abundance distribution resulting from the \emph{s}-process is characterized by three peaks corresponding to neutron-magic nuclei ($N=50$, 82, and 126): the first or ``light-\emph{s}" peak at $Z=38$--40, the ``heavy-\emph{s}" peak near $Z=56$, and the third peak at Pb.

This contrasts with \textit{rapid} \emph{n}-captures (the ``\emph{r}-process") that occur in cataclysmic events involving high-mass stars or their compact remnants. The high neutron densities associated with the \emph{r}-process (10--15 orders of magnitude larger than in the \emph{s}-process) lead to the formation of \emph{n}-rich nuclei \citep[][and references therein]{Cowan_Etal_2021, Arcones_Thielemann_2023}. For this reason, the peaks in the \emph{r}-process abundance pattern (near $Z=34$, 54, and 78) occur at lower atomic number than in the \emph{s}-process.

In more massive AGB stars ($\gtrsim3$--4~M$_{\odot}$), the high interior temperatures allow CNO-cycling to occur at the base of their convective envelopes \citep[``hot bottom burning," or HBB;][]{Iben_Renzini_1983, Cristallo_Etal_2015, Ventura_Etal_2015_HBB, Karakas_Lugaro_2016}. This creates the N and He enrichments that characterize ``Type~I" PNe \citep{Peimbert_1978, Kingsburgh_Barlow_1994, Ventura_Etal_2015_HBB}. The high temperatures can also activate a second neutron source, $^{22}$Ne($\alpha,n$)$^{25}$Mg \citep{Busso_Etal_1999}. The large neutron densities ($\sim$10$^{11}$~cm$^{-3}$) produced by the $^{22}$Ne source open branchings in the \emph{s}-process path that cause strong enrichments of Rb and Kr \citep[e.g.,][]{Busso_Etal_1988, VanRaai_etal_2012}. Because the $^{22}$Ne reaction operates \textit{during} the thermal pulses, over $\sim$10 year timescales, Fe-peak seed nuclei experience a low time-averaged neutron flux, and enrichments of elements heavier than Sr ($Z=38$) are minimal compared to when neutrons are chiefly produced by $\alpha$-captures onto $^{13}$C. Thus, the element-by-element patterns of \emph{s}-process enrichments produced by the $^{13}$C and $^{22}$Ne sources significantly differ from each other. However, it is not clear which dominates neutron production in intermediate-mass stars, due to the scarcity of empirical constraints \citep[e.g.,][]{Karakas_Etal_2012}. 

The \emph{s}-process abundance pattern is also sensitive to metallicity \citep[e.g.,][]{Busso_Etal_2001, Fishlock_Etal_2014, Karakas_Etal_2018}, since the neutron sources $^{13}$C and $^{22}$Ne are primary nuclear products, while the Fe-peak nuclei onto which neutrons are captured are secondary. 
Therefore, in lower metallicity stars, each Fe-peak seed nucleus experiences a higher neutron flux, resulting in larger enrichments at the heavy-\emph{s} peak than for elements near the light-\emph{s} peak. The origin of the third-peak element Pb was once uncertain, as it could not be explained by either the \emph{s}-process at near-solar metallicity nor by the \emph{r}-process, leading to the postulate of a separate ``strong" \emph{s}-process \citep{Clayton_Rassbach_1967} to account for the production of Pb. 
However, it is now considered that low-metallicity AGB stars are the synthesis sites for Pb via the ``main" \emph{s}-process described above \citep{Gallino_Etal_1998}, and possibly by a \emph{n}-capture process with intermediate neutron densities, the \emph{i}-process \citep[e.g.,][]{Hampel_Etal_2019, Choplin_Etal_2024}.

\subsection{The Need for New Observational Constraints} \label{sec:constraints}

Nebular spectroscopy provides critical new information to help resolve outstanding questions regarding AGB evolution. From a theoretical perspective, several uncertainties hinder predictions of nucleosynthetic yields in AGB stars. These include the treatment of mass-loss rates and histories, convection, mixing at convective boundaries, and other processes such as rotational shear mixing, thermohaline mixing, and magnetic buoyancy \citep[][and references therein]{Ventura_Etal_1998_ATON, Herwig_Etal_2003, Cristallo_Etal_2015, Karakas_Lugaro_2016, Battino_Etal_2019, Busso_Etal_2021}. 

Moreover, observations of AGB and post-AGB stars, particularly at subsolar metallicities, have unveiled new puzzles that cloud our understanding of AGB nucleosynthesis. For example, \citet{deSemdt_Etal_2015, deSmedt_Etal_2016} found unexpectedly low Pb abundances compared to those of lighter \emph{n}-capture elements in LMC and Galactic post-AGB stars with metallicities [Fe/H]~$<-0.7$~dex. This effect might be due to enrichments from an \emph{i}-process with neutron densities $\sim10^{14}$--10$^{15}$~cm$^{-3}$ \citep[\textit{intermediate} to the \emph{s}- and \emph{r}-processes; e.g.,][and references therein]{Hampel_Etal_2016, Hampel_Etal_2019}. 

As another example, \citet{Garcia-Hernandez_Etal_2009} measured large supersolar Rb abundances but little Zr enhancement in luminous Magellanic Cloud AGB stars with progenitor masses 4--8~M$_{\odot}$. Although the derived Rb abundances were later adjusted downward \citep{Zamora_Etal_2014, Perez-Mesa_Etal_2017}, the abundance patterns are consistent with the $^{22}$Ne source dominating \emph{s}-process neutron production \citep{VanRaai_etal_2012, Karakas_Etal_2012}. However, Galactic Type~I PNe, which are thought to arise from more massive progenitors, do not exhibit enhanced Kr abundances as predicted if $^{22}$Ne is indeed the neutron source, at least not in solar metallicity objects \citep{Sterling_Dinerstein_2008, Karakas_Etal_2009}. Additionally, \citet{Kamath_Etal_2022} showed that post-AGB stars with similar initial masses and metallicities (from [Fe/H]~=~--1.5 to --0.3~dex) display significant variations in \emph{n}-capture element abundances, suggesting that there is a stochastic component to \emph{s}-process nucleosynthesis in stars with otherwise similar properties.

New observations of subsolar-metallicity populations are essential to elucidate the most realistic computational treatments of mass loss, convection, and mixing mechanisms in AGB star models, as well as to identify the physical causes of the enigmatic observational findings described above. Spectroscopy of PNe can provide the necessary empirical data, but the few low-metallicity Galactic PNe largely reside in the older stellar populations of the thick disk and halo. These objects primarily descend from lower-mass stars that are unlikely to have experienced third dredge-up and \emph{s}-process enrichment during the AGB, and/or originate from dwarf galaxies accreted by the Milky Way \citep[e.g.,][]{Zijlstra_Etal_2006}. To draw conclusions regarding \emph{s}-process nucleosynthesis at subsolar metallicities over a significant range of progenitor masses, the sample must be drawn from a system that has experienced star formation within the last few Gyr.

\section{Survey of Planetary Nebulae in the Magellanic Clouds} 
\label{sec:sample}

\subsection{The PN Populations of the Magellanic Clouds and Previous IR Observations}

The gas-rich LMC and SMC are home to the nearest \citep[50 and 62~kpc, respectively;][]{Keller_Wood_2006, Graczyk_Etal_2014} and largest populations of subsolar metallicity PNe \citep{Reid_Parker_2010, Reid_2014, Dravskovic_Etal_2015}.  The LMC tidally captured the SMC $\sim$3--4 Gyr ago \citep{Bekki_Chiba_2005}, triggering vigorous star formation in both galaxies that continues to this day \citep[][and references therein]{Harris_Zaritsky_2009, Rubele_Etal_2018, Mazzi_Etal_2021, Massana_Etal_2022, Murray_Etal_2024}. Consequently, their PNe have progenitor star masses ranging from $\lesssim 1.0$~M$_{\odot}$ to as high as 6--8~M$_{\odot}$, as indicated by their chemical compositions \citep{Ventura_Etal_2015_LMC, Ventura_Etal_2016_SMC} and central star properties \citep{Villaver_Etal_2003, Villaver_Etal_2004, Villaver_Etal_2007}. Furthermore, studies of field giants in the Magellanic Clouds provide details of the likely initial abundances of the PN progenitor stars of the same populations \citep{Pompeia_Etal_2008, Lapenna_Etal_2012, VanDerSwaelmen_Etal_2013, Nidever_Etal_2020, Mucciarelli_Etal_2023}.

\citet{Mashburn_Etal_2016} conducted the first study of \emph{n}-capture elements in LMC and SMC PNe, using the moderate-resolution NIR spectrometers FIRE \citep{Simcoe_Etal_2013_FIRE} on the 6.5-meter Baade Telescope and GNIRS \citep{Elias_Etal_1998} on Gemini South. [\ion{Kr}{3}]~2.1986 and/or [\ion{Se}{4}]~2.2864~$\mu$m were detected in eight of the 10 targets, and [\ion{Kr}{6}]~1.2330~$\mu$m was identified in two high-ionization LMC PNe \citep{Sterling_Etal_2017}. \citet{Mashburn_Etal_2016} found that Se and Kr are enriched in about half of the targets in their sample, when scaled to the metallicities of their respective  progenitor stars as indicated by O and Ar. Specifically, Se is enhanced by factors of 3--8 (0.5--0.9~dex) relative to solar, while Kr is enriched by 4--20 times (0.6--1.3~dex). Aside from Mashburn et al., the only other investigations of \emph{n}-capture elements in extragalactic PNe were carried out for the few known PNe in the Sagittarius dwarf spheroidal galaxy \citep{Wood_Etal_2006_Hen2-436, Otsuka_Etal_BoBn1, Otsuka_Etal_Hen2-436, Otsuka_Wray16-423}. Those objects exhibit strong \emph{s}-process enrichments of Kr and Xe, but have similar progenitor masses of 1.5--2.0~M$_{\odot}$.

While \citet{Mashburn_Etal_2016} detected Se and Kr in seven LMC PNe and one SMC PN, rigorous tests of AGB evolutionary models require information about the abundances of additional \emph{n}-capture elements, particularly Rb and those beyond the first \emph{s}-process peak. The fact that several objects in the Mashburn et al.\ sample exhibit significant Se and Kr enrichments suggest that \emph{n}-capture elements near the heavy-\emph{s} peak should be even more strongly enriched, given the subsolar metallicities of these galaxies. This expectation, and the success of IGRINS in detecting faint \emph{n}-capture element lines -- including heavier species such as Cd ($Z=48$) and Te (52) -- in Galactic PNe \citep{Sterling_Etal_2016, Madonna_Etal_2018}, motivated the present work.

\subsection{Selection Criteria and Characteristics of the Sample} 

The observed targets (Table~\ref{tab:target_info}) were selected to sample a wide range of stellar populations and nebular characteristics (e.g., C/O ratio, central star temperature and initial mass, metallicity, and dust features). The PNe have central star temperatures spanning from 29,000~K to as high as 200,000~K, allowing for ions with a wide range of ionization potentials to be detected (\S\ref{sec:ids}).  Published O and Ar abundances \citep[][see also the Appendix]{Leisy_Dennefeld_2006, Shaw_Etal_2010} indicate that the targets have metallicities [(O, Ar)/H] ranging from --1.4~dex to approximately solar.

We primarily observed PNe with C/O~$>$~1, since elevated C abundances are strong indicators of \emph{s}-process enrichments in AGB stars and PNe \cite[e.g.,][]{Smith_Lambert_1990, Abia_Etal_2002, Sterling_Dinerstein_2008}.  Mid-infrared dust features also indicate whether the nebular chemistry is C- or O-rich. Two PNe in our sample, SMP~LMC~48\footnote{We use the \citet{Sanduleak_Etal_1978} catalog numbers for all targets in our sample, and from hereon abbreviate the nomenclature by omitting the ``SMP" prefix, e.g.\ LMC~48.} and SMC~13, exhibit emission from fullerenes, and LMC~99 and SMC~20 are candidate fullerene objects \citep{Garcia-Hernandez_Etal_2012, Sloan_Etal_2014}. \citet{Sterling_2020} noted that many -- but not all -- PNe exhibiting these C$_{60}$ and C$_{70}$ features are strongly enriched in \emph{n}-capture elements such as Kr and Te \citep{Sharpee_Etal_2007, Sterling_Dinerstein_2008, Mashburn_Etal_2016}. The C-rich objects in our sample likely derive from progenitors with initial masses of 1.2--3~M$_{\odot}$ in the LMC \citep{Villaver_Etal_2003, Villaver_Etal_2007, Ventura_Etal_2015_LMC, Ventura_Etal_2025} and      
1--2~M$_{\odot}$ in the SMC \citep{Villaver_Etal_2004, Ventura_Etal_2016_SMC}.

\begin{deluxetable*}{llcccccccccc}
\tablecaption{Properties of the Observed MCPNe} \label{tab:target_info}
\tabletypesize{\footnotesize}
\tablewidth{0pt}
\tablehead{ \\[-1.4pc]
\colhead{} & \colhead{} & \colhead{Diam.\tablenotemark{a}} & \colhead{} & \colhead{Dust} & \colhead{$T_{\rm{eff}}$\tablenotemark{c}} & \colhead{M$_{\rm{init}}$\tablenotemark{c}} & \colhead{} & \colhead{} & \colhead{} \\
\colhead{Galaxy} & \colhead{PN} & \colhead{($\prime\prime$)} & \colhead{Morph.\tablenotemark{a}} & \colhead{Type\tablenotemark{b}} & \colhead{($10^3$~K)} & \colhead{(M$_\odot$)} & \colhead{log$F_{\rm{H}\beta}$\tablenotemark{a}} & \colhead{C/O\tablenotemark{d}} & \colhead{[O/H]\tablenotemark{d}} & \colhead{[Ar/H]\tablenotemark{d}}
}
\startdata
LMC & SMP 29 & $0.51\times0.47$ & BC & \nodata & 200 & $2.0-2.8$ & $-12.74$ & 0.28 & $-0.64$ & $-0.11$ \\
& SMP 31 & 0.26 & R? & C & 28.6 & $1.0-1.5$ & $-12.92$ & \nodata & $-1.39$ & $-0.82$ \\
& SMP 47 & $0.45\times0.32$ & E & \nodata & 131 & $1.4-1.8$ & $-12.54$ & 2.0 & $-0.44$ & $-0.25$ \\
& SMP 48 & $0.40\times0.35$ & E & C, F & 59 & \nodata & $-12.48$ & 1.4 & $-0.50$ & $-0.50$ \\
& SMP 61 & $0.56\times0.54$ & E & C & 60 & $2.0-2.5$ & $-12.48$ & 2.8 & $-0.16$ & $-0.12$ \\
& SMP 73 & $0.31\times0.27$ & E(bc) & \nodata & 135 & \nodata & $-12.55$ & 1.3 & $-0.03$ & $-0.28$ \\
& SMP 99 & $0.85\times0.73$ & E(bc) & C, F? & 124 & \nodata & $-12.54$ & 1.3 & $-0.13$ & 0.24 \\
SMC & SMP 3 & $0.59\times0.48$ & E(bc) & C & 95 & 1.5 & $-13.13$ & 6.2 & $-0.71$ & $-0.89$ \\
& SMP 13 & 0.20 & R & C, F & 31.3 & \nodata & $-12.56$ & 4.7 & $-0.63$ & $-0.92$ \\
& SMP 14 & 0.83 & R & C & 116.8 & 1.0 & $-13.00$ & 7.1 & $-0.40$ & $-0.56$ \\
& SMP 20 & $0.20\times0.23$ & \nodata & C, F? & 86.5 & $1.0-1.5$ & $-12.42$ & 3.2 & $-0.95$ & $-1.35$ \\
& SMP 25 & \nodata & E & O & 74.8 & $2.5-3.5$ & $-13.20$ & 0.12 & $-1.13$ & $-0.90$ \\
\enddata
\tablenotetext{a}{Diameters, morphologies, and absolute H$\beta$ fluxes are from \citet{Shaw_Etal_2001}, \citet{Stanghellini_Etal_2002}, \citet{Stanghellini_Etal_2003}, \citet{Leisy_Dennefeld_2006}, and \citet{Shaw_Etal_2006}. The morphologies are abbreviated as E (elliptical), R (round), BC (bipolar core), and E(bc) (elliptical with a bipolar core).}
\tablenotetext{b}{C = C-rich, O = O-rich, F = Fullerene.  Dust compositions are from \citet{Stanghellini_Etal_2007}, \citet{Garcia-Hernandez_Etal_2012}, and \citet{Sloan_Etal_2014}.}
\tablenotetext{c}{References for central star temperatures and initial masses: \citet{Dopita_Meatheringham_1991b}, \citet{Dopita_Etal_1997}, \citet{Bianchi_Etal_1997}, \citet{Vassiliadis_Etal_1998}, \citet{Villaver_Etal_2003, Villaver_Etal_2004, Villaver_Etal_2007}, \citet{Herald_Bianchi_2007}, and \citet{Hajduk_2020}.}
\tablenotetext{d}{C, O, and Ar abundances from \citet{Pena_Etal_1997}, \citet{Stanghellini_Etal_2005, Stanghellini_Etal_2009}, \citet{Leisy_Dennefeld_2006}, and \citet{Shaw_Etal_2010}. The abundances of the metallicity references O and Ar are given relative to the solar abundances of \citet{Asplund_Etal_2021}, using the standard notation [X/H]~=~$\log(\rm{X/H})_{\rm PN} - \log(\rm{X/H})_\odot$..}
\end{deluxetable*}

In contrast, the O-rich PNe LMC~29 and SMC~25 have large N and low C abundances \citep{Leisy_Dennefeld_2006, Stanghellini_Etal_2009}, which indicate that HBB has substantially affected their compositions. \citet{Ventura_Etal_2015_LMC, Ventura_Etal_2016_SMC} found that the abundances agree well with models of progenitor mass $\geq 6$~M$_{\odot}$, but acknowledged that the declining initial mass function and short evolutionary timescales of such massive stars make their PNe statistically unlikely to be detected. Nevertheless, the properties of the central stars of these objects indicate that they have larger progenitor masses than the C-rich PNe \citep{Vassiliadis_Etal_1998, Villaver_Etal_2003, Villaver_Etal_2004, Villaver_Etal_2007}, and thus may exhibit \emph{s}-process enrichment patterns characteristic of the $^{22}$Ne neutron source \citep{Garcia-Hernandez_Etal_2006, Garcia-Hernandez_Etal_2009, VanRaai_etal_2012}.

Four of the targets in our sample (LMC~47, LMC~73, LMC~99, and SMC~20) were previously studied in the NIR by \citet{Mashburn_Etal_2016}. Those authors detected [\ion{Kr}{3}] and [\ion{Se}{4}] in LMC~47 and LMC~99, and Se in LMC~73. The IGRINS spectra supersede those data, both in terms of resolution and in the number of H and K band lines (including from \emph{n}-capture elements) that are detected (see \S\ref{sec:ids} and \S\ref{sec:compare}). 

\section{Observations and Analysis}\label{sec:obs_data}

\subsection{Observations and Data Reduction}

All targets were observed with the Gemini South Telescope, under GO program GS-2020B-Q-211 (P.I.\ Sterling), and University of Texas compensatory time programs GS-2020A-Q-238 and GS-2020B-Q-227 (P.I.\ Dinerstein). The observations were made with the Immersion GRating INfrared Spectrometer \citep[IGRINS;][]{Yuk_Etal_2010, Wang_Etal_2010, Gully_Etal_2012, Moon_Etal_2012, Han_Etal_2012, Park_Etal_2014, Oh_Etal_2014, Jeong_Etal_2014, Mace_Etal_2016, Mace_Etal_2018}.
IGRINS provides complete coverage of the H and K~bands (1.45--2.50~$\mu$m), outside of atmospherically blocked regions, in a single setting at resolving power $R\simeq 45,000$. The high resolution of IGRINS has proved invaluable for identifying emission lines of \emph{n}-capture elements \citep{Sterling_Etal_2016, Madonna_Etal_2018} and other species (see \S 4.2), as well as revealing the conditions and excitation mechanisms of H$_2$ emission \citep[e.g.,][]{Kaplan_Etal_2021, Kaplan_Etal_2026}. On Gemini South, the fixed slit has dimensions of 5\farcs0$\times$0\farcs34 and was oriented E-W for all observations.

Table~\ref{tab:obslog} provides details of the observations. With the exception of SMC~3 (thin clouds, seeing 1\farcs1--1\farcs4), each target was observed in clear conditions with seeing of 1\farcs0 or better. Because the PNe were not resolved under the seeing conditions, each target was observed in ABBA mode, nodding along the slit to maximize observing efficiency. An A0V star was observed immediately after each target at similar airmass, and these were used as standards for wavelength calibration, telluric correction, and relative flux calibration.

\begin{deluxetable}{llccc}
\tablecaption{Observing Log} \label{tab:obslog}
\tabletypesize{\footnotesize}
\tablewidth{0pt}
\tablehead{ \\[-1.4pc]
\colhead{Galaxy} & \colhead{PN} & \colhead{Program ID} & \colhead{Date Observed} & \colhead{Int.\ Time (Hr)}
}
\startdata
LMC & SMP~29 & GS-2020B-Q-211 & 2020-10-31 & 1.57 \\
    & SMP~31 & GS-2020B-Q-227 & 2020-11-06 & 1.32  \\
    & SMP~47 & GS-2020A-Q-238 & 2020-02-08 & 1.0 \\
    & SMP~48 & GS-2020B-Q-211 & 2020-10-30 & 1.06 \\
    & SMP~61 & GS-2020B-Q-211 & 2020-11-02 & 0.53 \\
    & SMP~73 & GS-2020B-Q-211 & 2020-10-30,31 & 1.06 \\
    & SMP~99 & GS-2020A-Q-238 & 2020-02-08 & 1.25 \\
SMC & SMP~3  & GS-2020B-Q-211 & 2020-11-13 & 1.83  \\
    & SMP~13 & GS-2020B-Q-211 & 2020-10-30 & 1.07 \\
    & SMP~14 & GS-2020B-Q-211 & 2020-10-30 & 1.82 \\
    & SMP~20 & GS-2020B-Q-227 & 2020-11-30 & 2.06 \\
    & SMP~25 & GS-2020B-Q-211 & 2020-11-19 & 2.79 \\
\enddata
\end{deluxetable}

The data were reduced with the IGRINS Pipeline Package (PLP) v2.2.0 \citep{igrins-plp-2.2.0}.  The IGRINS PLP performs standard data reduction steps including flat fielding, wavelength calibration using OH sky emission lines and telluric absorption lines in the standard stars, rectification of the echelle orders, and summing along the slit to extract the 1D spectrum of each target. 
Before reducing the data, cosmic rays were removed from the raw frames using Version 0.4 of the Python implementation of LA-Cosmic cosmics.py\footnote{Python implementation of LA-Cosmic \citep{dokkum2001} by Malte Tewes: \url{https://web.archive.org/web/20180624102811/http://obswww.unige.ch/~tewes/cosmics_dot_py/}}. The H and K~bands were reduced separately. The A0V standard star spectra were reduced the same way as the targets, except that optimal extraction \citep{horne86} was used to extract their 1D spectra.  

We used the custom python software plotspec\footnote{Plotspec: \url{https://github.com/kfkaplan/plotspec}}, which is designed for analyzing nebular IGRINS spectra, to process the data. Continuum emission from each nebula and central star was subtracted with a running median filter. This simplifies the scaling and merging of orders, and does not affect flux measurements. The telluric correction and relative spectrophotometric flux calibration was carried out by dividing the observed A0V standard star spectrum by an assumed synthetic A0V spectrum \citep[this technique is described in detail in][]{Kaplan_Etal_2017}. The result is an empirically telluric-corrected and relative flux-calibrated target spectrum. In cases where telluric subtraction is imperfect, the high-resolution of IGRINS allows for absorption and sky lines to be easily recognized. These features can be removed from nebular lines via Gaussian fitting, provided that the contamination is not severe. Finally, we combined the individual echelle orders into a single spectrum that spans the H and K bands. We correct for the extinction toward each target using the $c_{{\rm H}\beta}$ values from \citet{Leisy_Dennefeld_2006}, assuming the extinction law from \citet{reike85}.

\subsection{Line Measurements and Identifications} \label{sec:ids}

The IGRINS spectra reveal a rich array of nebular features (Table~\ref{table:lineids}), with 50--180 distinct emission lines detected in each LMC PN, and 40--120 in SMC targets other than SMC~3, in which only 14 lines were detected. For the targets that were previously observed by \citet{Mashburn_Etal_2016}, we detect 2--3 times the number of H and K~band emission lines, including new detections of Kr and Te in LMC~73; Rb, Cd, and Te in LMC~99; and of any \emph{n}-capture element (Se, Kr, and Te) in SMC~20. We also detect 50 H$_2$ emission lines in LMC~47, compared to 18 by Mashburn et al.

Fluxes were measured using custom IDL routines that perform a Riemann sum above a user-defined local continuum, without any assumptions regarding the shape of the line profile. In cases of blended nebular lines or those contaminated by telluric features or cosmic rays, we performed multi-component Gaussian fits to each component to determine the fluxes of the individual lines. We estimated systematic uncertainties in the fluxes by measuring each line three times, using different continuum placements within the envelope of continuum noise. These uncertainties were added in quadrature to the statistical uncertainties to produce the error bars to the line fluxes reported in Table~\ref{table:lineids}, which is available in its entirety as a supplemental file. In this table, the fluxes are given relative to an \ion{H}{1} reference line (11-4~1.6811~$\mu$m for the H~band, and Br$\gamma$~2.1661~$\mu$m for the K~band) with the flux of the \ion{H}{1} reference set to 100. The observed wavelengths are not corrected for the systemic velocity of the LMC/SMC. The reported fluxes are corrected for interstellar reddening, though this effect is small given the low extinction toward the Magellanic Clouds \citep[$c_{{\rm H}\beta} < 0.8$ for our targets;][]{Leisy_Dennefeld_2006} and the small differential extinction across the H and K~bands. As an example, in the most highly reddened target (LMC~31, with $c_{{\rm H}\beta} = 0.79$), the extinction correction is less than 7\% across the H~band relative to the \ion{H}{1} reference line, and $<4$\% in the K~band. The correction is less than 2.5\% for all \emph{n}-capture elements transitions relative to the \ion{H}{1} reference lines in this object.

\begin{deluxetable*}{lcccc}
\tablecaption{Line Identifications and Fluxes}
\tabletypesize{\footnotesize}
\tablewidth{0pt}
\tablehead{ \\[-1.4pc]
\colhead{Line ID} & \colhead{$\lambda_{\rm obs}$ ($\mu$m)} & \colhead{$\lambda_{\rm rest}$ ($\mu$m)} & \colhead{100$\times F/F$(\ion{H}{1})\tablenotemark{\footnotesize{a}}} & \colhead{Comments\tablenotemark{\footnotesize{b}}}
}
\startdata
\multicolumn{5}{c}{\textbf{LMC 29}} \\ 
\hline
\ion{He}{2} & 1.4776 & 1.4764 & 1.93E+02 $\pm$ 3.20E+01 & Fit \\
\ion{He}{2} & 1.4898 & 1.4886 & 3.33E+01 $\pm$ 4.41E+00 & Fit \\
\ion{He}{1} & 1.4942 & 1.4933 & 8.54E+00 $\pm$ 1.96E+00 & \\
\ion{H}{1} & 1.4954 & 1.4942 & 4.08E+00 $\pm$ 1.26E+00 & \\
\ion{H}{1} & 1.4984 & 1.4971 & 5.28E+00 $\pm$ 9.01E-01 & \\
~~~~\vdots & \vdots & \vdots & \vdots & \vdots \\
\enddata
\tablecomments{This table is available in its entirety in machine-readable form. A portion is reproduced here to illustrate its format.}
\tablenotetext{a}{Fluxes on the scale of $F$(\ion{H}{1})~=~100. \ion{H}{1} 11-4 is the reference line for the H band, and Br$\gamma$ for the K band. Upper limits for undetected \emph{n}-capture element lines are 3-$\sigma$ estimates computed from the rms noise of the continuum, assuming a typical line width of $\sim$50~km~s$^{-1}$.}
\tablenotetext{b}{Marginal detections are labeled with a colon, and lines whose fluxes were determined via multi-component Gaussian fits are indicated, as their uncertainties can be slightly larger than isolated features of comparable strength.}
\label{table:lineids}
\end{deluxetable*}

To identify the detected transitions, we utilized the Atomic Line List v3.00b4\footnote{\url{https://www.pa.uky.edu/~peter/newpage/}} 
\citep{VanHoof_2018} and the H$_2$ transition wavelengths of \citet{Roueff_Etal_2019}, as well as the identifications of \citet{Mashburn_Etal_2016}. In cases where the emitting ion was not apparent, we searched for transitions within 5~\AA\ of the estimated rest wavelength, and vetted each potential identification by searching for multiplet members and/or lines from the same upper level in the IGRINS spectral range. Tentative identifications that lack such supporting evidence are labeled as ``Uncertain ID" in Table~\ref{table:lineids}.

\begin{figure*}
    \centering
    \includegraphics[scale=0.75]{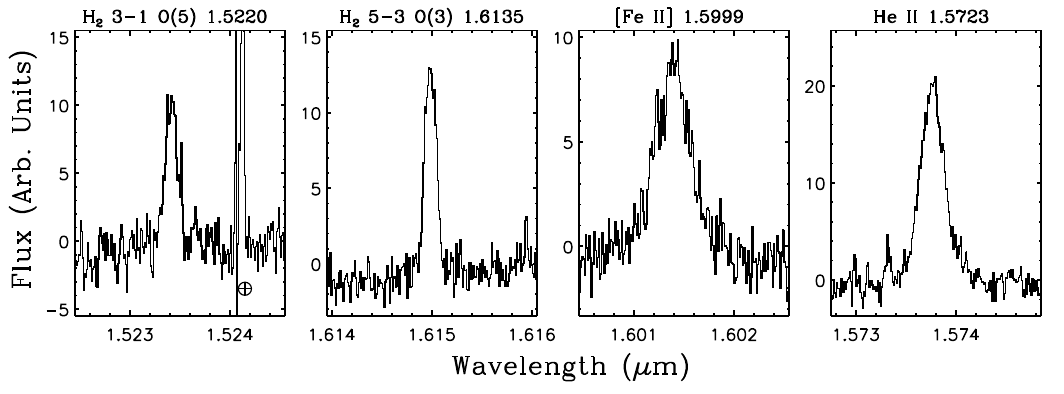}
    \caption{Profiles of selected [\ion{Fe}{2}], He, and vibrationally-excited H$_2$ lines in LMC~47. The H$_2$ features are narrower than atomic lines in this PN, aiding in their identification. Sky line residuals are indicated by circled plus signs.}
    \label{fig:feh2} 
\end{figure*}

The resolving power of IGRINS proved a valuable aid in identifying features, which can have distinct profiles depending on the emitting species. As an example, in Figure~\ref{fig:feh2} we show the profiles of two vibrationally-excited H$_2$ lines in the H~band spectrum of LMC~47, along with [\ion{Fe}{2}]~1.5999 and \ion{He}{2}~1.5723~$\mu$m. The H$_2$ lines have distinctly narrower profiles than low-ionization forbidden lines or permitted atomic transitions in this nebula. This allowed molecular features to be more easily recognized, leading us to identify lines as H$_2$ (e.g., 3-1~O(5)~1.5220~$\mu$m) that were labeled as [\ion{Fe}{2}] in the lower-resolution spectra of \citet{Mashburn_Etal_2016}. LMC~31 also exhibits a rich H$_2$ spectrum with very narrow vibrationally-excited lines. The H$_2$ spectra in these two PNe will be analyzed in a future paper.

\begin{figure*}
    \centering
    \includegraphics[scale=0.75]{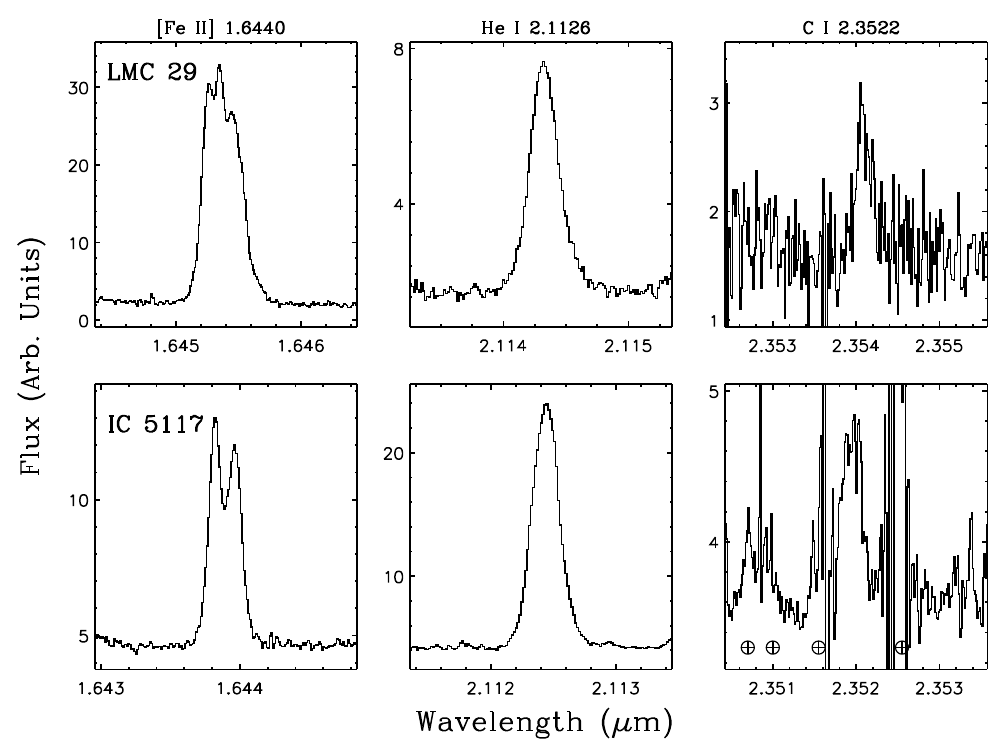}
    \caption{[\ion{Fe}{2}], \ion{He}{1}, and a feature at 2.3521~$\mu$m (rest wavelength) in LMC~29 and the Galactic PN IC~5117 \citep{Sterling_Etal_2016}. Telluric feature residuals are indicated with $\oplus$ symbols. The profile of the 2.3521~$\mu$m feature is more similar to permitted features such as \ion{He}{1} than low-ionization collisionally-excited lines, implying that it is unlikely to be due to [\ion{Sn}{2}]. We tentatively identify this line as \ion{C}{1}~2.3522~$\mu$m.}
    \label{fig:sn2_plots}
\end{figure*}

To further highlight the importance of high spectral resolution \citep[see also][]{Garcia-Rojas_Etal_2015}, the resolved line profiles allowed us to eliminate a potential \emph{n}-capture element identification, [\ion{Sn}{2}]~2.3521~$\mu$m, for a feature observed near 2.3541~$\mu$m in LMC~29, LMC~31, and LMC~47. This line is shown in Figure~\ref{fig:sn2_plots} for LMC~29, along with [\ion{Fe}{2}]~1.6440 and \ion{He}{1}~2.1126~$\mu$m. The profile is narrower than $[$\ion{Fe}{2}$]$ lines and resembles a permitted line, suggesting that this line does not arise from [\ion{Sn}{2}]. To determine whether the lower S/N of this feature (8.5) may mask the profile shape, we also reprocessed the IGRINS spectrum of the Galactic PN IC~5117 \citep{Sterling_Etal_2016} with the PLP~v2.2.0 and show the same three lines for this PN in the lower panels of the figure. In IC~5117, [\ion{Fe}{2}] is clearly double-peaked, while the 2.3521~$\mu$m line is singly-peaked like \ion{He}{1}. We conclude that this line is unlikely to be $[$\ion{Sn}{2}], and tentatively identify it as \ion{C}{1}~2.3522~$\mu$m, but lines from the same upper level at 1.47--1.50~$\mu$m are either undetected or are blended with \ion{H}{1} features.

\begin{figure*}
    \centering
    \includegraphics[scale=0.85]{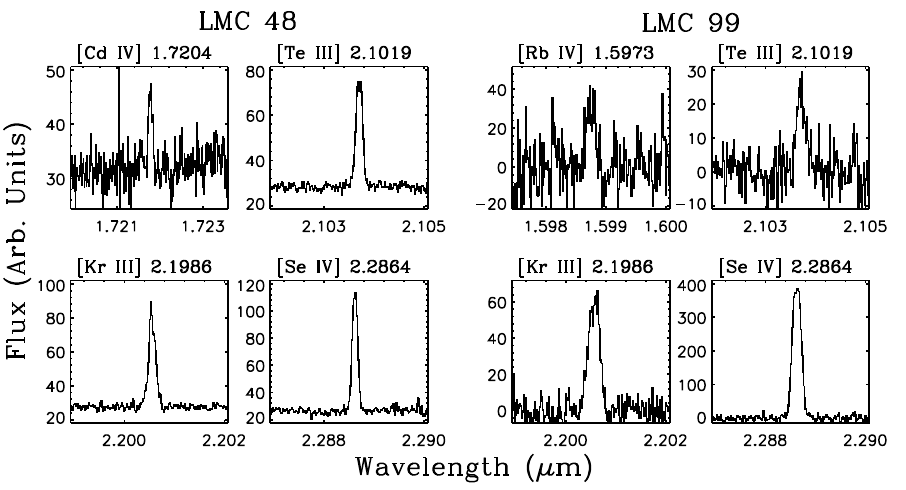}
    \caption{Neutron-capture element detections in LMC~48 and LMC~99.}
    \label{fig:lmc48_99} 
\end{figure*}

\begin{figure*}
    \centering
    \includegraphics[scale=0.75]{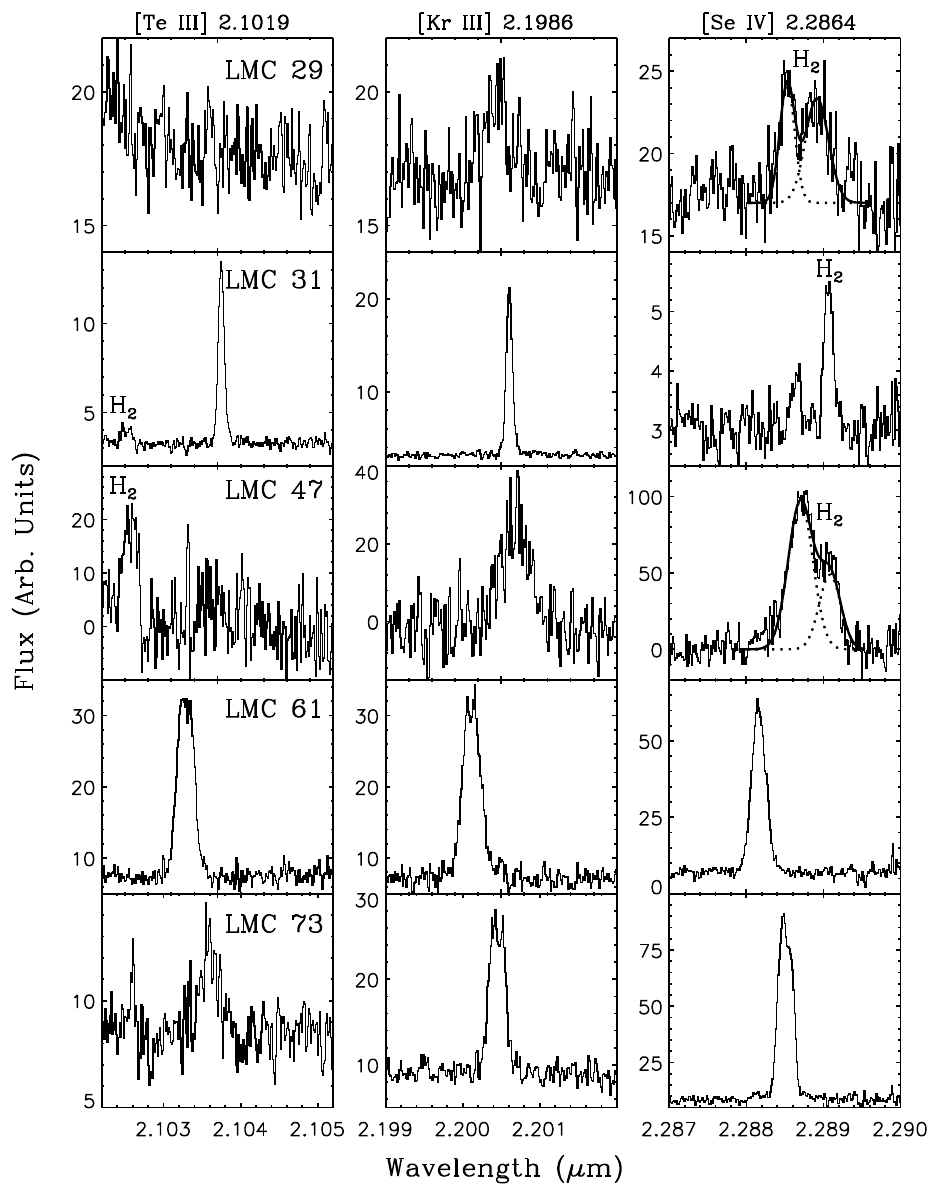}
    \caption{[\ion{Te}{3}], [\ion{Kr}{3}], and [\ion{Se}{4}] lines in LMC~29 (top panels) through LMC~73 (bottom panels). [\ion{Se}{4}]~2.2864~$\mu$m is blended with H$_2$~3-2~S(2)~2.2870~$\mu$m in LMC~29 and LMC~47, and we show as dotted lines components of the Gaussian fit used to deblend these features. Te is not detected in LMC~29 or LMC~47.}
    \label{fig:lmc_stack} 
\end{figure*}

\begin{figure*}
    \centering
    \includegraphics[scale=0.75]{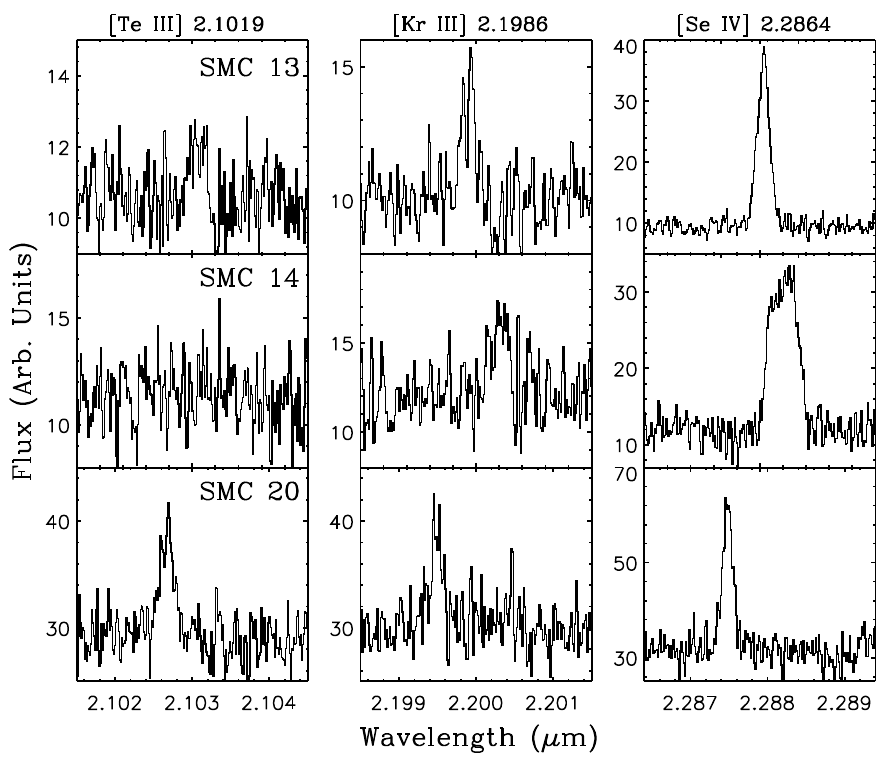}
    \caption{The same as Figure~\ref{fig:lmc_stack}, for three SMC~PNe.  No \emph{n}-capture element transitions were detected in SMC~3 or SMC~25.}
    \label{fig:smc_stack} 
\end{figure*}

We detect \emph{n}-capture elements in all of our targets except SMC~3 and SMC~25. [\ion{Kr}{3}]~2.1986 and [\ion{Se}{4}]~2.2864~$\mu$m are both detected in 10~PNe, and [\ion{Te}{3}]~2.1019~$\mu$m is observed in seven objects, often at remarkably high S/N. The high dispersion of IGRINS allowed us to deblend [\ion{Se}{4}] from H$_2$~3-2~S(2)~2.2870~$\mu$m in LMC~29 and LMC~47 via Gaussian fits, while in LMC~31 these two features are fully resolved due to the narrow line widths. LMC~99 exhibits the largest number of detected \emph{n}-capture elements, including [\ion{Rb}{4}]~1.5973 and (marginally) [\ion{Cd}{4}]~1.7204~$\mu$m in addition to Se, Kr, and Te, while [\ion{Cd}{4}] was also detected in LMC~48. To our knowledge, these are the first detections of Rb, Cd, and Te in extragalactic PNe. Figure~\ref{fig:lmc48_99} displays \emph{n}-capture element transitions in LMC~48 and LMC~99, while Figures~\ref{fig:lmc_stack} and \ref{fig:smc_stack} show [\ion{Te}{3}], [\ion{Kr}{3}], and [\ion{Se}{4}] in other LMC and SMC PNe, respectively, including non-detections of Te in LMC~29, LMC~47, and SMC~14. For those \emph{n}-capture element lines that were not detected, including [\ion{Br}{5}]~1.6429~$\mu$m, we report 3-$\sigma$ upper limits in Table~\ref{table:lineids}, estimated from the rms noise in the continuum.

The spectra also exhibit numerous other lines, including permitted transitions of C and O. In the Type~I PNe LMC~29 and SMC~25 we detect the high-ionization lines [\ion{Si}{6}]~1.9634 and [\ion{Ca}{8}]~2.3220~$\mu$m that have been observed in extreme Galactic Type~I PNe such as NGC~6302 and NGC~6537 \citep{Reconditi_Oliva_1993, Casassus_Etal_2000}. The presence of these features indicates a very hot central star \citep[$\gtrsim$~200~kK;][]{Casassus_Etal_2000}, as is expected for a PN from a higher-mass progenitor \citep{Blocker_1995, MillerBertolami_2016}.

\section{Abundance Determinations}\label{sec:abunds}

\subsection{Ionic Abundances} \label{sec:ionic}

We compute abundances with the PyNeb nebular analysis package \citep[v1.1.17;][]{Luridiana_Etal_2015_PyNeb, Morisset_Etal_2020}. To determine ionic abundances, we adopt the electron temperatures $T_{\rm e}$ and densities $n_{\rm e}$ reported by \citet{Shaw_Etal_2010} for SMC~13, SMC~14, and SMC~20, and those from \citet{Leisy_Dennefeld_2006} for other objects (see Table~\ref{tab:physcond}), with the exception of LMC~73 (see below). We recomputed the physical conditions of LMC~29, LMC~31, LMC~99, and SMC~14 to check that updated atomic data \citep[see Table~5 of][]{Garcia-Rojas_Etal_2015} do not significantly affect the derived $T_{\rm e}$ or $n_{\rm e}$, and find temperatures that agree with those from \citet{Leisy_Dennefeld_2006} and \citet{Shaw_Etal_2010} to within 500~K for all cases and densities that agree within 10\%. Based on this analysis, we conservatively estimate a 20\% uncertainty in $n_{\rm e}$, and $\pm$1000~K for $T_{\rm e}$ except for the SMC PNe studied by \citet{Shaw_Etal_2010}, who provided error bars to their $T_{\rm e}$ values. The adopted electron densities are from the low-ionization diagnostic lines [\ion{S}{2}]~$\lambda\lambda$6716, 6731 \citep{Leisy_Dennefeld_2006} or [\ion{O}{2}]~$\lambda\lambda$3726, 3729 \citep{Shaw_Etal_2010}. However, in the high-excitation PN SMC~14 \citet{Shaw_Etal_2010} also computed $n_{\rm e}$ from [\ion{Ar}{4}]~$\lambda$4711/4740. The [\ion{S}{2}] diagnostic may underestimate $n_{\rm e}$ \citep{Mendez-Delgado_Etal_2023}, impacting ionic abundances calculated from infrared lines. Additionally, the $[$\ion{O}{2}$]$ and $[$\ion{S}{2}$]$ diagnostics saturate at $n_{\rm e} \gtrsim 10^4$~cm$^{-3}$, so uncertainties in $n_{\rm e}$ are larger for higher-density nebulae. However, these are the only density diagnostics available for most of the targets in our sample. This demonstrates the need for higher-quality optical data for Magellanic Cloud PNe, as further discussed in \S\ref{sec:optical}. 

\begin{deluxetable*}{lcccc}
\tablecaption{Adopted Physical Conditions}
\tabletypesize{\footnotesize}
\tablewidth{0pt}
\tablehead{ \\[-1.4pc]
\colhead{PN} & \colhead{T$_{\rm e}$($[$\ion{N}{2}$]$ (K)} & \colhead{T$_{\rm e}$($[$\ion{O}{3}$]$) (K)} & \colhead{$n_{\rm e}$($[$\ion{S}{2}$]$) or $[$\ion{O}{2}$]$)\tablenotemark{a} (cm$^{-3}$)} & \colhead{$n_{\rm e}$($[$\ion{Ar}{4}$]$) (cm$^{-3}$)}
}
\startdata
LMC 29 & (19,800$\pm$1000) & 19,800$\pm$1000 & 4100$\pm$820 & (4100$\pm$820) \\
LMC 31 & 16,100$\pm$1000 & 14,000$\pm$1000 & 6600$\pm$1320 & (6600$\pm$1320) \\
LMC 47 & 13,400$\pm$1000 & 14,700$\pm$1000 & 4800$\pm$960 & (4800$\pm$960) \\
LMC 48 & 8800$\pm$1000 & 11,300$\pm$1000 & 1900$\pm$380 & (4100$\pm$380) \\
LMC 61 & 10,200$\pm$1000 & 11,100$\pm$1000 & 36,200$\pm$7240 & (36,200$\pm$7240) \\
LMC 73 & 11,580$\pm$540 & 12,140$\pm$150 & 9800$^{+5500}_{-3530}$ & 11,160$^{+1000}_{-910}$ \\
LMC 99 & 13,200$\pm$1000 & 12,200$\pm$1000 & 3600$\pm$720 & (3600$\pm$720) \\
SMC 3 & (13,600$\pm$1000) & 13,600$\pm$1000 & 21,500$\pm$4300 & (21,500$\pm$4300) \\
SMC 13 & (12,800$^{+1080}_{-940}$) & 12,800$^{+1080}_{-940}$ & 2900$\pm$580 & (2900$\pm$580) \\
SMC 14 & (13,100$^{+1090}_{-990}$) & 13,100$^{+1090}_{-990}$ & 2230$\pm$446 & 3890$\pm$778 \\
SMC 20 & (13,820$^{+1380}_{-1010}$) & 13,820$^{+1380}_{-1010}$ & 3900$\pm$780 & (3900$\pm$780) \\
SMC 25 & 21,100$\pm$1000 & 21,100$\pm$1000 & 9800$\pm$1960 & (9800$\pm$1960) \\
\enddata
\tablecomments{Physical conditions \citep[taken from][Mampaso et al., in prep.]{Leisy_Dennefeld_2006, Shaw_Etal_2010} used for NIR and optical (see Appendix) abundance calculations. When some lines in a diagnostic ratio are not observed, the value from the other diagnostic ratio is shown in parentheses. We adopt uncertainties of $\pm$1000~K in $T_{\rm e}$ values except for SMC~13, SMC~14, and SMC~20 \citep[error bars from][]{Shaw_Etal_2010}, and 20\% in $n_{\rm e}$ (see discussion in \S\ref{sec:ionic}). For LMC~73, error bars on both $T_{\rm e}$ and $n_{\rm e}$ are from Mampaso et al., in prep.}
\tablenotetext{a}{Electron densities are from the $[$\ion{O}{2}$]$~$\lambda$3726/$\lambda$3729 ratio for SMC~13, SMC~14, and SMC~20 \citep{Shaw_Etal_2010}, and $[$\ion{S}{2}$]$~$\lambda$6716/$\lambda$6731 for all other PNe.}
\label{tab:physcond}
\end{deluxetable*}

LMC~73 was recently observed in the optical with FORS2 on the Very Large Telescope. For this object, we utilize the physical conditions determined from that spectrum (Mampaso et al., in prep.; see Table~\ref{tab:physcond}). Further details of the FORS2 spectrum and its analysis, including the light element abundances that we utilize to compute \emph{n}-capture element abundances, will be given in Mampaso et al.\ (in prep). The only difference in the analysis of Mampaso et al.\ compared to our light element abundance determinations (see Appendix) are in the atomic data adopted for Ne$^{2+}$. Those authors use the calculations of \citet{McLaughlin_Etal_2011}, whereas we utilize the \citet{Galavis_Etal_1997} transition probabilities and \citet{McLaughlin_Bell_2000} effective collision strengths for this ion. This has no effect on our analysis of \emph{n}-capture element abundances.

We assume a two-zone model for ionic abundance calculations, with $T_{\rm e}$([\ion{N}{2}]) used for low-ionization species (ionization potential $<39$~eV) such as Kr$^{2+}$ and Te$^{2+}$, and the [\ion{O}{3}] temperature for higher ions. For LMC~73 and SMC~14, we use the [\ion{Ar}{4}] density for the high-ionization zone. Although \citet{Shaw_Etal_2010} note that their $[$\ion{Ar}{4}$]$ densities are unreliable due to the weakness of the lines, this diagnostic is the most appropriate one for the high ionization zone of SMC~14, and adopting it affects the abundances by less than 10\% compared to when the $[$\ion{O}{2}$]$ density is used for all ions. References for the transition probabilities and effective collision strengths used for \emph{n}-capture element ions are provided in Table~\ref{tab:atomic_data}, while those for optical diagnostic lines and other light element ions are the same as in \citet[][see their Table~5]{Garcia-Rojas_Etal_2015}.

\begin{deluxetable*}{lcc}
\tablecaption{Atomic Data for Neutron-Capture Elements}
\tabletypesize{\footnotesize}
\tablewidth{0pt}
\tablehead{ \\[-1.4pc]
\colhead{Ion} & \colhead{Transition Probabilities} & \colhead{Collision Strengths} 
}
\startdata
Se$^{3+}$ & \citet{Biemont_Hansen_1987} & K.\ Butler 2007, priv.\ comm. \\
Br$^{4+}$ & \citet{Madonna_Etal_2018} & \citet{Madonna_Etal_2018} \\
Kr$^{2+}$ & \citet{Biemont_Hansen_1986a} & \citet{Schoning_1997_Kr} \\
Rb$^{3+}$ & \citet{Sterling_Etal_2016} & \citet{Sterling_Etal_2016} \\
Cd$^{3+}$\tablenotemark{a} & \citet{Biemont_Hansen_1987} & K.\ Butler 2007, priv.\ comm. \\
Te$^{2+}$ & \citet{Madonna_Etal_2018} & \citet{Madonna_Etal_2018} \\
\enddata
\tablenotemark{a}{Following \citet{Sterling_Etal_2016}, we adopt the Ge$^{5+}$ transition probability and Se$^{3+}$ collision strength for Cd$^{3+}$.}
\label{tab:atomic_data}
\end{deluxetable*}

We report ionic abundances (or 3-$\sigma$ upper limits) of \emph{n}-capture elements for each PN in Table~\ref{tab:ncap_abunds}. The error bars include uncertainties in the line fluxes and physical conditions, propagated via Monte Carlo simulations. Specifically, for each line 1500 random values of the flux were drawn from a Gaussian distribution centered on the measured flux, with FWHM equal to the error bar given in Table~\ref{table:lineids}. In a similar manner, $T_{\rm e}$ and $n_{\rm e}$ were sampled. Ionic abundances were determined from each value, and the standard deviation is taken as the error bar in the abundance. A larger number of Monte Carlo simulations has a negligible affect on the reported uncertainties \citep{Garcia-Rojas_Etal_2015}. Upper limits to the Br$^{4+}$/H$^+$ abundance are given only for high-excitation PNe that exhibit \ion{He}{2} emission, as Br$^{4+}$ is negligibly populated in objects with cooler central stars. 

\startlongtable
\begin{deluxetable*}{cccc|cccc}
\tablecolumns{9}
\tablewidth{0pc}
\tabletypesize{\scriptsize}
\tablecaption{Neutron-Capture Element Ionic and Total Abundances}
\tablehead{ \\[-0.6pc]
\colhead{Ion} & \colhead{X$^{i+}$/H$^+$} & \colhead{ICF} & \multicolumn{1}{c|}{$12+\log{\mathrm{(X/H)}}$} & \hspace{2pc} & \colhead{X$^{i+}$/H$^+$} & \colhead{ICF} & \colhead{$12+\log{\mathrm{(X/H)}}$}
}
\startdata
\multicolumn{1}{c}{} & \multicolumn{3}{c|}{\textbf{LMC SMP 29}} & \hspace{2pc} & \multicolumn{3}{c}{\textbf{LMC SMP 31}} \\
\hline
Se$^{3+}$ & (9.90 $\pm$ 2.25)E--11 & 2.80 $\pm$ 1.02 & 2.44 $\pm$ 0.16 & \hspace{2pc} & (9.24 $\pm$ 1.48)E--11 & 3.60 $\pm$ 0.77 & 2.52 $\pm$ 0.10 \\
Br$^{4+}$ & $<3.23$E--10 & 1.42 $\pm$ 0.83 & $<2.66$ & \hspace{2pc} & \nodata & \nodata & \nodata \\
Kr$^{2+}$ & (1.26 $\pm$ 0.31)E--10 & 5.22 $\pm$ 1.81 & 2.82 $\pm$ 0.15 & \hspace{2pc} & (1.92 $\pm$ 0.10)E--09 & 1.56 $\pm$ 0.37 & 3.48 $\pm$ 0.09 \\
Rb$^{3+}$ & $<3.76$E--10 & 1.71 $\pm$ 0.49 & $<2.81$ & \hspace{2pc} & $<1.14$E--09 & 2.09 $\pm$ 0.65 & $<3.38$ \\
Cd$^{3+}$ & $<2.98$E--10 & 1.71 $\pm$ 0.49 & $<2.71$ & \hspace{2pc} & $<4.27$E--10 & 2.09 $\pm$ 0.65 & $<2.95$ \\
Te$^{2+}$ & $<3.28$E--11 & 6.09 $\pm$ 4.45 & $<2.30$ & \hspace{2pc} & (1.59 $\pm$ 0.13)E--09 & 1.60 $\pm$ 2.20 & 2.41 $\pm$ 0.38 \\
\hline
\multicolumn{1}{c}{} & \multicolumn{3}{c|}{\textbf{LMC SMP 47}} & \hspace{2pc} & \multicolumn{3}{c}{\textbf{LMC SMP 48}} \\
\hline
Se$^{3+}$ & (2.09 $\pm$ 0.32)E--10 & 2.10 $\pm$ 1.15 & 2.64 $\pm$ 0.20 & \hspace{2pc} & (3.73 $\pm$ 0.35)E--10 & 1.81 $\pm$ 4.46 & 2.83 $\pm$ 0.54 \\
Br$^{4+}$ & $<2.24$E--10 & 2.54 $\pm$ 1.91 & $<2.76$ & \hspace{2pc} & \nodata & \nodata & \nodata \\
Kr$^{2+}$ & (7.13 $\pm$ 1.38)E--10 & 3.65 $\pm$ 1.82 & 3.44 $\pm$ 0.19 & \hspace{2pc} & (1.80 $\pm$ 0.14)E--09 & 1.47 $\pm$ 0.51 & 3.44 $\pm$ 0.13 \\
Rb$^{3+}$ & $<7.41$E--10 & 1.39 $\pm$ 0.49 & $<3.01$ & \hspace{2pc} & $<8.76$E--10 & 1.24 $\pm$ 0.86 & $<3.04$ \\
Cd$^{3+}$ & $<3.51$E--10 & 1.39 $\pm$ 0.49 & $<2.69$ & \hspace{2pc} & (1.23 $\pm$ 0.28)E--10 & 1.24 $\pm$ 0.86 & 2.18 $\pm$ 0.24 \\
Te$^{2+}$ & $<4.91$E--11 & 4.59 $\pm$ 3.53 & $<2.35$ & \hspace{2pc} & (1.92 $\pm$ 0.12)E--10 & 3.04 $\pm$ 3.13 & 2.77 $\pm$ 0.31 \\
\hline
\multicolumn{1}{c}{} & \multicolumn{3}{c|}{\textbf{LMC SMP 61}} & \hspace{2pc} & \multicolumn{3}{c}{\textbf{LMC SMP 73}} \\
\hline
Se$^{3+}$ & (1.14 $\pm$ 0.13)E--09 & 4.52 $\pm$ 4.83 & 3.71 $\pm$ 0.32 & \hspace{2pc} & (1.81 $\pm$ 0.06)E--09 & 2.08 $\pm$ 0.49 & 3.58 $\pm$ 0.09\\
Br$^{4+}$ & $<3.51$E--10 & 3.13 $\pm$ 3.03 & $<3.04$ & \hspace{2pc} & $<2.05$E--09 & 1.63$\pm$0.98 & $<3.31$ \\
Kr$^{2+}$ & (2.90 $\pm$ 0.28)E--09 & 2.40 $\pm$ 2.01 & 3.84 $\pm$ 0.27 & \hspace{2pc} & (2.15 $\pm$ 0.16)E--09 & 1.26 $\pm$ 0.18\tablenotemark{a} & 4.01 $\pm$ 0.06\tablenotemark{a} \\
Rb$^{3+}$ & $<2.30$E--09 & 2.54 $\pm$ 1.64 & $<3.77$ & \hspace{2pc} & $<2.81$E--09 & 1.37 $\pm$ 0.24 & $<3.59$ \\
Cd$^{3+}$ & $<6.23$E--10 & 2.54 $\pm$ 1.64 & $<3.20$ & \hspace{2pc} & $<1.09$E--09 & 1.37 $\pm$ 0.24 & $<3.18$ \\
Te$^{2+}$ & (3.96 $\pm$ 0.33)E--10 & 1.96 $\pm$ 2.73 & 2.89 $\pm$ 0.38 & \hspace{2pc} & (6.26 $\pm$ 1.10)E--11 & 4.13 $\pm$ 2.98 & 2.41 $\pm$ 0.25 \\
\hline
\multicolumn{1}{c}{} & \multicolumn{3}{c|}{\textbf{LMC SMP 99}} & \hspace{2pc} & \multicolumn{3}{c}{\textbf{SMC SMP 3}} \\
\hline
Se$^{3+}$ & (1.93 $\pm$ 0.20)E--09 & 1.98 $\pm$ 3.27 & 3.58 $\pm$ 0.45 & \hspace{2pc} & $<6.73$E--10 & 1.55 $\pm$ 0.71 & $<3.02$ \\
Br$^{4+}$ & $<3.34$E--10 & 17.1 $\pm$ 15.4 & $<3.76$ & \hspace{2pc} & $<2.36$E--09 & \nodata & \nodata \\
Kr$^{2+}$ & (1.65 $\pm$ 0.12)E--09 & 3.30 $\pm$ 1.95 & 3.74 $\pm$ 0.21 & \hspace{2pc} & $<3.18$E--09 & 1.62 $\pm$ 2.08 & $<3.71$ \\
Rb$^{3+}$ & (6.31 $\pm$ 0.78)E--10 & 1.32 $\pm$ 0.79 & 2.92 $\pm$ 0.21 & \hspace{2pc} & $<1.75$E--08 & 1.11 $\pm$ 0.30 & $<4.29$ \\
Cd$^{3+}$ & (1.10 $\pm$ 0.37)E--10 & 1.32 $\pm$ 0.79 & 2.16 $\pm$ 0.23 & \hspace{2pc} & $<4.40$E--09 & 1.11 $\pm$ 0.30 & $<3.69$ \\
Te$^{2+}$ & (6.10 $\pm$ 0.86)E--11 & 4.56 $\pm$ 3.96 & 2.44 $\pm$ 0.28 & \hspace{2pc} & $<4.88$E--10 & 3.65 $\pm$ 2.91 & $<3.25$ \\
\hline
\multicolumn{1}{c}{} & \multicolumn{3}{c|}{\textbf{SMC SMP 13}} & \hspace{2pc} & \multicolumn{3}{c}{\textbf{SMC SMP 14}} \\
\hline
Se$^{3+}$ & (3.85 $\pm$ 0.44)E--10 & 1.43 $\pm$ 1.58 & 2.74 $\pm$ 0.33 & \hspace{2pc} & (1.12 $\pm$ 0.14)E--09 & 2.01 $\pm$ 2.06 & 3.35 $\pm$ 0.31 \\
Br$^{4+}$ & $<2.27$E--10 & \nodata & \nodata & \hspace{2pc} & $<7.04$E--10 & 1.79 $\pm$ 1.48 & $<3.10$ \\
Kr$^{2+}$ & (2.60 $\pm$ 0.41)E--10 & 1.62 $\pm$ 2.56 & 2.62 $\pm$ 0.41 & \hspace{2pc} & (6.76 $\pm$ 1.12)E--10 & 4.00 $\pm$ 2.25 & 3.44 $\pm$ 0.20 \\
Rb$^{3+}$ & $<1.13$E--09 & 1.06 $\pm$ 0.44 & $<3.08$ & \hspace{2pc} & $<4.43$E--09 & 1.34 $\pm$ 0.63 & $<3.78$ \\
Cd$^{3+}$ & $<3.90$E--10 & 1.06 $\pm$ 0.44 & $<2.61$ & \hspace{2pc} & $<1.45$E--09 & 1.34 $\pm$ 0.63 & $<3.29$ \\
Te$^{2+}$ & (1.63 $\pm$ 0.43)E--11 & 4.04 $\pm$ 5.66 & 1.82 $\pm$ 0.39 & \hspace{2pc} & $<$1.24E--10 & 4.94 $\pm$ 4.06 & $<2.87$ \\
\hline
\multicolumn{1}{c}{} & \multicolumn{3}{c|}{\textbf{SMC SMP 20}} & \hspace{2pc} & \multicolumn{3}{c}{\textbf{SMC SMP 25}} \\
\hline
Se$^{3+}$ & (6.83 $\pm$ 0.86)E--11 & 1.40 $\pm$ 1.23 & 1.98 $\pm$ 0.28 & \hspace{2pc} & $<3.30$E--10 & 2.37 $\pm$ 1.04 & $<2.89$ \\
Br$^{4+}$ & \nodata & \nodata & \nodata & \hspace{2pc} & $<5.26$E--10 & \nodata & \nodata \\
Kr$^{2+}$ & (8.08 $\pm$ 2.29)E--11 & 1.46 $\pm$ 1.61 & 2.07 $\pm$ 0.33 & \hspace{2pc} & $<1.78$E--09 & 3.47 $\pm$ 1.78 & $<3.79$ \\
Rb$^{3+}$ & $<4.44$E--10 & 1.04 $\pm$ 0.43 & $<2.66$ & \hspace{2pc} & $<5.25$E--09 & 1.51 $\pm$ 0.47 & $<3.90$ \\
Cd$^{3+}$ & $<3.34$E--10 & 1.04 $\pm$ 0.43 & $<2.54$ & \hspace{2pc} & $<2.10$E--09 & 1.51 $\pm$ 0.47 & $<3.50$ \\
Te$^{2+}$ & (1.89 $\pm$ 0.27)E--11 & 4.20 $\pm$ 5.63 & 1.90 $\pm$ 0.37 & \hspace{2pc} & $<1.93$E--10 & 5.23 $\pm$ 4.05 & $<3.00$ \\
\enddata
\tablecomments{ICFs calculated from the formulae of \citet{Sterling_Etal_2015} for Se and Kr. We adopt the Se ICF Eq.~7 from \citet{Sterling_Etal_2015} for Te, based on the similarities in the Se$^{2+}$ and Te$^{2+}$ ionization potential ranges. Following \citet{Sterling_Etal_2016} and \citet{Madonna_Etal_2018}, we utilize the approximate corrections ICF(Br)~=~Ar/Ar$^{3+}$ and ICF(Rb, Cd)~=~O/O$^{2+}$ based on similarities in ionization potential ranges. Upper limits to the Br$^{4+}$/H$^+$ and Br/H abundance are given only for high-excitation PNe, i.e.\ those exhibiting \ion{He}{2}~2.1891~$\mu$m. No Br ICF or elemental Br abundance is listed if [\ion{Ar}{4}] was not detected in published optical spectra \citep{Leisy_Dennefeld_2006, Shaw_Etal_2010}.}
\tablenotetext{a}{The elemental Kr abundance in LMC~73 is calculated from Kr$^{2+}$/H$^+$ (this work) and Kr$^{3+}$/H$^+=(6.43\pm0.43)\times10^{-9}$ (Mampaso et al., in prep.), using Eq.\ 4 from \citet{Sterling_Etal_2015} as the ICF.}
\label{tab:ncap_abunds}
\end{deluxetable*}

\subsection{Elemental Abundances} \label{sec:elemental}

To convert ionic abundances to elemental abundances, it is necessary to estimate the fraction of a given element that resides in the detected ions. The most accurate corrections for unobserved ionization states (``ionization correction factors," or ICFs) are obtained from numerical simulations tailored to an individual object that reproduce the observed line intensities and ionization equilibrium of each element. In practice, it is common to use analytic ICF formulae derived from large grids of photoionization models that encompass the central star properties and nebular conditions encountered in PNe. Such model-based ICF schemes have been determined for light elements \citep[e.g.,][]{Delgado-Inglada_Etal_2014, Amayo_Etal_2020}. However, the accuracy of the modeled ionization balance solutions, and thus ionization corrections, rely on the availability of accurate atomic data for processes that affect the ionization equilibrium, namely photoionization cross sections and rate coefficients for recombination and charge transfer. For \emph{n}-capture elements, photoionization and recombination data have been determined for Se and Kr \citep{Sterling_Witthoeft_2011_Se, Sterling_2011_Kr, Sterling_Stancil_2011_CT}, but are incomplete or unknown for other \emph{n}-capture elements. Recently, \citet{Banerjee_Etal_2025} computed photoionization and recombination data for the first four charge states of Rb and four other heavy elements, and \citet{Singh_Etal_2025} calculated dielectronic recombination data for four \emph{n}-capture element ions, including Te$^{2+}$. However, ionization equilibrium models have not yet been computed for these elements. \citet{Sterling_Etal_2015} incorporated the Se and Kr atomic data into a test branch of Cloudy \citep{Ferland_Etal_2013, Chatzikos_Etal_2023} to derive ICF formulae for these elements from grids of models. Other \emph{n}-capture elements lack detailed atomic data, and ICFs must be approximated by using, as proxies, commonly-detected ions with ionization potential ranges similar  to those of the detected ions. 

The ICF formulae for \emph{n}-capture elements \citep{Sterling_Etal_2015, Sterling_Etal_2016, Madonna_Etal_2018} utilize the ionic fractions of light elements such as O, S, and Ar derived from optical spectra. \citet{Leisy_Dennefeld_2006} provide only elemental abundances, and therefore we computed ionic abundances from their reported line intensities. Elemental abundances from the optical data were then recalculated using the ICF schema of \citet{Delgado-Inglada_Etal_2014}. We present the recalculated light element abundances in Appendix~A. 

We use the ICF formulae of \citet{Sterling_Etal_2015} for Se and Kr, opting for their Eq.\ 1 as the Kr ICF instead of Eq.\ 2. The reason for this choice is that Eq.\ 2 relies on the Ar$^{2+}$/Ar fraction, and Ar abundances derived with the \citet{Delgado-Inglada_Etal_2014} ICF have relatively large uncertainties since they are computed from a single ion (Ar$^{2+}$). In the case of LMC~73, we use the Kr$^{3+}$/H$^+$ abundance of ($6.03\pm0.43$)$\times10^{-9}$ from Mampaso et al.\ (in prep.) in addition to our Kr$^{2+}$/H$^+$ value to determine the Kr abundance, with Eq.\ 4 of \citet{Sterling_Etal_2015}. To compute Se abundances from Se$^{3+}$/H$^+$, we applied Eq.\ 8 of \citet{Sterling_Etal_2015}. Eq.\ 7 of \citet[][for Se/Se$^{2+}$]{Sterling_Etal_2015} is adopted as the Te ICF, given the similarities of the ionization potential range of Te$^{2+}$ (18.6--27.8~eV) and Se$^{2+}$ (21.2--31.7~eV). Following \citet{Sterling_Etal_2016}, we approximate ICF(Rb, Cd) as O/O$^{2+}$, and use Ar/Ar$^{3+}$ as the Br ICF \citep{Madonna_Etal_2018}. The numerical ICFs and elemental abundances are given in Table~\ref{tab:ncap_abunds}. Uncertainties in ionic abundances and light element abundances were propagated via Monte Carlo simulations to estimate error bars for the ICFs and \emph{n}-capture element abundances.

When the detected ion is a minority species, the resulting elemental abundance may have larger systematic uncertainties due to the large ionization correction, and should be treated with caution. For example, the Se ionization equilibrium solutions of \citet{Sterling_Etal_2015} likely overestimate the Se$^{3+}$/Se ionic fraction -- and thus underestimate the Se abundance -- in low- and moderate-excitation PNe, as first suggested by \citet{Sterling_Etal_2017} and corroborated by additional observations of multiple Se ions in Galactic PNe \citep[][Sterling et al., in prep.]{Sterling_Etal_2025_AAS}. Thus the Se abundances in LMC~31 and LMC~48 in Table~\ref{tab:ncap_abunds} are likely underestimated. Similarly, Rb$^{3+}$ and Cd$^{3+}$ are minority species in these low-excitation PNe, while Te$^{2+}$ is expected to be a trace ion in high-ionization PNe (e.g., LMC~29, LMC~73, LMC~99, and SMC~14). In these objects, the Rb, Cd, and Te abundances may have larger systematic uncertainties than for the other PNe in our sample. In contrast, the Kr ICF schema are well-tested against observations of PNe in which multiple Kr ions have been detected \citep{Sterling_Etal_2015, Garcia-Rojas_Etal_2015, Madonna_Etal_2017}. In particular, the Kr/Kr$^{2+}$ ICFs \citep[Eqs.\ 1 and 2 of][]{Sterling_Etal_2015} yield Kr abundances in good agreement with those derived with ICFs that incorporate multiple ionization states.

\subsection{Elemental Enrichments} \label{sec:enrich} 

To determine whether a \emph{n}-capture element has been enriched by the \emph{s}-process in a PN, knowledge of the progenitor star's initial composition is needed. Specifically, the enhancement in the abundance must be calculated relative to a reference element that reliably indicates the star's initial metallicity.

In stellar abundance studies, Fe is generally  used as the indicator of overall metallicity. However, this is inappropriate for PNe, in which Fe is heavily depleted into dust \citep[e.g.,][]{Shields_1978, Perinotto_Etal_1999,Delgado-Inglada_Etal_2009, Delgado-Inglada_Rodriguez_2014}. Instead, the $\alpha$-elements O, S, and Ar, or Cl \citep[which tends to track these; e.g.,][]{Orte-Garcia_Etal_2025} are used as reference elements that indicate the metallicity of photoionized nebulae. Fortunately, abundance studies of cluster and field giants in the Magellanic Clouds indicate that [$\alpha$/Fe]~$\sim 0.0$ to within 0.1--0.2~dex, down to metallicities [Fe/H]~$=-1.5$~dex in the LMC and --1.3~dex in the SMC \citep{Lapenna_Etal_2012, Nidever_Etal_2020, Hasselquist_Etal_2021, Mucciarelli_Etal_2023}. This suggests that the abundances of these elements are representative of the initial [Fe/H] metallicities of the PN progenitor stars, which is valuable for making comparisons to evolutionary models that are indexed by their [Fe/H] values. 

The O abundance is usually the most accurately determined of the $\alpha$-elements in PNe. However, theory and observations indicate that O may be enriched by AGB nucleosynthesis \citep{Delgado-Inglada_Etal_2015}, particularly at low metallicity \citep{Pequignot_Etal_2000}, and destroyed by efficient HBB in the higher mass progenitors of Type~I PNe  \citep[e.g.,][]{Leisy_Dennefeld_2006, Ventura_Etal_2015_LMC}. It has been proposed that S abundances are systematically  underestimated in PNe \citep[the ``S anomaly," e.g.,][but see also \citealt{Tan_Parker_2024}]{Henry_Etal_2012}. For these reasons, \citet{Delgado-Inglada_Etal_2015} concluded that Cl is the best metallicity tracer for PNe, followed by Ar (which has larger uncertainties). Unfortunately, Cl has not been detected in any PN in our sample other than LMC~73 and LMC~99.

In Table~\ref{tab:ncap_enrich} we give the \emph{n}-capture elemental abundance ratios to H, O, S, and Ar in each object, relative to the solar composition \citep{Asplund_Etal_2021}. The [X/S] values are markedly higher than [X/(O, Ar)] in some objects, particularly in the SMC PNe, an outcome that we attribute to the S anomaly \citep[or systematic uncertainties in the S ICF;][]{Tan_Parker_2024}. The abundances relative to O and to Ar generally agree within the error bars, with the exceptions of LMC~29, LMC~31, and LMC~99, suggesting that both O and Ar are good metallicity indicators. We therefore choose O as the primary metallicity indicator for our sample, since its abundance is more accurately determined than that of Ar. As discussed by \citet{Leisy_Dennefeld_1996, Leisy_Dennefeld_2006}, the quality of published optical data (see \S\ref{sec:optical}) may affect the accuracy of the Ar abundances in LMC~31 and LMC~99. For example, in LMC~99 we derive an unusually large Ar abundance (2.5 times larger than solar) compared to O and Cl ($\sim$0.7 solar). For this reason, we believe O is the better reference element for these two PNe. However, in LMC~29 our calculated [O/H] is 2-3 times lower than [S/H] and [Ar/H], which suggests that O destruction may have occurred during HBB in its relatively high-mass progenitor \citep{Leisy_Dennefeld_2006, Ventura_Etal_2015_LMC}. We therefore adopt Ar as the metallicity reference in LMC~29.

Additionally, because the chemical evolution histories of the Magellanic Clouds differ from that of the Milky Way, the initial abundances of the \emph{n}-capture elements do not necessarily follow the solar distribution scaled by metallicity. Indeed, the compositions of field giants indicate that AGB stars contributed more to the chemical evolution of the LMC than in the Galaxy \citep[e.g.,][]{Pompeia_Etal_2008, Hasselquist_Etal_2021, Minelli_Etal_2021}, and the same may be true for the SMC \citep[e.g.,][but see \citealt{Mucciarelli_Etal_2023}]{Hasselquist_Etal_2021}. These studies found that the abundances of light-\emph{s} elements such as Zr are approximately solar relative to Fe in the LMC and SMC, to within 0.2~dex for the metallicity range of our sample. However, heavy-\emph{s} species, as indicated by [Ba/Fe], [La/Fe], and [Ce/Fe], may have initial abundances 0.2--0.4~dex (1.6--2.5 times) larger than solar. Thus, we assume that the initial abundances of light \emph{n}-capture elements (Se, Kr, Rb) are scaled solar, while our [Cd/O] and [Te/O] values may need to be reduced by $\sim$0.3~dex to determine the actual enrichment level.


\startlongtable
\begin{deluxetable*}{ccccc|cccc}
\tablecolumns{9}
\tablewidth{0pc}
\tabletypesize{\scriptsize}
\tablecaption{Enrichment Factors Relative to H and Different Metallicity References}
\tablehead{ \\[-0.6pc]
\colhead{Element} & \colhead{$[$X/H$]$} & \colhead{$[$X/O$]$} & \colhead{$[$X/S$]$} & \multicolumn{1}{c|}{$[$X/Ar$]$} &  \colhead{$[$X/H$]$} & \colhead{$[$X/O$]$} & \colhead{$[$X/S$]$} & \colhead{$[$X/Ar$]$}
}
\startdata
\multicolumn{1}{c}{} & \multicolumn{4}{c|}{\textbf{LMC 29}} & \multicolumn{4}{c}{\textbf{LMC 31}} \\
\hline
Se & --0.90 $\pm$ 0.16 & --0.16 $\pm$ 0.18 & --0.47 $\pm$ 0.19 & --0.67 $\pm$ 0.26 & --0.82 $\pm$ 0.10 & 0.55 $\pm$ 0.11 & 0.57 $\pm$ 0.11 & 0.14 $\pm$ 0.24 \\
Br & $<0.12$ & $<0.86$ & $<0.74$ & $<0.54$ & \nodata & \nodata & \nodata & \nodata \\
Kr & --0.30 $\pm$ 0.15 & 0.44 $\pm$ 0.18 & 0.12 $\pm$ 0.19 & --0.08 $\pm$ 0.26 & 0.36 $\pm$ 0.09 & 1.73 $\pm$ 0.11 & 1.75 $\pm$ 0.10 & 1.32 $\pm$ 0.24 \\
Rb & $<0.44$ & $<1.18$ & $<0.86$ & $<0.66$ & $<1.01$ & $<2.38$ & $<2.40$ & $<1.97$ \\
Cd & $<1.00$ & $<1.74$ & $<1.42$ & $<1.22$ & $<1.24$ & $<2.61$ & $<2.63$ & $<2.20$ \\
Te & $<0.12$ & $<0.86$ & $<0.54$ & $<0.35$ & 0.23 $\pm$ 0.38 & 1.59 $\pm$ 0.38 & 1.62 $\pm$ 0.38 & 1.19 $\pm$ 0.41 \\
\hline
\multicolumn{1}{c}{} & \multicolumn{4}{c|}{\textbf{LMC 47}} & \multicolumn{4}{c}{\textbf{LMC 48}} \\
\hline
Se & --0.70 $\pm$ 0.20 & --0.20 $\pm$ 0.21 & --0.08 $\pm$ 0.24 & --0.39 $\pm$ 0.29 & --0.51 $\pm$ 0.54 & --0.10 $\pm$ 0.55 & --0.02 $\pm$ 0.54 & 0.09 $\pm$ 0.56 \\
Br & $<0.22$ & $<0.71$ & $<0.83$ & $<0.53$ & \nodata & \nodata & \nodata & \nodata \\
Kr & 0.30 $\pm$ 0.19 & 0.79 $\pm$ 0.20 & 0.91 $\pm$ 0.23 & 0.61 $\pm$ 0.28 & 0.30 $\pm$ 0.13 & 0.71 $\pm$ 0.20 & 0.80 $\pm$ 0.15 & 0.91 $\pm$ 0.30 \\
Rb & $<0.64$ & $<1.14$ & $<1.26$ & $<0.95$ & $<0.67$ & $<1.08$ & $<1.17$ & $<1.28$ \\
Cd & $<0.98$ & $<1.47$ & $<1.59$ & $<1.29$ & 0.47 $\pm$ 0.24 & 0.88 $\pm$ 0.27 & 0.97 $\pm$ 0.24 & 1.08 $\pm$ 0.34 \\
Te & $<0.17$ & $<0.67$ & $<0.79$ & $<0.48$ & 0.59 $\pm$ 0.31 & 0.99 $\pm$ 0.33 & 1.08 $\pm$ 0.31 & 1.19 $\pm$ 0.38 \\
\hline
\multicolumn{1}{c}{} & \multicolumn{4}{c|}{\textbf{LMC 61}} & \multicolumn{4}{c}{\textbf{LMC 73}} \\
\hline
Se & 0.37 $\pm$ 0.32 & 0.44 $\pm$ 0.34 & 0.95 $\pm$ 0.32 & 0.66 $\pm$ 0.38 & 0.24 $\pm$ 0.09 & 0.37 $\pm$ 0.11 & 0.73 $\pm$ 0.15 & 0.53 $\pm$ 0.23 \\
Br & $<0.50$ & $<0.57$ & $<1.08$ & $<0.79$ & $<0.77$ & $<0.90$ & $<1.34$ & $<1.14$ \\
Kr & 0.72 $\pm$ 0.27 & 0.79 $\pm$ 0.29 & 1.30 $\pm$ 0.27 & 1.01 $\pm$ 0.35 & 0.89 $\pm$ 0.06 & 1.03 $\pm$ 0.09 & 1.39 $\pm$ 0.13 & 1.19 $\pm$ 0.23 \\
Rb & $<1.40$ & $<1.46$ & $<1.97$ & $<1.69$ & $<1.22$ & $<1.35$ & $<1.71$ & $<1.51$ \\
Cd & $<1.49$ & $<1.56$ & $<2.06$ & $<1.78$ & $<1.47$ & $<1.60$ & $<1.96$ & $<1.76$ \\
Te & 0.71 $\pm$ 0.38 & 0.78 $\pm$ 0.39 & 1.29 $\pm$ 0.38 & 1.00 $\pm$ 0.43 & 0.23 $\pm$ 0.25 & 0.36 $\pm$ 0.25 & 0.73 $\pm$ 0.26 & 0.53 $\pm$ 0.31 \\
\hline
\multicolumn{1}{c}{} & \multicolumn{4}{c|}{\textbf{LMC 99}} & \multicolumn{4}{c}{\textbf{SMC 3}} \\
\hline
Se & 0.24 $\pm$ 0.45 & 0.41 $\pm$ 0.46 & 0.89 $\pm$ 0.47 & --0.16 $\pm$ 0.48 & $<-0.32$ & $<0.33$ & $<1.04$ & $<0.57$ \\
Br & $<1.22$ & $<1.38$ & $<1.87$ & $<0.81$ & \nodata & \nodata & \nodata & \nodata \\
Kr & 0.62 $\pm$ 0.21 & 0.78 $\pm$ 0.24 & 1.27 $\pm$ 0.26 & 0.21 $\pm$ 0.32 & $<0.59$ & $<1.25$ & $<1.95$ & $<1.49$ \\
Rb & 0.55 $\pm$ 0.21 & 0.72 $\pm$ 0.24 & 1.20 $\pm$ 0.26 & 0.15 $\pm$ 0.32 & $<1.92$ & $<2.57$ & $<3.28$ & $<2.81$ \\
Cd & 0.45 $\pm$ 0.23 & 0.62 $\pm$ 0.26 & 1.11 $\pm$ 0.28 & 0.05 $\pm$ 0.32 & $<1.98$ & $<2.63$ & $<3.34$ & $<2.87$ \\
Te & 0.26 $\pm$ 0.28 & 0.43 $\pm$ 0.30 & 0.91 $\pm$ 0.32 & --0.14 $\pm$ 0.36 & $<1.07$ & $<1.72$ & $<2.43$ & $<1.96$ \\
\hline
\multicolumn{1}{c}{} & \multicolumn{4}{c|}{\textbf{SMC 13}} & \multicolumn{4}{c}{\textbf{SMC 14}} \\
\hline
Se & --0.60 $\pm$ 0.33 & 0.03 $\pm$ 0.34 & 0.82 $\pm$ 0.33 & 0.13 $\pm$ 0.38 & 0.01 $\pm$ 0.31 & 0.44 $\pm$ 0.32 & 0.91 $\pm$ 0.33 & 0.62 $\pm$ 0.36 \\
Br & \nodata & \nodata & \nodata & \nodata & $<0.56$ & $<0.99$ & $<1.45$ & $<1.16$ \\
Kr & --0.50 $\pm$ 0.41 & 0.13 $\pm$ 0.42 & 0.92 $\pm$ 0.42 & 0.23 $\pm$ 0.45 & 0.32 $\pm$ 0.20 & 0.78 $\pm$ 0.23 & 1.24 $\pm$ 0.29 & 0.96 $\pm$ 0.33 \\
Rb & $<0.71$ & $<1.33$ & $<2.12$ & $<1.43$ & $<1.41$ & $<1.84$ & $<2.31$ & $<2.01$ \\
Cd & $<0.90$ & $<1.53$ & $<2.32$ & $<1.63$ & $<1.58$ & $<2.01$ & $<2.47$ & $<2.18$ \\
Te & --0.36 $\pm$ 0.39 & 0.27 $\pm$ 0.40 & 1.05 $\pm$ 0.40 & 0.37 $\pm$ 0.43 & $<0.60$ & $<1.03$ & $<1.50$ & $<1.21$ \\
\hline
\multicolumn{1}{c}{} & \multicolumn{4}{c|}{\textbf{SMC 20}} & \multicolumn{4}{c}{\textbf{SMC 25}}\\
\hline
Se & --1.36 $\pm$ 0.28 & --0.42 $\pm$ 0.29 & 0.35 $\pm$ 0.29 & --0.20 $\pm$ 0.35 & $<-0.45$ & $<0.55$ & $<0.27$ & $<0.48$ \\
Kr & --1.05 $\pm$ 0.33 & --0.10 $\pm$ 0.34 & 0.67 $\pm$ 0.34 & 0.11 $\pm$ 0.38 & $<0.67$ & $<1.66$ & $<1.39$ & $<1.59$ \\
Rb & $<0.29$ & $<1.24$ & $<2.01$ & $<1.45$ & $<1.53$ & $<2.52$ & $<2.24$ & $<2.45$ \\
Cd & $<0.83$ & $<1.77$ & $<2.54$ & $<1.98$ & $<1.79$ & $<2.78$ & $<2.51$ & $<2.71$ \\
Te & --0.28 $\pm$ 0.37 & 0.67 $\pm$ 0.37 & 1.44 $\pm$ 0.37 & 0.88 $\pm$ 0.41 & $<0.82$ & $<1.82$ & $<1.54$ & $<1.75$ \\
\enddata
\tablecomments{Elemental abundances are given relative to the solar abundances of \citet{Asplund_Etal_2021}.} 
\label{tab:ncap_enrich}
\end{deluxetable*}

\section{Comparison of Abundances to Theoretical Predictions} \label{sec:models}

Nebular abundances of \emph{n}-capture elements, together with those of more common species such as He, C, N, and O, provide crucial constraints for models of stellar evolution and nucleosynthesis. In this section, we compare our results with theoretical predictions for the absolute enrichments of \emph{n}-capture elements in low- and intermediate-mass stars. The abundances of PNe reflect the outcomes of nucleosynthetic processes operating during the thermally-pulsing AGB phase (see \S2.1), superimposed on those occurring in earlier evolutionary stages. Specifically, low- and intermediate-mass stars can undergo up to three stages of convective dredge-up. First dredge-up occurs during the red giant branch, while second dredge-up takes place during the early AGB in stars more massive than 3--4~M$_{\odot}$. Both enrich the envelope with He and $^{14}$N, and modify C isotopic ratios \citep{Iben_Renzini_1983, Busso_Etal_1999, Karakas_Lattanzio_2014}. Third dredge-up (TDU) occurs during the thermally-pulsing AGB evolutionary stage for stars with M~$\geq$1.0--1.5~M$_{\odot}$, depending on the metallicity, and leads to C and \emph{s}-process enrichments observed in PNe and AGB stars.

We compare our abundances with predictions from three different AGB nucleosynthesis codes: those from the Monash group \citep{Karakas_Lugaro_2016, Karakas_Etal_2018}, the FRUITY database \citep{Cristallo_Etal_2015}, and the NuGrid collaboration \citep{Pignatari_Etal_2016, Battino_Etal_2019, Battino_Etal_2021}. Such comparisons with \emph{s}-process enrichments in PNe had previously only considered the light \emph{n}-capture elements Ge, Se, and Kr in Galactic PNe against predictions from the Monash code \citep{Karakas_Etal_2007, Karakas_Etal_2009}. These evolutionary model codes differ in their treatment of physical processes during the AGB phase as well in the adopted nuclear reaction rates. The predicted \emph{s}-process enrichments depend on nucleosynthesis in the intershell, and the efficiency and number of TDU episodes. Key factors influencing these enrichments include the treatment of convection and/or alternate mixing mechanisms (which govern the transport of material from the intershell to the surface), the mass-loss rate (which sets the AGB lifetime and number of TDU events), and the size of the $^{13}$C pocket (which controls the amount of \emph{s}-processed material produced).

\subsection{Methodology: Nearest Neighbor Classification}

We use a methodology similar to that of \citet{denHartogh_Etal_2023}, but adapted to find the best-fit models to empirical abundances of PNe rather than of Ba stars. This led to two changes in approach: 1) we use the Nearest Neighbor Classificator (NNC) only, and 2) we use the surface abundances at different times during the AGB evolution instead of diluting the surface abundances of the final models (as would be appropriate for Ba stars, where enriched material is transferred to the envelope of a binary companion). The reason for the first modification is that PN abundances have larger uncertainties than those of Ba stars, with more elements having only upper limits to their abundances, and the NNC is more resilient to noisy data (as opposed to the Neural Network Ensemble Classificator, or NNEC). The second change was made to allow for variance in the predicted enrichments, which is justified given the possibility that binary interactions may truncate the AGB phase \citep[e.g.,][]{DeMarco_2009, Jones_Boffin_2017, Osborn_Etal_2025}. 

The NNC algorithm calculates a goodness-of-fit (GoF) distance between the models and the empirical abundances (Table~\ref{tab:ncap_enrich}), where the measured logarithmic abundances are taken as random, normally-distributed variables. To compute GoF values, we use a modified $\chi^2$ test between the observed ($O_i$) and predicted abundances ($X_i$), where the subscript $i$ denotes element $i$ \citep[Eq.~2 of][]{denHartogh_Etal_2023}:
\begin{equation}
\chi^2_m = \sum\frac{(X_i - O_i)^2}{\sigma(O_i)}. \label{eq1}
\end{equation}
In this equation $\sigma(O_i)$ is the empirical abundance uncertainty. In this way, $\chi^2_m$ values are determined for the surface abundance predictions of different AGB models (\S\ref{sec:modelsets}), and elements with the largest measured uncertainties are given the least weight.

Next, a Monte~Carlo method is used to generate 10$^5$ sets of abundances $X_i$ from the empirical logarithmic abundances, using the uncertainty as the standard deviation. The GoF is then calculated from the related $1 - \mathrm{CDF}$ distribution (or `tail distribution'), where CDF is the cumulative distribution function of the $\chi^2_m$ values generated from the Monte~Carlo method. The GoF is the point where the modified $\chi^2$ score of the tail distribution is equal to that of a given model versus observations. For example, a GoF above 50\% corresponds to a $\chi^2_m$ below the point at which the CDF exceeds 0.5. We take GoF~$>$~50\% to be a good fit to the measured abundances.

For upper limits we also use a normal distribution, assigning the value of the upper limit to the mean and an uncertainty of 1~dex for the standard deviation. Although in principle this allows for values higher than the upper limit, in practice we do not observe values above it in the model predictions. We conducted a test where we increased the assigned standard deviation to 2~dex for upper limits, and we observed no significant difference in the classification.

\subsection{Model Sets Considered in the Analysis} \label{sec:modelsets}

Table~\ref{tab:models} summarizes the models considered for fitting the observational results (hereafter, M = Monash, F = FRUITY, and Nu = NuGrid). For most metallicities we consider results from AGB models from 1~M$_{\odot}$ to the maximum mass for a CO-core AGB star, which is around 7-8~M$_{\odot}$, depending on metallicity \citep[e.g., see discussions in][]{Karakas_Lugaro_2016, Karakas_Etal_2018}. Each model is calculated from the zero-age main sequence to the tip of the AGB. The models are evolved through a number of thermal pulses, where the total number depends on the rate of mass loss on the AGB. Since not all thermal pulses are followed by a TDU mixing event, the TDU counter starts when dredge-up begins. The surface abundances (elemental and isotopic) are provided as a function of thermal pulse number, which is a proxy for time. 

\begin{deluxetable*}{cc|cc|cc}
\tablecolumns{6}
\tablewidth{0pc}
\tabletypesize{\footnotesize}
\tablecaption{Summary of the AGB models considered for the comparison to the observations. We used a total of 128 FRUITY models, 226 Monash models and 12 NuGrid models. Because each model includes several TDU episodes, we compared to each PN a total of 8703 different abundance patterns.} \label{tab:models}
\tablehead{ \\[-0.6pc]
\multicolumn{2}{c|}{FRUITY (F)} & \multicolumn{2}{c|}{Monash (M)} & \multicolumn{2}{c}{NuGrid (Nu)} \\
\cline{1 - 2} \cline{3 - 4} \cline{5 - 6}
\colhead{Mass (M$_\odot$)} & \multicolumn{1}{c|}{$Z$ range} & \colhead{Mass range (M$_\odot$)} & \multicolumn{1}{c|}{$Z$} & \colhead{Mass (M$_\odot$)} & \colhead{$Z$ $\times10^3$}
}
\startdata
$1.3$ & $2\times10^{-5}$ - $0.014$ & $0.9$ - $6.0$ & $10^{-4}$ & $2.0$ & $1, 2, 6, 10, 20, 30$\\
$1.5$ & $2\times10^{-5}$ - $0.014$ & $1.7$ - $6.0$ & $6\times10^{-4}$ & $3.0$ & $1,10, 20, 30$\\
$2.0$ & $2\times10^{-5}$ - $0.014$ & $1.0$ - $7.0$ & $10^{-3}$ & & \\
$2.5$ & $2\times10^{-4}$ - $0.014$ & $1.0$ - $7.0$ & $2.8\times10^{-3}$ & & \\
$2.5$ & $10^{-4}$ - $0.014$ & $1.0$ - $7.5$ & $7\times10^{-3}$ & & \\
$3.0$ & $10^{-4}$ - $0.014$ & $2.0$ - $3.0$ & $10^{-2}$ & & \\
$4.0$ & $10^{-4}$ - $0.014$ & $1.0$ - $8.0$ & $0.014$ & & \\
$5.0$ & $10^{-4}$ - $0.014$ & $1.0$ - $8.0$ &  $3\times10^{-2}$ & & \\
$6.0$ & $10^{-4}$ - $0.014$ & & & & \\
\hline
\multicolumn{6}{c}{Definition of Flags in Model Names} \\
\hline
\multicolumn{2}{c|}{FRUITY} & \multicolumn{4}{c}{Monash} \\
Flag & IRV\tablenotemark{a} & Flag & \multicolumn{1}{c}{PMZ Mass (M$_{\odot}$)} & Flag & \multicolumn{1}{c}{PMZ Mass (M$_{\odot}$)}\\ \hline
- & 0 & a & \multicolumn{1}{c}{0} & i & $2\times10^{-3}$ \\
a & 10 & b & \multicolumn{1}{c}{$5\times10^{-5}$} & j & $4\times10^{-3}$ \\
b & 30 & c & \multicolumn{1}{c}{$1\times10^{-4}$} & k & $6\times10^{-3}$\\
c & 60 & d & \multicolumn{1}{c}{$2\times10^{-4}$} & & \\
d & T00 & e & \multicolumn{1}{c}{$5\times10^{-4}$} & & \\
e & T10 & f & \multicolumn{1}{c}{$6\times10^{-4}$} & & \\
f & T30 & g & \multicolumn{1}{c}{$8\times10^{-4}$} & & \\
g & T60 & h & \multicolumn{1}{c}{$1\times10^{-3}$} & & \\
\hline
\enddata
\tablenotetext{a}{For FRUITY models, the flags indicate the initial rotational velocity (IRV) in km~s$^{-1}$. Those marked with a ``T" use an extended $^{13}$C~pocket \citep[see][]{Cristallo_Etal_2015}.}
\end{deluxetable*}

We considered a total of 128 FRUITY models, 226 Monash models and 12 NuGrid models. Each model includes several TDU episodes, the number of which depends on factors that include the mass-loss rate, mass, and metallicity. By considering surface abundances at the end of each TDU episode, we compared a total of 8703 unique abundance patterns (corresponding to an average of $\sim$24 TDUs per model) to each PN's measured abundances. The naming scheme for models in the figures is as follows: the model series (F, M, or Nu), initial mass in solar masses (after ``m''), initial overall metallicity (where, e.g., ``z003'' indicates $Z=0.003$), TDU number (for the F and Nu sets) or thermal pulse (TP) number (M set) where applicable, and a flag (see Table~\ref{tab:models}) for variations in the rotation rate and mass of the partial mixing zone (PMZ; see \S\ref{sec:Monash}) and $^{13}$C pockets.

In comparing model predictions to the measured enrichment factors (Table~\ref{tab:ncap_enrich}), it should be noted that each code adopts a different reference for the solar composition: \citet{Asplund_Etal_2009} for M, \citet{Lodders_2003} for F, and \citet{Grevesse_Noels_1993} for Nu models, while we utilize the abundances of \citet{Asplund_Etal_2021} in our empirical analysis (see Table~\ref{tab:solar} in the Appendix). For all but the noble gases, the first two references agree with \citet{Asplund_Etal_2021} to within 10\% (0.04~dex). The Kr abundance of \citet{Asplund_Etal_2021} is 0.13~dex lower than \citet{Lodders_2003} and \citet{Asplund_Etal_2009}, while Ar is 0.17~dex lower than cited by \citet{Lodders_2003}, and hence the model abundances [Kr/H] and (in the case of F models) [Ar/H] are too low by the same amount compared to the derived PN abundances. In contrast, \citet{Asplund_Etal_2021} adopted the solar Ne abundance to be 0.13 and 0.19~dex (35\% and 55\%) larger than \citet{Asplund_Etal_2009} and \citet{Lodders_2003}, respectively. The solar abundances of \citet{Grevesse_Noels_1993} differ considerably from the other references \citep[e.g., see discussion in][]{Asplund_Etal_2009}, with N and O 0.2 and 0.24~dex higher than that of \citet{Asplund_Etal_2021}, and Cl and Br larger by 0.33 and 0.42~dex. Other elements also have higher initial abundances by 0.07 (Se) to 0.17~dex (Kr), and hence the enrichment factors of the Nu models are smaller than implied in the figures shown below. For non-solar metallicities, each model uses abundances scaled to the solar distribution, except for Nu models with 0.001~$\leq Z \leq 0.006$ that include $\alpha$-element enhancements \citep[see][]{Battino_Etal_2021}. In our analysis, the theoretical enrichments have not been modified for the different solar compositions adopted, since abundance changes can affect nucleosynthetic predictions. When comparing the M and F models to our measured abundances, the differences in the initial (scaled solar) abundances are smaller than the uncertainties in the empirical abundances in most cases (except for Ne and, in LMC~73, Kr). Therefore the different solar compositions used in the empirical versus theoretical abundances do not significantly affect the comparisons below or our conclusions. In contrast, the larger solar abundances of \citet{Grevesse_Noels_1993} may partially contribute to poor fits of Nu models with the nebular abundances.

\subsubsection{Monash models} \label{sec:Monash}

In all M models for stars with initial mass $M \lesssim 5$ M$_{\odot}$ that experience TDU, a $^{13}$C-pocket is included by inserting a PMZ of protons into the top of the intershell region at the deepest extent of each TDU episode. The resulting $^{13}$C pocket is smaller than the PMZ, and is roughly one-tenth the mass of the He-rich intershell \citep[for details, see][]{Karakas_Lugaro_2016}. One caveat with these models is that the size of the PMZ is constant for a given mass, while the He intershell zone decreases in mass with evolution. For each initial mass there is a {\em standard} PMZ mass: 0.002~M$_{\odot}$ for 1.5--3.0~M$_{\odot}$, 0.001~M$_{\odot}$ for 3--4~~M$_{\odot}$, and 10$^{-4}$~M$_{\odot}$ for 4--4.75~M$_{\odot}$ models. No PMZ is inserted for higher mass progenitors. The decrease in PMZ mass as initial stellar mass increases reflects a decrease in the intershell mass. The M models use the mass-loss prescription of \citet{Vassiliadis_Wood_1993} on the AGB, which is slower than that adopted by FRUITY \citep{Straniero_Etal_2006} or NuGrid \citep{Blocker_1995}. This leads to a larger number of thermal pulses and TDU events, and generally larger \emph{s}-process enrichments (though the treatment of mixing across convective boundaries, which produces the PMZ and $^{13}$C~pocket, is also important). Rotation and magnetic fields are not included in the M models. For further details, see \citet{Lugaro_etal_2012} for metallicity $Z= 10^{-4}$; \citet{Fishlock_Etal_2014} for $Z=0.001$; and \citet{Karakas_Lugaro_2016} and \citet{Karakas_Etal_2018} for $Z=0.0028, 0.007, 0.014$, and 0.03 models. 

\subsubsection{FRUITY models}

The F models also cover an extensive range in mass and metallicity (Table~\ref{tab:models}). For these models, convective boundary mixing at the base of the convective envelope is modeled via an exponentially decreasing velocity function, with the exponent treated as a free parameter. This drives the formation of the $^{13}$C pocket and $s$-process nuclei \citep[for more details see][]{Cristallo_Etal_2009, Cristallo_Etal_2015}. In addition to the method used to form the $^{13}$C pocket, the F models differ from M models in other respects, including the treatment of mass loss \citep{Straniero_Etal_2006} and rotation. The most significant disparity is for more massive models ($>4$~M$_{\odot}$), as these intermediate-mass F models have lower interior temperatures than M models of similar mass and do not experience HBB or significant activation of the $^{22}$Ne neutron source except at the lowest metallicities \citep[e.g., see \S5 of][]{Karakas_Lugaro_2016}. The best-fitting F models for some objects tend to be less massive than M models for this reason (\S\ref{sec:goodfits}).

The F models are available via the FRUITY database\footnote{\tt http://fruity.oa-abruzzo.inaf.it/}, which provides surface composition data as a function of TDU number. For less massive models that do not experience TDU, only the post-first dredge-up composition is provided. We use the tabulated data as a function of TDU number from the database to compare to the PNe. 

\subsubsection{NuGrid models}

The Nu models that we utilize \citep{Battino_Etal_2016, Battino_Etal_2019, Battino_Etal_2021} have a much narrower mass and metallicity range, and less dense sampling of each parameter than the M and F models. The \citet{Ritter_Etal_2018} set is more complete in mass and metallicity, however it was calculated including a single exponential function to mimic diffusive overshoot. This is similar to the FRUITY models, but uses the diffusion coefficient instead of the velocity for the formation of the PMZ. This results in $^{13}$C-pocket masses too low to produce significant $s$-process enrichment. In contrast, the Battino et al.\ models utilize a double exponential function (to add the effect of gravity waves on top of diffusive overshoot) for the formation of the PMZ. These models produce significant \emph{s}-process nucleosynthesis, but only the 2~M$_{\odot}$ models cover metallicities between $Z=0.002$ and $Z=0.01$ that are the most relevant for the observed PNe. 

While we include these in the figures below, they do not provide good fits to the empirical abundances, primarily due to the limited number of models available. An additional factor that worsens the GoF is that subsolar metallicity Nu models include $\alpha$-element enhancements, as seen in Galactic thick disk and halo stars but not in the Magellanic Clouds, at least not for the metallicity range of our targets. Even taking into account the larger initial abundances (by 0.1--0.25~dex) from the \citet{Grevesse_Noels_1993} solar composition, the $\alpha$-enhancements result in predicted O and Ne abundances much larger than observed. These issues are not due to the input physics, and demonstrate the need for Nu model sets that cover a wider range of initial masses, metallicities, and without $\alpha$-element enhancements, so that their ability to reproduce empirical data can be better assessed. We nevertheless feel that the comparison of the predicted abundances is useful for PNe with progenitor masses near 2~M$_{\odot}$ and those with near-solar metallicity, and include Nu models in the figures. 

\subsection{Analysis of Individual Nebulae} \label{sec:goodfits}

For five PNe in the sample, good fits (GoF~$>50$\%) are found to the empirical abundances, as presented below. The figures for each PN show the best-fitting models, though for some objects there are several high-GoF models.\footnote{Additional fits to the empirical abundances are available at \dataset[doi:10.5281/zenodo.19700217]{https://doi.org/10.5281/zenodo.19700217}. The tables in this repository list all models with GoF~$>50$\% for the choice of light element anchor(s), and the figures compare the best-fitting models against the observations.}  We tested using different light elements, or combinations thereof, as metallicity anchors, but as discussed in \S\ref{sec:enrich}, O has the most accurately determined abundance and therefore is the best reference element for our targets, except for LMC~29 for which we use Ar. Therefore the fits utilize O (Ar, for LMC~29) as one of the anchor elements. In all of these models, we find that standard choices for the PMZ mass (\S\ref{sec:Monash}, Table~\ref{tab:models}) reproduce the measured \emph{s}-process enrichments. 

\begin{figure*}
    \centering
    \includegraphics[scale=0.35]{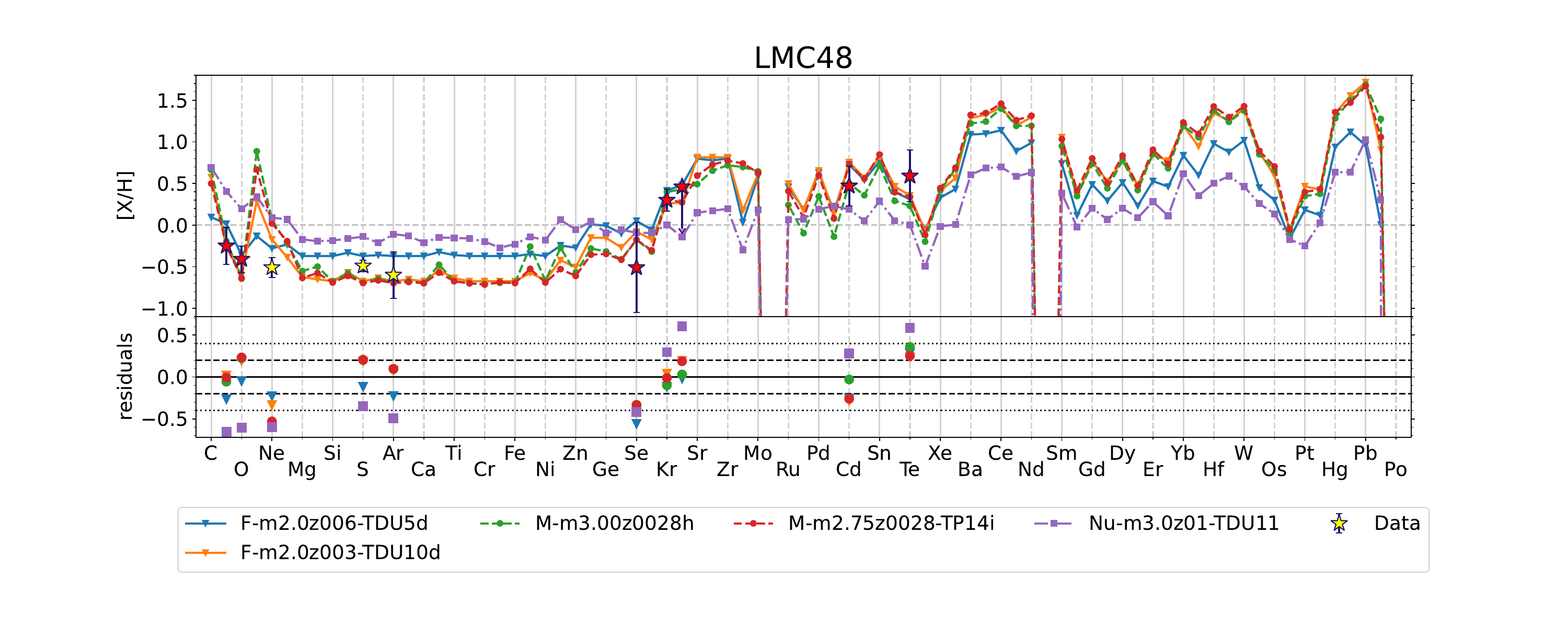} 
    \caption{Comparison between observed abundances and best-fitting F, M, and Nu models. Red stars indicate the empirical abundances used in the fits (upper limits, such as Rb here, have one-sided error bars), and yellow stars are elements whose abundances are not included in the fits. 
    The model names indicate the model set, the initial mass, initial metallicity $Z$, TDU or TP number at which the composition is read out, and a flag that distinguishes between different rotation rates (F models) or implementations of the PMZ (M models; see \S 6.2 and Table~\ref{tab:models}).
    The F and M models shown here have GoF between 61\% and 76\%, while that for the Nu model is 1.4\%. The GoFs of the best-fitting models to the empirical abundances, including those in the following figures, are available at \dataset[doi:10.5281/zenodo.19700217]{https://doi.org/10.5281/zenodo.19700217}.}
    \label{fig:lmc48}
\end{figure*}

\textit{LMC~48.} The abundances of LMC~48 are fit well by 2.0~M$_{\odot}$ F models and 2.5--3.0 M$_{\odot}$ M models with metallicities between 0.003 and 0.006 ($[$Fe/H$]$~=~--0.67 and --0.37~dex, respectively), with GoF ranging from 61\% to 76\% (Figure~\ref{fig:lmc48}). The good agreement between predicted and empirical abundances is due to two factors: (i) the light elements all indicate a similar metallicity, and (ii) at these metallicities the $s$-process behaves in such a way in the M and F models that the observed Kr, Cd, and Te abundances are matched. Including the Ne abundance reduces the GoF to 50--55\%, as this element is predicted to be somewhat higher than observed, due to the production and dredge-up of $^{22}$Ne. Nu models provide a poor fit to the measured abundances, as the GoF is only 1.4\% for the 3.0~M$_{\odot}$, $Z=0.01$ model shown in Figure~\ref{fig:lmc48}. The reason that a lower-metallicity 2.0~M$_{\odot}$ Nu model is not selected is due to the $\alpha$-enhancements included in the overall metallicity. These models produce much larger O, N, and Ne abundances than observed, while Ar and S, which are less enhanced, are well below the observed abundances. Because Magellanic Cloud stars in the metallicity range of our targets do not exhibit $\alpha$-enhanced compositions \citep{Lapenna_Etal_2012, Hasselquist_Etal_2021, Mucciarelli_Etal_2023}, lower-metallicity Nu models produce a lower GoF than the $Z=0.01$ models that do not have $\alpha$-enhancements. This is also the case for most of the other targets. 

\begin{figure*}
    \centering
    \includegraphics[scale=0.35]{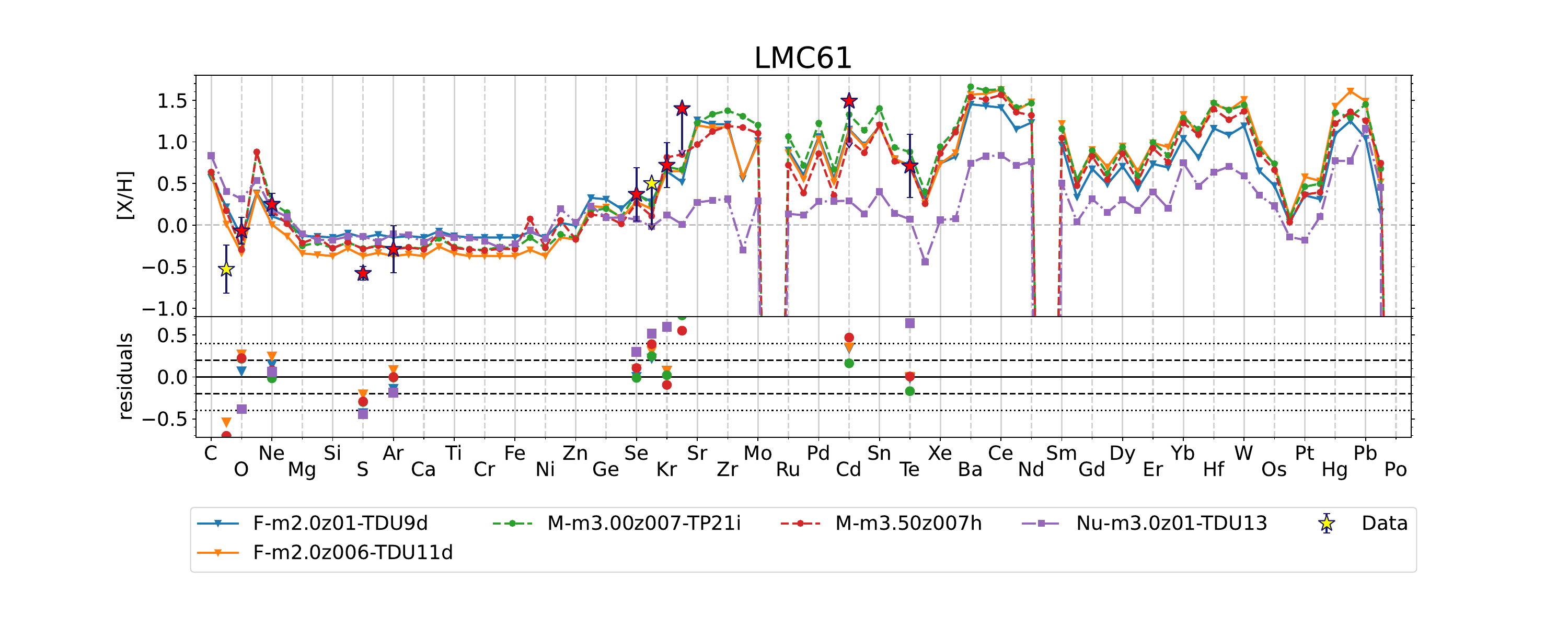}
    \caption{Same as Figure~\ref{fig:lmc48} but for LMC~61. 
    The F and M models shown here have GoF between 81\% and 87\%, except for the F models with $Z=0.01$, with GoF 55\%. The Nu model has a GoF of 5\%.}
    \label{fig:lmc61} 
\end{figure*}

\textit{LMC~61.} This PN is also very well fitted (Figure~\ref{fig:lmc61}), by models between 2 (F) and $\sim$3 M$_{\odot}$ (M) and metallicities between 0.006 ($[$Fe/H$]=-0.37$) and 0.01 (($[$Fe/H$]=-0.15$). The GoF of the F and M models is over 80\%, except for the F model with $Z=0.01$ which has GoF 55\%. The [Ne/H] abundance in this PN is the highest among those of all the observed elements lighter than Fe, and it is in agreement with Ne enhancement due to production of $^{22}$Ne in the models. The \emph{n}-capture elements are also well fitted by the M and F models, while the Nu model (3~M$_{\odot}$, $Z=0.01$) does not predict significant \emph{s}-process enrichment for elements lighter than Ba. Additionally, the $Z=0.01$ Nu models include convective boundary mixing beneath the He~shell, leading to larger N and O abundances (by 0.4--0.8~dex) than observed here and in other targets.

\textit{LMC 73.} The light element abundances of this nebula are generally consistent with a $Z=0.006$--0.007 progenitor. The best fits are M models of $\simeq$4~M$_{\odot}$ and Z=0.007 ({$[$Fe/H$]=-0.30$; Figure~\ref{fig:lmc73}), which have GoFs reaching 80\% (and up to 97\% when only O and Ar are metallicity anchors). The progenitor mass is strongly influenced by the measured Kr abundance. The high temperature of the 4~M$_{\odot}$ model activates the $^{22}$Ne neutron source, which produces Kr to the observed level. In this model, the $^{22}$Ne neutron source has a more significant effect than that of the $^{13}$C pocket, which results in an enhancement of Kr higher than that of Te, as observed. N is also relatively well fitted at this mass due to the effects of the first and second dredge-up. The F models do not experience a strong activation of the $^{22}$Ne neutron source (due to the lower intershell temperature) even at higher masses, and therefore are unable to simultaneously reproduce the large Kr enhancement and the abundances of Se and Te. The best-fitting F model is 2 M$_{\odot}$ with $Z=0.006$ at the end of its evolution, but the GoF is only 36\%. More massive F models result in a worse GoF despite the higher interior temperature, as less \emph{s}-processed material is dredged-up as a result of the lower intershell and $^{13}$C pocket mass \citep{Cristallo_Etal_2015}. Nu models, such as the one shown, do not produce a sufficiently large enhancement of the light \emph{n}-capture elements Se and Kr, resulting in a poor fit. 

\begin{figure*}
    \centering
    \includegraphics[scale=0.35]{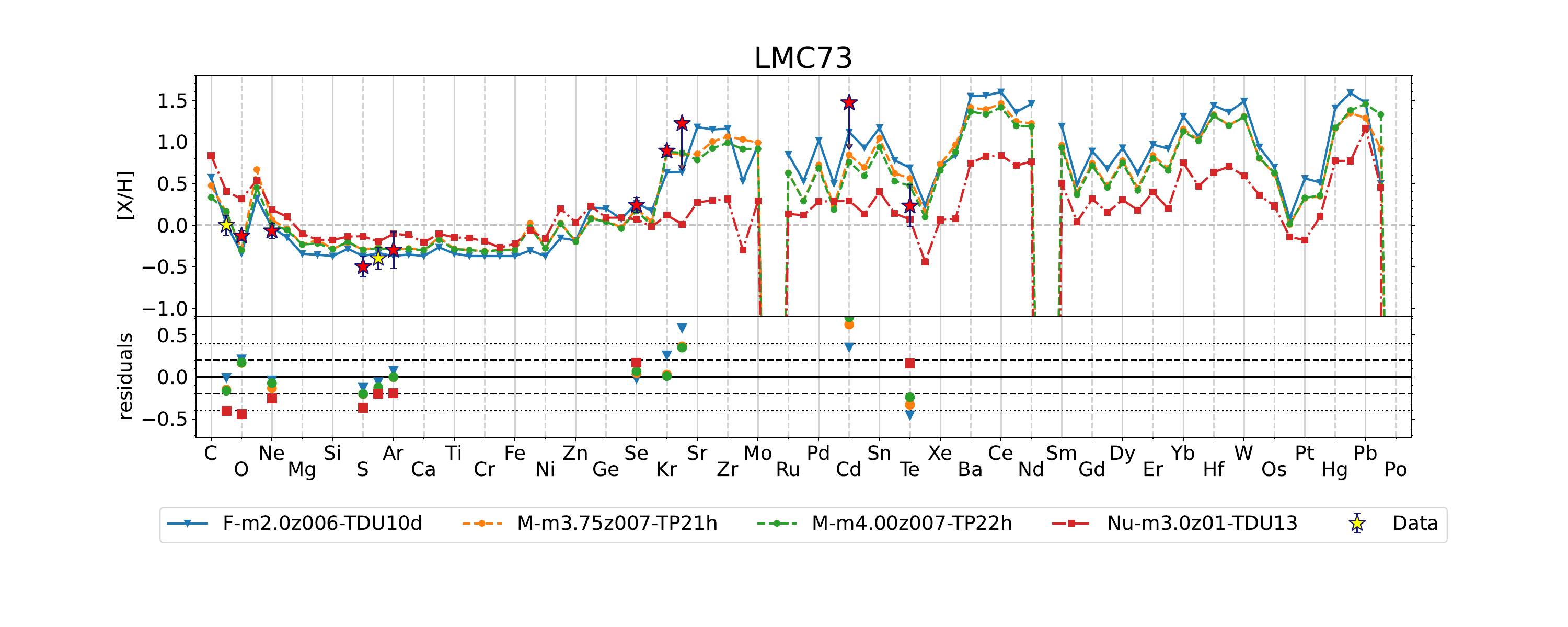}
    \caption{Same as Figure~\ref{fig:lmc48} but for LMC~73. 
    The M models shown here have GoF between 72\% and 80\%, the F model 36\%, while that of the Nu model is close to zero.}
    \label{fig:lmc73} 
\end{figure*}

\textit{LMC 99.} Our model fits for this PN are similar to those of LMC~73. Again, the high Kr/Te ratio favors M models with mass around 4 M$_{\odot}$ for metallicities from 0.006 down to 0.0028 ($[$Fe/H$]=-0.37$ to --0.70~dex; Figure~\ref{fig:lmc99}) with GoFs $>70$\%. These metallicities agree with the light element abundances, with the exception of the likely spurious Ar abundance (more than twice solar; see Table~\ref{tab:append_lmc} and \S\ref{sec:optical}). For the F models, the 2 M$_{\odot}$, $Z=0.006$ model is selected as the best fit, here for an earlier evolutionary time than for LMC~73 (the 5$^{\rm th}$ rather than the 10$^{\rm th}$ out of 11 TDU events). The low-mass F model has a higher GoF than in the case of LMC 73 (82\% versus 36\%), due to the larger error bars on the Kr abundance (recall that for LMC~73, both [\ion{Kr}{3}] and optical [\ion{Kr}{4}] lines were used to compute the elemental abundance, resulting in a more precise and accurate abundance determination). Kr is the key to distinguishing between the solutions provided by the higher-mass M models and the lower-mass F models, as it generally has the best-determined abundance of the observed \emph{n}-capture elements and its production is affected by the activation of the $^{22}$Ne neutron source.

\begin{figure*}
    \centering
    \includegraphics[scale=0.35]{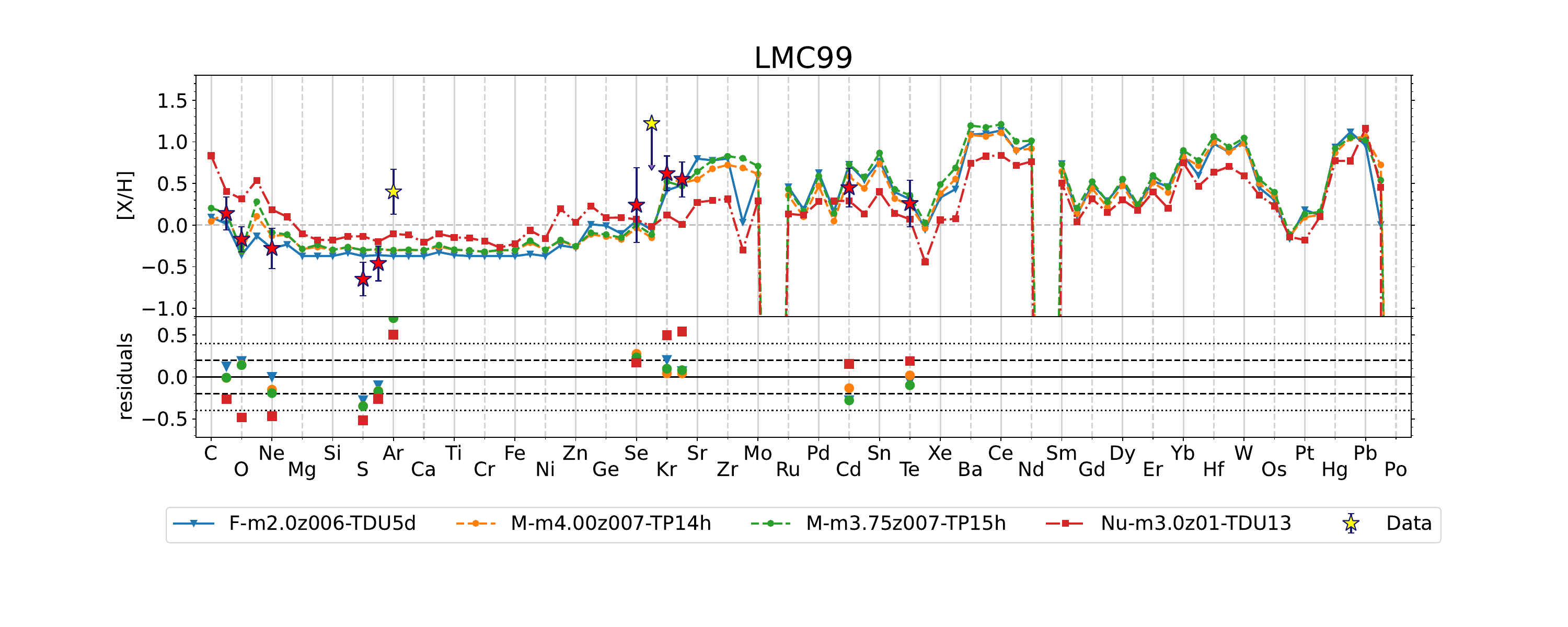}
    \caption{Same as Figure~\ref{fig:lmc48} but for LMC~99. 
    The M models shown here have GoF between 73\% and 87\%. The F model also has a high GoF of 82\%, while that of the Nu model is close to zero.}
    \label{fig:lmc99} 
\end{figure*}

\textit{SMC 14.} The empirical abundances of SMC~14 agree well with F and M models with 2--2.5~M$_{\odot}$ and metallicity 0.0028--0.006 (Figure~\ref{fig:smc14}). These models produce a GoF between 68\% and 83\%, while that of the Nu model shown is 4\%.

\begin{figure*}
    \centering
    \includegraphics[scale=0.35]{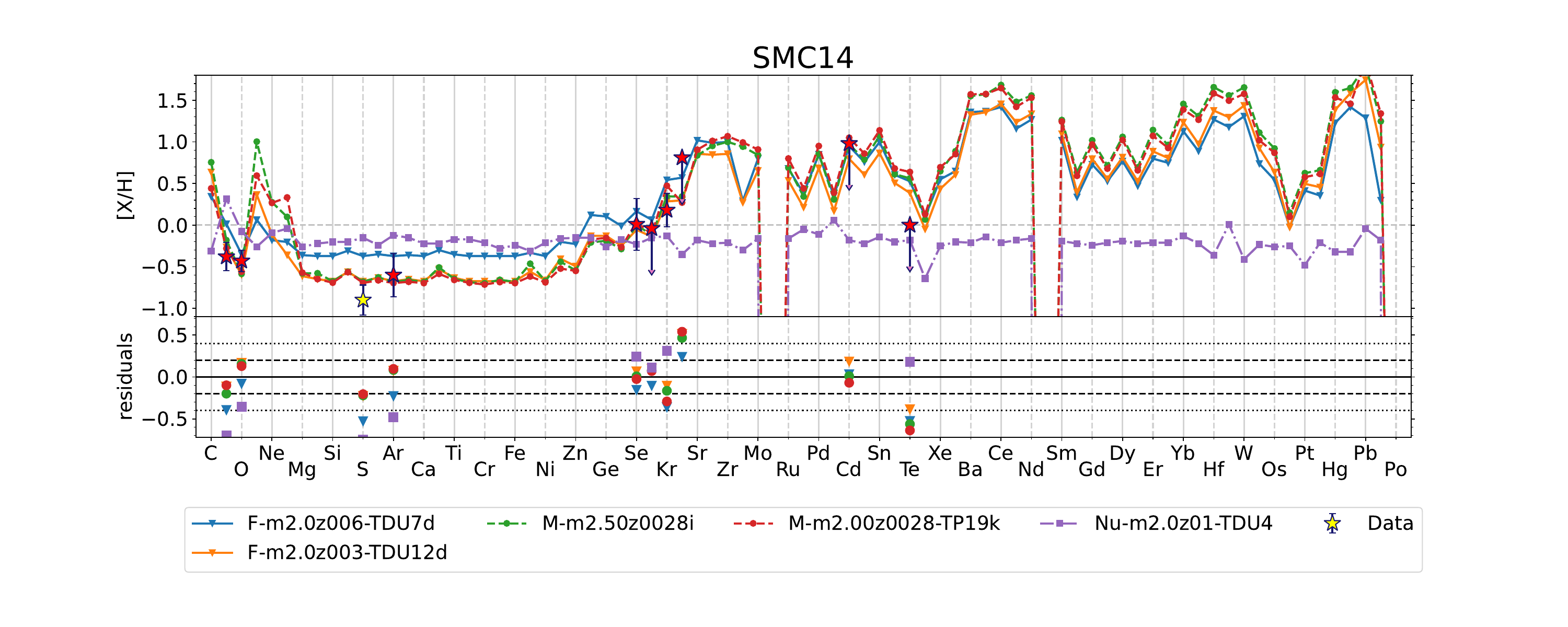}
    \caption{Same as Figure~\ref{fig:lmc48} but for SMC~14. 
    The M and F models shown here have GoF between 68\% and 83\%, while that of the Nu model is 4\%.}
    \label{fig:smc14} 
\end{figure*}

\subsubsection{Other PNe} \label{sec:othermodels}

Of the remaining PNe, we did not attempt to fit the abundances of SMC~3 or SMC~25, in which no \emph{n}-capture elements were detected, or of SMC~13 since it does not exhibit \emph{s}-process enrichments and thus its composition likely reflects the initial composition. We did not find models that provide good fits to the empirical abundances of LMC~29, LMC~31, LMC~47, or SMC~20. Notably, LMC~31 and SMC~20 are the lowest-metallicity targets in our sample (Tables~\ref{tab:append_lmc} and \ref{tab:append_smc} in the Appendix). 

\textit{LMC~29.} This object is a Type~I PN based on its elevated He and N abundances \citep{Peimbert_1978, Kingsburgh_Barlow_1994}, which indicates that its progenitor has a higher mass than the other targets. This is supported by its high central star temperature \citep[200,000~K;][]{Vassiliadis_Etal_1998}. Ar is used as the metallicity reference for this PN, since O may have been destroyed during HBB. Relative to Ar, Se and Kr do not show evidence of \emph{s}-process enhancement. The 6.0~M$_{\odot}$ F and 4.5~M$_{\odot}$ M models reproduce the S, Ar, and Kr abundances with a metallicity $Z=0.006$, but only the M model (which has stronger HBB) predicts a sufficiently large N abundance (Figure~\ref{fig:lmc29}). However, the GoF is only 9.5\% for the M model and 4\% for F, because the Kr abundance cannot be reproduced with the subsolar Se abundance. If O were instead chosen as the metallicity anchor, $Z=0.003$ models are selected, and better reproduce the O, Ne, Se, and Kr (which is enriched relative to O) abundances, but Ar is discrepant and causes a low GoF. Unfortunately, Nu models of the appropriate mass are not available.

\begin{figure*}
    \centering
    \includegraphics[scale=0.35]{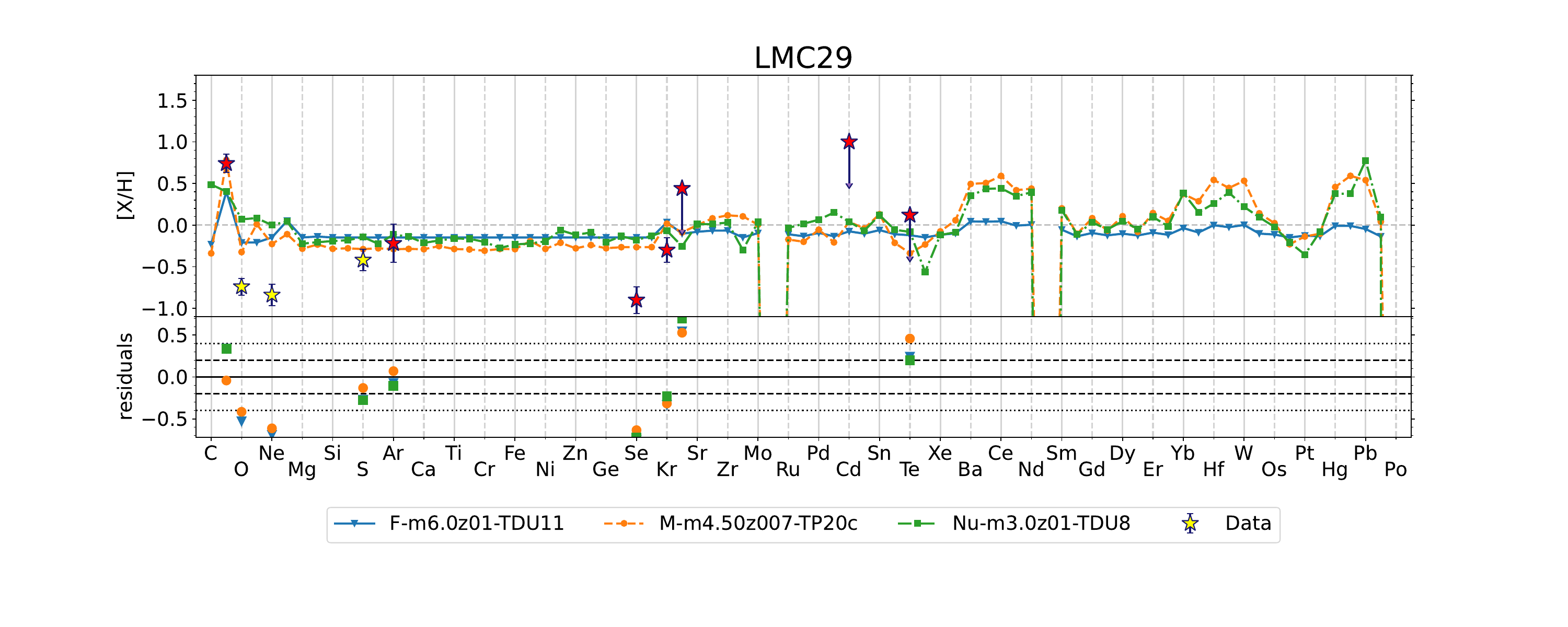}
    \caption{Same as Figure~\ref{fig:lmc48} but for LMC~29. 
    The M, F, and Nu models shown here have GoF of 9\%, 4\%, and 5\% respectively.}
    \label{fig:lmc29} 
\end{figure*}
 
\textit{LMC~31.} The abundances of LMC~31 indicate a low-metallicity progenitor, given that N, O, Ne, Ar, and S have [X/H] values on the order of $-$1.0~dex. Therefore, M and F models with $Z\approx 0.001$ ($[$Fe/H$]$~=~--1.15) are the preferred solutions, but the maximum GoF is only 17\% for a 1.15~M$_{\odot}$ M model and 4\% for a 2~M$_{\odot}$ F model (Figure~\ref{fig:lmc31}). The Se abundance in the M model is predicted to be higher than measured \citep[although this may be due to systematic uncertainties in the Se ICF;][Sterling et al., in prep.]{Sterling_Etal_2017, Sterling_Etal_2025_AAS} while the model Te abundance is too low. The F model agrees better with the observed Se abundance, but the predicted Kr abundance is not sufficiently large. The Nu model is the same mass and metallicity (2.0~M$_{\odot}$ and $Z=0.001$) as the F model, but does not predict significant \emph{s}-process enrichments. This is because the model is truncated after one TDU, to prevent enrichment of O (the initial abundance in this model is higher than the observed value).

\begin{figure*}
    \centering
    \includegraphics[scale=0.35]{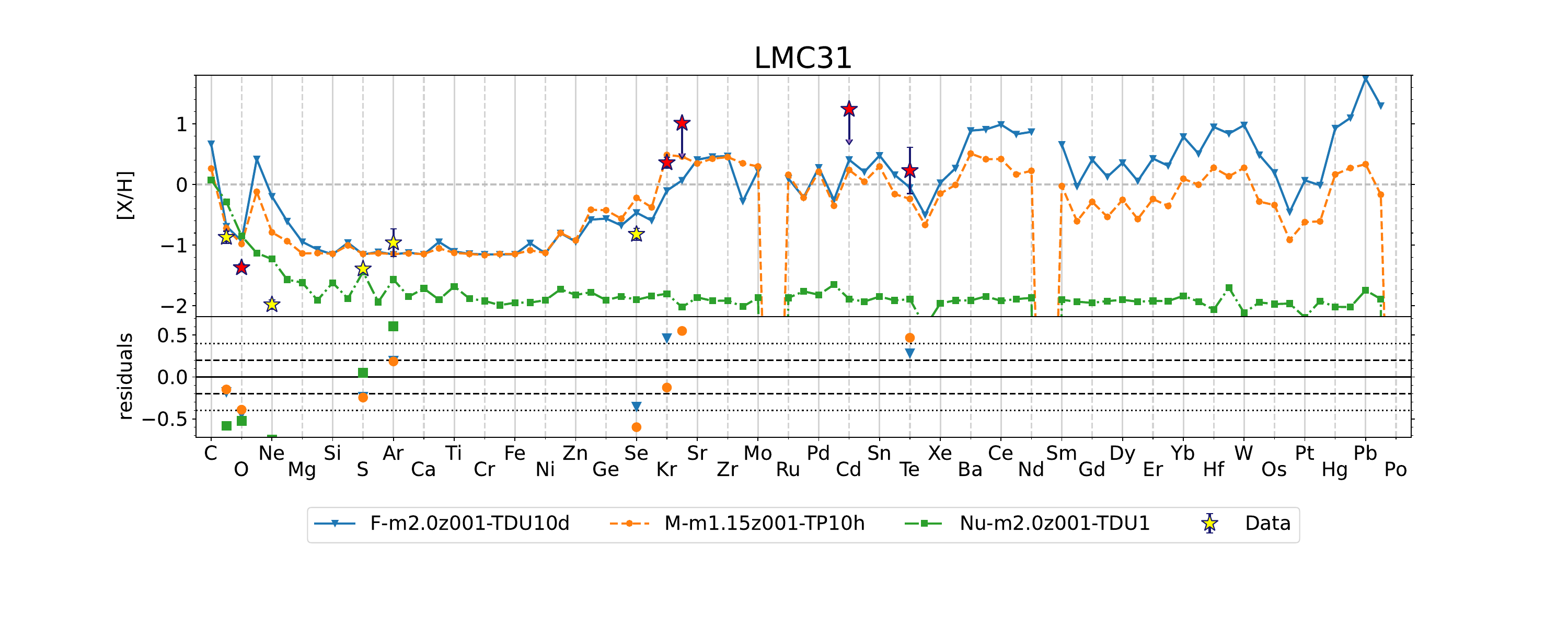}
    \caption{Same as Figure~\ref{fig:lmc48} but for LMC~31. The M model shown has a GoF of 17\%, while that of the F model is 4\%, and the Nu model is close to zero.}
    \label{fig:lmc31} 
\end{figure*}
 
\textit{LMC~47.} In this PN, the O, Ne, and S abundances indicate a metallicity of $\sim$0.003, and the best fits are obtained by models of mass 2~M$_{\odot}$ (F) and 3~M$_{\odot}$ (M). These models reproduce the measured Kr abundance, but as found for LMC~31, the predicted Se abundance is larger than measured (Figure~\ref{fig:lmc47}). For this reason, the maximum GoF found for F models is only 42\%, and 35\% for M models. However, LMC~47 is a Type~I PN, and if N is included in the fit then higher-mass models are needed for HBB activation. This also activates the $^{22}$Ne source in the M models, leading to a stronger Kr enrichment than observed. To lower the Kr enrichment, the metallicity would then need to be increased, and the Ne, S, and Se abundances would be larger than measured, decreasing the GoF to 20\% or less. This is the reason that higher-mass M models are not selected for this PN.

\begin{figure*}
    \centering
    \includegraphics[scale=0.35]{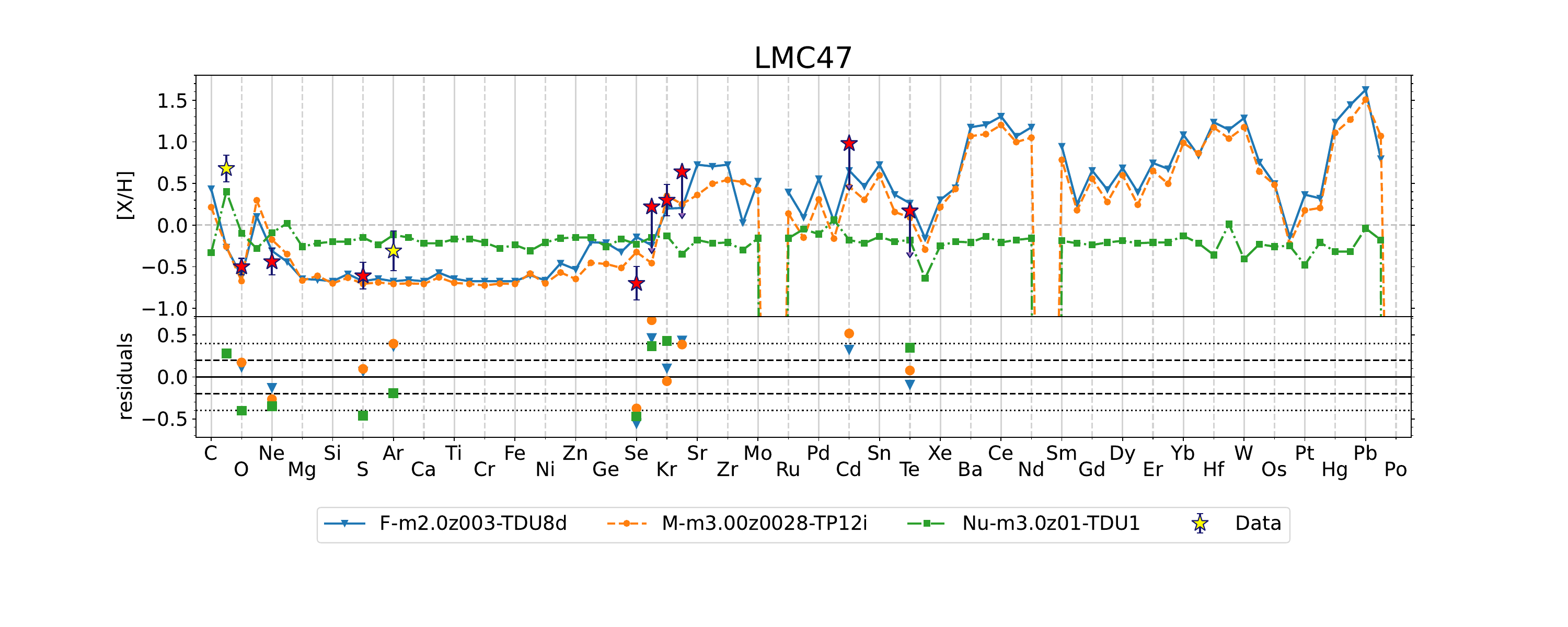}
    \caption{Same as Figure~\ref{fig:lmc48} but for LMC~47. 
    The M, F, and Nu models shown have GoFs of 35\%, 42\%, and 3\%, respectively.}
    \label{fig:lmc47} 
\end{figure*}

\textit{SMC~20.} Due to the low light element abundances, this PN is best represented by models with $Z=0.0003$--0.0006 ($[$Fe/H$]$~=~--1.7 to --1.4) and mass between 1.5 (F) and 2.5~M$_{\odot}$ (M) (Figure~\ref{fig:smc20}). The Ne, Se, and Kr abundances are lower than predicted by these models, and this results in a maximum GoF of 21\%. The $Z=0.002$ ($[$Fe/H$]=-0.85$) Nu models predict much larger N, O, and Ne abundances than are observed because of the $\alpha$-enhanced composition, and this results in a GoF~$\approx 0$\% (although the predicted abundances otherwise generally agree with the measured abundances). The relatively high Te abundance is consistent with $s$-process production at low metallicity, but Se and Kr would also be produced in that case. However, the F and M models are scaled by metallicity to the solar distribution. If the initial [Te/O] abundance is elevated by $\sim$0.3~dex, as found for heavy-\emph{s} species in SMC field giants of similar metallicity \citep[e.g.,][]{Mucciarelli_Etal_2023}, then the Te abundance in SMC~20 would be consistent within the error bars with no enrichment. In that case, a less massive progenitor that does not experience TDU would be favored. The poor fit to this PN may therefore be due to the assumption of scaled solar abundances in the F and M models rather than a physical cause. 

\begin{figure*}
    \centering
    \includegraphics[scale=0.35]{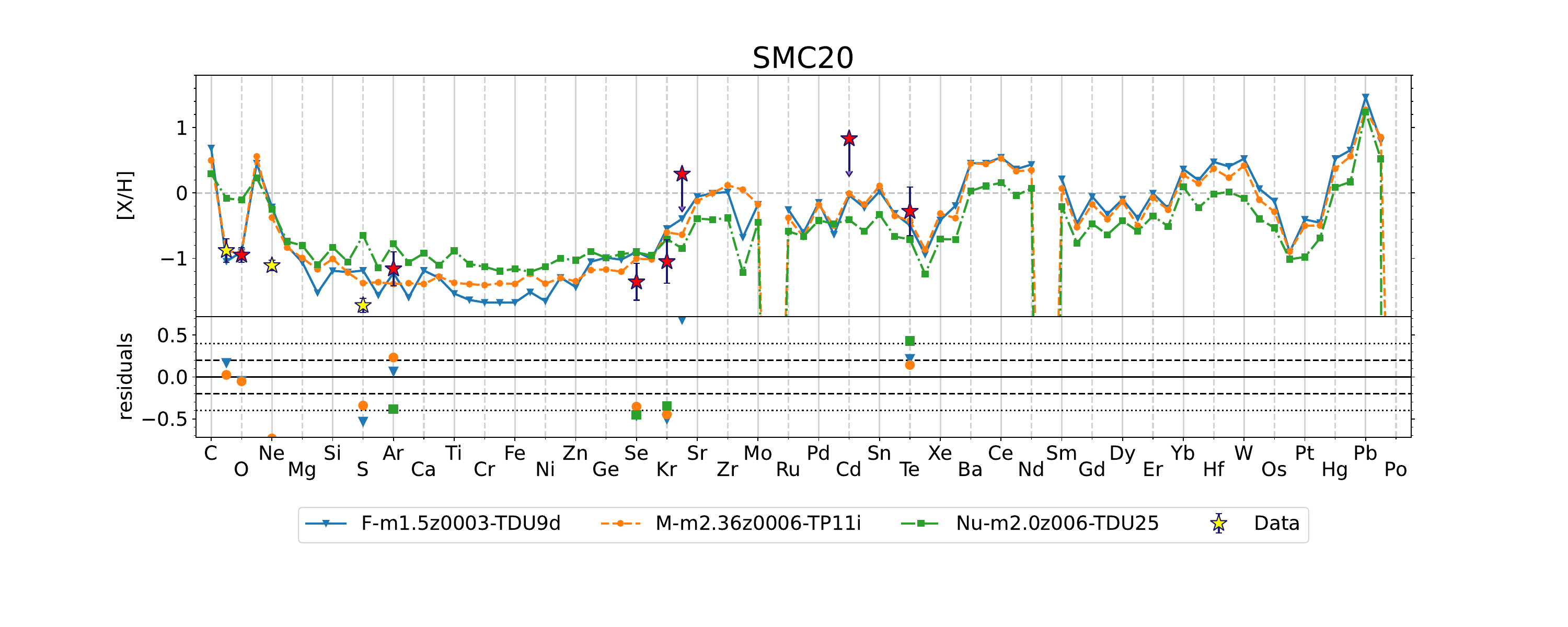}
    \caption{Same as Figure~\ref{fig:lmc48} but for SMC~20. 
    The M, F, and Nu models shown have GoFs of 18\%, 21\%, and approximately zero, respectively.}
    \label{fig:smc20} 
\end{figure*}

\section{Discussion} \label{sec:discuss}

Of the 10 PNe in which we have detected \emph{n}-capture element transitions\footnote{Because \emph{n}-capture elements were not detected in SMC~3 and SMC~25 and the upper limits do not tightly constrain the abundances, we do not discuss these two objects further.}, we find clear abundance enhancements in seven, and possibly eight depending on the initial Te abundance of SMC~20 (see \S\ref{sec:enrich}, \ref{sec:othermodels} and below). LMC~29, SMC~13, and possibly SMC~20 are the exceptions. We remind the reader that systematic uncertainties in the ICFs (\S\ref{sec:elemental}) are larger for some elements if the detected ion has a low fractional population. In particular, the Se abundance may be underestimated in low-ionization PNe \citep[][Sterling et al., in prep.]{Sterling_Etal_2017, Sterling_Etal_2025_AAS} such as LMC~31 and LMC~48, while the Te abundance in high-ionization PNe (e.g., LMC~73 and LMC~99), in which Te$^{2+}$ is a minority ion, has larger systematic uncertainties than in targets with cooler central stars. 

[Se/O] is solar or subsolar in about half of our sample, although this may be due in part to systematic uncertainties in the ICF. Se is enriched by factors of 2.5--4 in LMC~31, LMC~61, LMC~73, LMC~99, and SMC~14. Kr is more strongly enriched (factors of 5, to as large as 50 in LMC~31), as is Te (2.5--40 times solar). In LMC~99, [Rb/O] is about five times the solar value, comparable to [Kr/O], and Cd is enriched by factors of 4 and 8 in LMC~99 and LMC~48, respectively.  

SMC~20 shows no enhancement in Se or Kr, but the [Te/O] abundance of 0.67$\pm$0.37~dex suggests that it may be enriched in this element. This abundance pattern can be attributed to the low metallicity ([O/H]~=~--0.95, [Ar/H]~=~--1.16; Table~\ref{tab:append_smc}), as more neutrons are captured per Fe-peak seed nucleus, driving the \emph{s}-process to elements beyond the first peak at $Z=38$--40 \citep{Busso_Etal_1999, Karakas_Etal_2018}. The [Te/Kr] ratio (0.77$\pm$0.48) in SMC~20 is similar to the [hs/ls] ratio of a 2~M$_{\odot}$ star with SMC metallicities \citep[Figure~\ref{fig:smc20};][]{Karakas_Etal_2018}, where [hs/ls] quantifies the relative abundances of elements at the second (heavy) \emph{s}-process peak near $Z=56$ and the first (light) peak. LMC~31 also has a low metallicity ([O/H]~=~--1.37, [Ar/H]~=~--0.96), but in this object the enrichment of Kr and Te is very similar, 40--50 times solar. Given the low degree of ionization of these PNe, the Te and Kr abundances are expected to be accurate since Kr$^{2+}$ and Te$^{2+}$ comprise a significant fraction of the elemental abundances. Indeed, in LMC~31 the Kr$^{2+}$ and Te$^{2+}$ \textit{ionic} abundances (relative to O) are 25--35 times the solar elemental value. At face value, the substantial difference in [Te/Kr] among these two low-metallicity PNe is puzzling. One reason for the poor model fits to SMC~20 is that models that produce enhanced Te abundances also produce enrichments in Se and Kr. However, it is possible that the initial [Te/O] abundance of this PN was supersolar by 0.2--0.4~dex, as found for heavy-\emph{s} species such as Ba and La in SMC field giants \citep{Mucciarelli_Etal_2023}. If that is the case, the Te abundance would be consistent with no \emph{s}-process enrichment, within the error bars, as found for Se and Kr. In contrast, the [Te/Kr] ratio in LMC~31 (--0.13$\pm$0.38~dex) agrees with FRUITY model predictions (\S\ref{sec:othermodels}, Figure~\ref{fig:lmc31}) for a progenitor of $\sim$2~M$_{\odot}$ and $Z=0.001$ ([Fe/H]~=~-1.15). In other \emph{s}-process enriched PNe in which both elements are detected, the [Te/Kr] ratio ranges from --0.66$\pm$0.27~dex (LMC 73) to 0.28$\pm$0.32~dex (LMC~48), values that are consistent within the error bars with predictions from M and F models (Figures~\ref{fig:lmc48}--\ref{fig:lmc99}). There is no apparent trend with metallicity (Fig.\ \ref{fig:tekr}), though the uncertainties in [Te/Kr] may obscure such a correlation.

\begin{figure}
    \centering
    \includegraphics[scale=0.85]{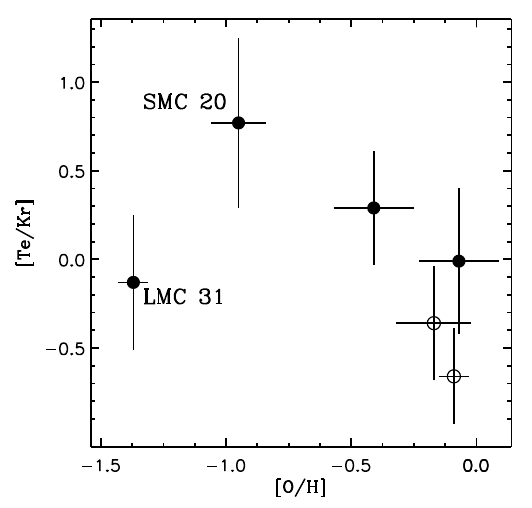}
    \caption{[Te/Kr] plotted against metallicity [O/H] for the six \emph{s}-process enriched PNe in which both Kr and Te are detected. Open symbols indicate high-ionization PNe, in which the Te abundance may be less accurate. Error bars are indicated for each object.}
    \label{fig:tekr} 
\end{figure}

The largest overall \emph{s}-process enhancements are found in LMC~31 and LMC~48, which are young PNe based on the low temperatures of their central stars (Table~\ref{tab:target_info}). Both exhibit C-rich dust \citep{Bernard-Salas_Etal_2009}, and fullerene emission has been found in LMC~48 \citep{Garcia-Hernandez_Etal_2012}. Fullerene-bearing PNe tend to be young PNe with central star effective temperatures less than 40,000--50,000~K \citep{Otsuka_Etal_FullerenePN, Sloan_Etal_2014}, and the presence of these large C molecules indicates a strongly C-rich -- and therefore \emph{s}-process enriched -- environment. \citet{Sterling_2020} noted that PNe exhibiting fullerene emission are among the most strongly \emph{s}-process enhanced PNe known, although exceptions exist such as IC~2501 \citep{Sharpee_Etal_2007}. Fullerenes have also been detected in SMC~13, and perhaps in LMC~99 and SMC~20 \citep{Garcia-Hernandez_Etal_2012, Sloan_Etal_2014}. LMC~99 is \emph{s}-process enhanced, albeit less so than LMC~48, but we do not find elevated \emph{n}-capture element abundances in SMC~13. This is especially puzzling given its C/O ratio of 4.7 \citep{Stanghellini_Etal_2009, Shaw_Etal_2010}. Thus fullerene emission in the mid-infrared spectra of PNe tends to correlate with, but is not necessarily indicative of, large \emph{s}-process enrichments.

A significant correlation is expected between \emph{s}-process and C enrichments, since these nuclei form in the same regions and are brought to the stellar envelope by TDU \citep[e.g.,][]{Busso_Etal_2001, Karakas_Lattanzio_2014}. Such correlations have been found in AGB \citep[e.g.,][]{Abia_Etal_2002, VanEck_Etal_2017, Shetye_Etal_2018} and post-AGB stars \citep{VanWinckel_2003, Menon_Etal_2025}, as well as with Kr in Galactic PNe \citep{Sterling_Dinerstein_2008, Garcia-Rojas_Etal_2015}. However, we find no statistically significant correlation between the Se, Kr, or Te abundances (relative to the metallicity reference O or Ar) of our targets and the C/O ratio. This is likely due to a combination of the magnitude of the uncertainties in our abundance determinations and those of the temperature-sensitive C lines in the UV \citep{Pena_Etal_1997, Stanghellini_Etal_2005, Stanghellini_Etal_2009}, which could mask trends.

In addition to SMC~13, LMC~29 does not show signs of \emph{s}-process enrichment. LMC~29 is a Type~I PN, with large enhancements of N and He as may be produced by HBB during the AGB phase of a $>$3--4~M$_{\odot}$ progenitor \citep{Cristallo_Etal_2015, Karakas_Lugaro_2016, Ventura_Etal_2015_LMC, MillerBertolami_2016, Karakas_Etal_2018}. Based on its CNO abundances, \citet{Ventura_Etal_2015_LMC} suggested that LMC~29 may originate from a star with initial mass as high as 6--8~M$_{\odot}$, though they acknowledge that detecting PNe from such massive progenitors is statistically unlikely, given the short evolutionary timescales and declining initial mass function. We did not find a good fit to the abundances of LMC~29 with Monash, FRUITY, or NuGrid models (\S\ref{sec:othermodels}), and thus are unable to constrain its progenitor mass, although the large $T_{\rm eff}$ of its central star \citep{Vassiliadis_Etal_1998} indicates that it is more massive than other targets in our sample. Its morphology, listed as bipolar core and possibly bipolar by \citet{Shaw_Etal_2006}, also suggests a more massive progenitor, as Galactic bipolar PNe tend to have smaller scale heights than other PNe, and hence are associated with a younger, more massive population \citep[e.g.,][]{Corradi_Schwarz_1995, Stanghellini_Haywood_2010}.

\citet{Garcia-Hernandez_Etal_2009} found large \emph{s}-process enrichments of Rb, but not Zr, in optically obscured AGB stars with masses 4--8~M$_{\odot}$ in the Magellanic Clouds. This result is consistent with enrichments with $^{22}$Ne($\alpha$,$n$)$^{25}$Mg producing \emph{s}-process neutrons \citep[e.g.,][]{Busso_Etal_1988, VanRaai_etal_2012}, particularly after revisions to the Rb abundances that accounted for circumstellar absorption \citep{Zamora_Etal_2014} and utilized updated model atmospheres \citep{Perez-Mesa_Etal_2017}. The $^{22}$Ne neutron source is expected to be increasingly important in more massive AGB stars, since their intershell temperatures are sufficiently high to overcome the Coulomb barrier between $\alpha$-particles and $^{22}$Ne, while the intershell and $^{13}$C~pocket mass decreases. If $^{22}$Ne is the dominant \emph{s}-process neutron source, then Kr is also expected to be significantly enriched along with Rb \citep{VanRaai_etal_2012}. However, as found for Galactic Type~I PNe \citep{Sterling_Dinerstein_2008, Karakas_Etal_2009, Sterling_Etal_2015}, the [Kr/Ar] abundance is not enhanced in LMC~29, and the 3-$\sigma$ upper limit to the Rb abundance allows for at most modest enrichment. This statement depends on the metallicity reference ([Kr/O]~=~$0.44\pm0.18$), but as noted in \S\ref{sec:enrich}, the low O abundance in LMC~29 compared to S and Ar suggests that O may have been destroyed during HBB.

LMC~47 is also a Type~I PN, though with a less extreme N/O ratio than LMC~29 that, along with its lower central star temperature \citep{Dopita_Meatheringham1991a}, suggests its progenitor is not as massive \citep[\S\ref{sec:othermodels}, Table~\ref{tab:target_info};][]{Ventura_Etal_2015_LMC}. In LMC~47, [Kr/O] and [Kr/Ar] are supersolar, but the upper limits to the Rb, Cd, and Te abundances do not allow conclusions to be drawn regarding the neutron source. Unfortunately, the abundance upper limits to the third Type~I PN in our sample, SMC~25, are large and do not meaningfully constrain its heavy element composition.

The lack of \emph{s}-process enrichments in Type~I PNe in the Galaxy and the Magellanic Clouds may be due to heavy dilution of nucleosynthesized material into the massive envelope \citep[e.g.,][]{Karakas_Etal_2009, Karakas_Lugaro_2016, Karakas_Etal_2018}, as conjectured by \citet{Sterling_Dinerstein_2008}. But the discovery of Rb-rich, Zr-poor OH/IR stars with initial masses 4--8~M$_{\odot}$ \citep{Garcia-Hernandez_Etal_2006, Garcia-Hernandez_Etal_2009} seems to contradict this possibility. Instead, the rapid evolutionary time of Rb-rich 4--8~M$_{\odot}$ stars after leaving the AGB \citep{MillerBertolami_2016} may make their PNe unlikely to be detected, or render them too faint for \emph{n}-capture elements to be detected. Alternatively, N enrichments might occur in less massive AGB stars, perhaps due to a lower mass threshold for HBB than is typically found in models as suggested by \citet{Henry_Etal_2018} and \citet{Gallera-Rosillo_Etal_2022}, or due to an extra-mixing mechanism during or before the AGB \citep[][and references therein]{Karakas_Lattanzio_2014} that possibly results from binary interactions. In that case, relatively bright Type~I PNe such as LMC~29 and the Galactic objects studied by \citet{Sterling_Dinerstein_2008} may descend from less massive progenitors than the Rb-rich AGB stars in the samples of \citet{Garcia-Hernandez_Etal_2006, Garcia-Hernandez_Etal_2009}.

\subsection{Comparison to Previous Results} \label{sec:compare}

Four targets in our sample (LMC~47, LMC~73, LMC~99, and SMC~20) were previously observed in the NIR by \citet[][hereafter M16]{Mashburn_Etal_2016}, using the FIRE spectrograph on the 6.5-m Baade Telescope. Those authors detected at least one of [\ion{Kr}{3}]~2.1986 and [\ion{Se}{4}]~2.2864~$\mu$m in the three LMC objects in common, and derived ionic and elemental abundances of each following a similar methodology to ours. The primary difference is that M16 adopted light element abundances from the literature rather than re-computing them.

We compare our abundances to those found by M16 in Table~\ref{tab:compare}. Overall, the agreement is excellent for lines that are detected in both studies. Note that M16 did not account for the line widths in their upper limits to the fluxes of undetected lines, and hence their abundance upper limits are underestimated by factors of $\sim$5 (0.7~dex). It is for this reason that we do not compare our Rb and Cd abundances to the upper limits of M16. Additionally, M16 did not compute upper limits to the Te abundance, since [\ion{Te}{3}]~2.1019~$\mu$m was not identified until later \citep{Madonna_Etal_2018}.

The most significant difference that we find in the ionic abundances is that of Kr$^{2+}$/H$^+$ in LMC~47, as our value is approximately half that found by M16. This may be due to the relatively low S/N of our [\ion{Kr}{3}] detection (Fig.\ \ref{fig:lmc_stack}). In the case of LMC~73, we find a larger Kr abundance than the upper limit of M16, even after accounting for a line width of $\sim$50~km~s$^{-1}$ in their flux upper limit, because we also include Kr$^{3+}$/H$^+$ (Mampaso et al., in prep.) in our abundance determination. The elemental abundance derived from two Kr ions is more accurate than that from Kr$^{2+}$/H$^+$ alone, which we find to be $\sim$60\% lower but with larger error bars ($\pm$0.18~dex, or 50\%).

\begin{deluxetable*}{l|cc|cc|cc|cc}
\tablecaption{Comparison of Se and Kr Abundances with \citet{Mashburn_Etal_2016}} \label{tab:compare}
\tabletypesize{\scriptsize}
\tablewidth{0pt}
\tablehead{\\
\colhead{} & \multicolumn{2}{|c|}{LMC~47} & \multicolumn{2}{c|}{LMC~73} & \multicolumn{2}{c|}{LMC~99} & \multicolumn{2}{c}{SMC~20} \\
\colhead{Abund.} & \multicolumn{1}{|c}{This Work} & \colhead{M16} & \multicolumn{1}{|c}{This Work} & \colhead{M16} & \multicolumn{1}{|c}{This Work} & \colhead{M16} & \multicolumn{1}{|c}{This Work} & \colhead{M16}}
\startdata
Se$^{3+}$/H$^+$ & (2.1$\pm$0.3)E-10 & (1.9$\pm$0.6)E-10 & (1.8$\pm$0.1)E-09 & (1.9$\pm$0.3)E-09 & (1.9$\pm$0.2)E-09 & (2.0$\pm$0.2)E-09 & (6.8$\pm$0.9)E-11 & $\leq2.4$E-11 \\
$[$Se/H$]$ & --0.70$\pm$0.21 & --0.74$\pm$0.30 & 0.24$\pm$0.09 & 0.21$\pm$0.29 & 0.24$\pm$0.45 & 0.25$\pm$0.24 & --1.36$\pm$0.28 & $\leq -1.82$ \\
$[$Se/O$]$ & --0.20$\pm$0.21 & --0.24$\pm$0.32 & 0.37$\pm$0.11 & 0.30$\pm$0.31 & 0.41$\pm$0.46 & 0.53$\pm$0.26 & --0.42$\pm$0.29  & $\leq -0.87$ \\
Kr$^{2+}$/H$^+$ & (7.1$\pm$1.4)E-10 & (1.5$\pm$0.3)E-09 & (2.2$\pm$0.2)E-09 & $\leq5.8$E-10 & (1.7$\pm$0.1)E-09 & (2.2$\pm$0.2)E-09 & (8.1$\pm$2.3)E-11 & $\leq2.4$E-10 \\
$[$Kr/H$]$ & 0.30$\pm$0.19 & 0.62$\pm$0.20 & 0.89$\pm$0.06 & $\leq 0.21$ & 0.62$\pm$0.21 & 0.65$\pm$0.21 & --1.05$\pm$0.33 & $\leq -0.54$ \\
$[$Kr/O$]$ & 0.79$\pm$0.20 & 1.12$\pm$0.23 & 1.03$\pm$0.09 & $\leq 0.30$ & 0.78$\pm$0.24 & 0.93$\pm$0.23 & --0.10$\pm$0.34 & $\leq0.41$ \\
\enddata
\tablecomments{The logarithmic Kr abundances of \citet[][M16]{Mashburn_Etal_2016} have been increased by 0.13~dex to account for the lower solar abundance found by \citet{Asplund_Etal_2021} compared to earlier estimates \citep{Asplund_Etal_2009}. M16 did not account for line widths in upper limit measurements of undetected lines, and thus their abundance upper limits are underestimated by factors of $\sim$5 (0.7~dex).}
\end{deluxetable*}

\subsection{Sensitivity to Parameters Derived from Optical Spectra} \label{sec:optical}

Our derived \emph{n}-capture element abundances rely on light element ionic and total gas phase abundances from optical spectroscopy. In particular, the ICFs \citep{Sterling_Etal_2015, Sterling_Etal_2016, Madonna_Etal_2018} require the ionic fractions O$^{2+}$/O (Se, Rb, Cd, and in LMC~73, Kr), S$^{2+}$/S (Kr), Ar$^{2+}$/Ar (Te), and Ar$^{3+}$/Ar (Br). Furthermore, the enrichment factors are computed relative to a metallicity reference, O or Ar. 

For SMC~3, SMC~25, and the LMC PNe, we use \citet[][hereafter LD06]{Leisy_Dennefeld_2006} for extinction coefficients, physical parameters $T_{\rm e}$ and $n_{\rm e}$, and optical line intensities. The LD06 data are a compilation that include their own observations as well as spectra from the literature. The light element abundances for LMC~29, LMC~31, and LMC~61 were derived from the spectra of \citet{Meatheringham_Dopita_1991a, Meatheringham_Dopita_1991b}, while the others were observed by \citet{Leisy_Dennefeld_1996} or LD06. While LD06 did exceptional work in analyzing optical spectra for a large number of LMC and SMC PNe, using consistent criteria to determine physical conditions and abundances, their results were sometimes limited by the quality of the data they utilized. 

For objects in common to the samples observed by Leisy \& Dennefeld and Meatheringham \& Dopita, the intensities disagree for lines at the edges of the spectra, affecting important lines such as [\ion{O}{2}]~$\lambda$3727 and [\ion{Ar}{3}]~7136 and 7751 \AA. Leisy \& Dennefeld attributed these discrepancies to imperfect corrections for extinction or the spectral response in one of the datasets \citep[see also][]{Shaw_Etal_2010}. This may be the cause of the anomalously high Ar abundance (2.5 times solar) for LMC~99 that we compute (\citealt{Leisy_Dennefeld_1996} found Ar/H to be 1.7 times larger than solar). In LMC~31, \citet{Meatheringham_Dopita_1991a} did not find a consistent extinction coefficient from different Balmer lines, with $c_{\rm H\beta}$ ranging from 0.05 (H$\delta$/H$\beta$) to 1.06 (H$\alpha$/H$\beta$). LD06 used a weighted average of $c_{\rm H\beta}=0.79$, and this may be the culprit for the larger [Ar/H] abundance (--0.96 dex), derived from [\ion{Ar}{3}]~$\lambda$7136, compared to [O/H] (--1.37~dex). The temperature diagnostic [\ion{N}{2}]~$\lambda$5755 was not detected in LMC~29 \citep{Meatheringham_Dopita_1991b}, and thus $T_{\rm e}$([\ion{O}{3}]) was used for both low- and high-ionization species.

The adopted values of the electron densities also affect ionic abundances. For the PNe in our sample other than LMC~73 (Mampaso et al., in prep.) and SMC~14 \citep{Shaw_Etal_2010}, $n_{\rm e}$ was computed from [\ion{S}{2}]~$\lambda$6731/$\lambda$6716 or (in SMC~13, SMC~14, and SMC~20) $[$\ion{O}{2}$]$~$\lambda$3726/$\lambda$3729. Other density diagnostic lines, such as [\ion{Cl}{3}]~$\lambda\lambda$5518, 5538 and [\ion{Ar}{4}]~$\lambda\lambda$4711, 4740 are weaker and were not detected in most objects in our sample. Recent work has shown that $n_{\rm e}$([\ion{S}{2}]) is systematically lower than $n_{\rm e}$([\ion{Ar}{4}]) in PNe \citep{Delgado-Inglada_Etal_2020} and \ion{H}{2} regions \citep{Mendez-Delgado_Etal_2023}. Whether this result is due to density stratification or to the decreasing sensitivity of the [\ion{S}{2}]~$\lambda$6731/$\lambda$6716 diagnostic ratio at densities $>10^4$~cm$^{-3}$, the [\ion{S}{2}] density may not be representative of the entire nebula. Uncertainties in $n_{\rm e}$ can introduce biases in the $T_{\rm e}$ determination \citep{Mendez-Delgado_Etal_2023} and in abundances derived from NIR lines.

A deep optical spectrum of LMC~73 was recently observed by Mampaso et al.\ (in prep.) with VLT/FORS2. Those authors found $n_{\rm e}\sim 10^4$~cm$^{-3}$ from [\ion{S}{2}], [\ion{Cl}{3}], and [\ion{Ar}{4}], compared to 4500~cm$^{-3}$ by LD06. While the [\ion{O}{3}] temperatures in the two studies agree well, Mampaso et al.\ found a $T_{\rm e}$([\ion{N}{2}]) $\sim$2000~K higher than LD06's value. The differences result in Kr$^{2+}$/H$^+$ and Te$^{2+}$/H$^+$ abundances 20\% lower than if the physical conditions of LD06 are used. However, the Kr/H (derived from Kr$^{2+}$) and Te/H abundances differ by only $\sim$5\%, due to the lower S$^{2+}$ and Ar$^{2+}$ ionic fractions that produce larger ICF values than found from the LD06 spectrum. Therefore the effects of optical data quality on \emph{n}-capture element abundances are not always large, but this is not expected to be generally true.

Finally, \citet{Delgado-Inglada_Etal_2015} argued that Cl is the best metallicity tracer in PNe, since it is unaffected by AGB nucleosynthesis (whereas O may be) and has smaller abundance uncertainties than Ar \citep{Delgado-Inglada_Etal_2014}. Unfortunately, existing optical spectra of our targets \citep[][and references therein]{Leisy_Dennefeld_2006, Shaw_Etal_2010} are not sufficiently deep for Cl lines to have been detected, with the exceptions of LMC~73 and LMC~99.

Deep, high-quality optical spectra provide more accurate physical conditions and light element abundances that are needed to interpret NIR \emph{n}-capture element lines. Moreover, the detection of [\ion{Kr}{4}] and other optical transitions allow more accurate and precise \emph{n}-capture element abundances to be determined, and hence more tightly constrain model predictions. It is possible that the poor model fits to the abundances of some of our targets (\S\ref{sec:othermodels}) are due in part to the quality of the optical spectra. To mitigate these issues, improved optical data are needed for PNe in the Magellanic Clouds, particularly the LMC.

\subsection{Implications for the \emph{s}-process in AGB Models}
\label{sec:generalcons}

We found good fits with M and F models for five out of the ten PNe in which \emph{n}-capture elements were detected. For these objects the main advantage is that elements belonging to both the first and second $s$-process peaks are detected, except for SMC 14, for which the fit was based only on the first-peak elements Se and Kr. In general, the goodness-of-fit to the observed abundances is not appreciably better for M or F models, and the best fits are found for masses in the range $2-4~M_{\odot}$. These masses are typical of models that experience efficient TDU and therefore significantly produce \emph{s}-process nuclei. The best fits also tend toward standard choices of the $^{13}$C pocket mass in the considered models. The largest difference between the M and F sets is found for LMC~73 and LMC~99, where M models of mass around 4~M$_{\odot}$ provide the best fit to the relatively high Kr abundance, while for the F set less massive stars ($\sim$2~M$_{\odot}$) are selected as the best fits. The main discriminant here is the abundance of Kr, which is more enhanced in the higher-mass M models due to a strong activation of the $^{22}$Ne neutron source. In the case of LMC~73 specifically, the small error bars on the Kr abundance favor the higher GoF of the more massive M models over the lower-mass F models. The reason that more massive F models are not selected for LMC~73 and LMC~99 is because they do not reach as high of interior temperatures as M models, and therefore the $^{22}$Ne neutron source is not strongly activated. If the F model mass were increased, the predicted \emph{s}-process enrichments would be smaller due to the less massive intershell region (and hence PMZ and $^{13}$C pocket), which would worsen the agreement with the empirical abundances. 

We were unable to find good fits to the compositions of LMC~29, LMC~31, LMC~47, and SMC~20 (\S\ref{sec:othermodels}). These objects generally have less data available: typically only Se and Kr are detected in these nebulae, and some are not clearly enhanced by the $s$-process (LMC~29 and, depending on the initial Te abundance, SMC~20). The poor fits are partially due to the low Se abundance relative to Kr in LMC~29, LMC~31, and LMC~47, which may be due to systematic uncertainties in the Se ICF \citep[][Sterling et al., in prep.]{Sterling_Etal_2025_AAS}. In SMC~20, the lack of Se and Kr enhancements is inconsistent with the $[$Te/O$]$ abundance of 0.67$\pm$0.37. However, the initial Te abundance may have been supersolar, as found for other heavy-\emph{s} species \citep{Mucciarelli_Etal_2023}, which is not accounted for in the models. For the LMC objects in this group, particularly LMC~29 and LMC~31, the quality of the optical data may also contribute to the poor fits (see \S\ref{sec:optical}). 

The observed abundances of the \emph{n}-capture species are not well fit by NuGrid models. For example, the best GoF found (for LMC 29) was only 5\%. However, this is due in part to the limited number and scope of available NuGrid models (a factor of $\sim$10 and 20 times fewer than for the F and M databases, respectively), including a less extensive and dense grid of initial mass and metallicity than those of the F and M sets. A further complication is that the relative abundances of some of the light species used as proxies for metallicity differ between the NuGrid models on the one hand, and the M and F models on the other, with the NuGrid models assuming $\alpha$-enhanced compositions for subsolar metallicities (as discussed further below).

However, for PNe with progenitor masses near 2~M$_{\odot}$, for which subsolar metallicity Nu models are available, or those with nearly solar metallicities, the comparison with observations is more meaningful. Specifically, LMC~61, LMC~73, and LMC~99 are best fit by 3~M$_{\odot}$, $Z=0.01$ models, slightly higher metallicity than the best-fitting M and F models. The Nu models do not produce sufficiently large \emph{s}-process enrichments for these PNe, especially for the first-peak elements Se, Kr, and Rb, compared to the observed abundances or M and F models. This is partially due to the less massive PMZ of the Nu models (2--3 times lower than in the M and F models of the same mass), resulting in a smaller amount of enriched material that is dredged up. For the same reason, the low-metallicity 2~M$_{\odot}$ Nu models do not match the observed enrichments of SMC~14 or LMC~31. Another factor contributing to the poor fits of the Nu models to LMC~61, LMC~73, and LMC~99 is that for 0.01~$<Z<$~0.03, Nu models include convective mixing into the core at the bottom of the He-intershell during thermal pulses \citep{Battino_Etal_2019}. This core overshoot is required to match the surface abundances of PG~1159 stars, but leads to large N and O enhancements that are not observed in these PNe. With the caveat that the metallicities of these PNe are not a perfect match to the $Z=0.01$ Nu models, the fact that M and F models are able to reproduce the N and O abundances suggests that significant amounts of boundary mixing of the convective thermal pulse into the CO core does not occur. Core overshoot is not included in the $Z=0.001$--0.006 models of \citet{Battino_Etal_2021}, but the GoF of the 2~M$_{\odot}$ Nu models are low because of the $\alpha$-enhanced initial compositions. Definitive conclusions regarding Nu models thus require a grid over a larger range of initial mass and metallicity, as well as models without $\alpha$-element enhancement or convective boundary mixing at the base of the thermal pulses.

\subsection{Implications for the \emph{r}-process and Galactic Chemical Evolution}

Se and Te have been tentatively detected in the \emph{r}-process enriched kilonovae AT2017gfo and GRB~230307A / AT2023vfi \citep{Hotokezaka_Etal_2022, Hotokezaka_Etal_2023, Levan_Etal_2023}. The large expansion velocities of these objects lead to very broad lines that, combined with modeling uncertainties, hinder accurate abundance determinations of individual elements \citep[e.g.,][]{Watson_Etal_2019}. Observations of these species in PNe constrain their \emph{s}-process production, and in turn their \emph{r}-process production, although $s$-process Se is believed to be mostly a product of the weak \emph{s}-process in massive stars \citep[e.g.,][]{Prantzos_Etal_1990, The_Etal_2007, Pignatari_Etal_2010, Kobayashi_Etal_2020} and its $r$-process component may have contributions from sources other than binary neutron star mergers \citep[e.g.,][]{Cowan_Etal_2021}. The good abundance matches we find using $s$-process models to four PNe in which Te is observed (LMC~48, LMC~61, LMC~73, and LMC~99) indicate that the $s$-process contribution to this element from AGB stars is of the order predicted by galactic chemical evolution \citep[$\sim$15-20\%;][]{Arlandini_Etal_1999, Bisterzo_Etal_2014, Prantzos_Etal_2018, Kobayashi_Etal_2020}. Its $r$-process residual is thus constrained to be 80-85\% of the solar Te abundance. If kilonova observations can be used to determine Te abundances \citep[e.g., see][]{Jerkstrand_Etal_2026}, it will be interesting to establish whether they are consistent with this contribution to the solar value or if there is variability. The abundance of at least one other \emph{n}-capture element in kilonovae would be needed to establish the relative abundances.

\section{Summary and Future Work} 
\label{sec:Conclusions}

Our major findings are as follows:
\begin{enumerate}

\item We have presented H and K~band IGRINS spectra of 12 PNe in the LMC and SMC, obtained to investigate \emph{s}-process enrichments of \emph{n}-capture elements including Se, Kr, Rb, Cd, and Te. The high resolution of IGRINS proved invaluable to the identification of spectral features, by means of their precise wavelengths and line profiles. In PNe hosting vibrationally-excited H$_2$, we resolved [\ion{Se}{4}]~2.2864~$\mu$m from the adjacent H$_2$~3-2~S(2)~2.2870~$\mu$m line. The resolved line profiles enabled us to identify numerous vibrationally-excited H$_2$ transitions in LMC~31 and LMC~47, as well as to eliminate a potentially observable \emph{n}-capture element transition  ([\ion{Sn}{2}]~2.3521~$\mu$m) as an identification.

\item We detect two or more \emph{n}-capture element transitions in 10 of the 12 PNe. [\ion{Kr}{3}]~2.1986 and [\ion{Se}{4}]~2.2864~$\mu$m were detected in each of these 10 objects, and [\ion{Te}{3}]~2.1019~$\mu$m in seven. The weaker H~band lines, [\ion{Rb}{4}]~1.5973 and [\ion{Cd}{4}]~1.7204~$\mu$m, were detected in 1--2 PNe (LMC 48, LMC~99). We derive ionic abundances using published values of electron temperatures and densities. Total elemental abundances for Se and Kr are computed with model-based ICFs from \citet{Sterling_Etal_2015}. To determine abundances of the other \emph{n}-capture elements, for which atomic data affecting the ionization balance are incomplete or unknown, we use approximate ICFs based on similarities in the ionization potential ranges of the detected ions and those of widely-observed O, S, and Ar ions.  We recompute ionic fractions of the latter from published optical line intensities using more recent atomic data and ICF formulations. Our \emph{n}-capture element abundance determinations can have significant systematic uncertainties in cases where the observed ions contain small fractions of their elements, e.g., Se$^{3+}$ in low-ionization PNe and Te$^{2+}$ in high-ionization objects. 

\item Determining elemental \emph{s}-process enrichments requires a metallicity reference that reliably indicates the initial composition. The abundances of field giants in the Magellanic Clouds indicate that $\alpha$-elements closely track [Fe/H] in the metallicity range of our targets \citep[e.g.,][]{Hasselquist_Etal_2021, Mucciarelli_Etal_2023}. Since O generally has the most accurately determined abundance among the light elements in PNe, we choose [O/H] as the reference element in most cases. The exception is the extreme Type~I PN LMC~29, for which we use [Ar/H], since its O abundance may have been modified by HBB. An additional consideration is that \emph{n}-capture elements near the heavy-\emph{s} peak have been found to have supersolar abundances relative to Fe in LMC and SMC field giants, while light-\emph{s} species follow the solar distribution. Therefore our derived [Te/(O, Ar)] values may need to be revised downward by $\sim$0.3~dex, in order to correctly determine \emph{s}-process enhancements.

\item Seven of the ten PNe in which we detect \emph{n}-capture element lines exhibit \emph{s}-process enrichments, ranging from factors of two, to as high as 40-50 times solar for Kr and Te. SMC~20 may be enriched in Te ([Te/O]~=~$0.67\pm0.37$), but the uncertainty is sufficiently large that we cannot eliminate the possibility of no enrichment if the progenitor star had an initial [Te/O] of $\sim$0.3~dex. We find solar or subsolar [Se/O] in half the targets, although this might result from systematic uncertainties in the ICF for low-ionization PNe \citep{Sterling_Etal_2017, Sterling_Etal_2025_AAS}. All the observed LMC PNe except LMC~29 appear to be  \emph{s}-process enriched, although this is likely a selection effect, since our sample of primarily C-rich PNe was chosen to quantitatively measure the magnitude of such enrichments. In contrast, SMC~14 is the only observed SMC PN that clearly displays elevated \emph{n}-capture element abundances. 

\item We compare our observational results to enrichments predicted by three AGB evolutionary codes: Monash \citep[M;][]{Karakas_Lugaro_2016, Karakas_Etal_2018}; FRUITY \citep[F;][]{Cristallo_Etal_2015}; and NuGrid \citep[Nu;][]{Pignatari_Etal_2016, Battino_Etal_2019, Battino_Etal_2021}. We use a machine-learning algorithm to search abundance patterns from a large parameter space of initial mass, metallicity, and number of thermal pulses during the AGB evolution. The goal is to find which models (and progenitor star properties) provide the best fits to the empirical compositions. For 5 PNe, we find good matches (goodness-of-fit or GoF~$>50$\%) with models of 2--4~M$_{\odot}$ and metallicities $Z=0.003$--0.010 ($[$Fe/H$]=-0.7$ to --0.15). Models from both the M and F sets fit the measured abundances well, although only the M models reproduce the high Kr abundances of LMC~73 and LMC~99. The large Kr enrichments of these two PNe, relative to other \emph{n}-capture elements such as Te, require $\sim4$~M$_{\odot}$ models in which the $^{22}$Ne neutron source is activated. This mechanism increases the production of both Kr and Rb compared to other \emph{n}-capture species \citep{VanRaai_etal_2012, Karakas_Etal_2012}. In contrast, the intershell temperatures of F models are too low to significantly activate the $^{22}$Ne source. As a result, the preferred F models for these PNe have lower initial masses and poorer GoF. We were unable to find high GoF models for the other targets in which \emph{n}-capture elements were detected, which include the most metal-poor PNe in our sample (LMC~31 and SMC~20). For LMC~29, LMC~31, and LMC~47 the low GoF is due in part to low [Se/(O, Ar)] abundances, which may be attributed to a systematic effect that applies mainly to low-ionization PNe such as LMC~31. Uncertainties in the light element abundances derived from optical studies (Tables~\ref{tab:append_lmc} and \ref{tab:append_smc}), and/or gaps in parameter-space coverage of the available models, may also contribute to our inability to fit these PNe. Nu models do not fit the abundances of any of the observed PNe, but this is likely due to the narrow range and low density of masses and metallicities in the available Nu models. For PNe with progenitors of mass $\sim$2~M$_{\odot}$, for which appropriate metallicities are available, Nu models have larger O and Ne abundances than observed due to the $\alpha$-enhanced initial compositions, while the \emph{s}-process enrichments are generally lower than observed, as a result of the smaller $^{13}$C pocket mass compared to M and F models. The light element compositions of the Nu models are the primary factor leading to poor fits to the observed abundances.

\item The progenitor of the Type~I PN LMC~29 was likely the most massive star in the sample, based on its composition and high central star temperature. A similar lack of significant \emph{s}-process enrichment has been found in Galactic Type~I PNe \citep{Sterling_Dinerstein_2008, Karakas_Etal_2009, Sterling_Etal_2015}. This suggests that the 4--8~M$_{\odot}$ Rb-rich AGB stars observed by \citet{Garcia-Hernandez_Etal_2006, Garcia-Hernandez_Etal_2009} produce PNe that fade so rapidly \citep{MillerBertolami_2016} that their faint \emph{n}-capture element lines fall below current detection limits.

\end{enumerate}

Reducing the remaining uncertainties and producing precise and robust values for \emph{s}-process enrichments in PNe calls for progress in several areas. From an observational standpoint, the quality of existing optical data for Magellanic Cloud PNe represent one of the larger sources of systematic uncertainties, due to the reliance of ICFs on the fractional abundances of light element ions and the need for a metallicity reference to determine \emph{s}-process enrichment factors. These abundances in turn depend on values of $n_{\rm e}$ and $T_{\rm e}$ reported in the literature. The only density diagnostic lines previously detected in most targets in our sample are [\ion{S}{2}]~$\lambda\lambda$6716, 6731 and [\ion{O}{2}]~$\lambda\lambda$3726, 3729. These diagnostic ratios saturate at $n_{\rm e}>10^4$~cm$^{-3}$, and may systematically underestimate the density in some PNe \citep{Delgado-Inglada_Etal_2020}. Inaccurate $n_{\rm e}$ values affect the derived $T_{\rm e}$ \citep{Mendez-Delgado_Etal_2023} and abundances. Some of our derived [Ar/H] values are either implausibly large (LMC~99) or disagree with [O/H] (LMC~31), which might be due to inaccurate extinction and/or response corrections, as noted by \citet{Leisy_Dennefeld_2006}. Finally, arguably the best metallicity reference element, Cl, has not been detected in 10 of the observed PNe.

Deep, high-quality optical data can correct the issues noted above. Moreover, detecting optical lines of \emph{n}-capture element ions such as [\ion{Kr}{4}] can produce more accurate and precise abundances than from NIR spectra alone, as we have shown for LMC~73. This is due both to the detection of additional \emph{n}-capture element ions, and better-constrained values of $n_{\rm e}$ and $T_{\rm e}$ that employ additional diagnostic ratios (some of them based on higher-excitation light-element ions that are co-extensive with high-ionization \emph{n}-capture element species). Furthermore, the measurement of multiple ions of a given \emph{n}-capture element  element can constrain the ionization balance of that element and enable improved formulations of its ICFs \citep{Sterling_Etal_2015, Sterling_Etal_2025_AAS}. Such ICFs can then be applied to sources where only a single ion is observed, allowing accurate abundance determinations for these objects. Aiming at these goals, we will present new optical data for PNe in our sample and that of \citet{Mashburn_Etal_2016} in a forthcoming paper.

From the point of view of theoretical predictions, it would be interesting to investigate whether the Se and Kr abundances could be a signature of the $intermediate$ \emph{n}-capture ($i$) process instead of the \emph{s}-process. Evidence for the \emph{i}-process has been observed in the abundance patterns of post-AGB stars in  the LMC and SMC \citep{Lugaro_Etal_2015, Hampel_Etal_2019}, and one of its characteristics is a relatively high production of Kr. This may be in line with the compositions of some of the observed PNe (e.g., LMC~31, LMC~47, LMC~73, and LMC~99). In this context, the reported Rb abundances, even if only upper limits, represent a crucial constraint, as the $i$-process also produces large Rb enrichments. Both the \emph{s}-process, with significant neutron production by $\alpha$-captures onto $^{22}$Ne, and the \emph{i}-process predict large enhancements of Kr and Rb. However, the source of enrichment can be distinguished by the relative abundances of elements near the second \emph{s}-process peak, such as Te and Xe. In particular, \emph{i}-process nucleosynthetic models \citep{Hampel_Etal_2016, Hampel_Etal_2019} predict Xe to have enrichments 3--4 times (0.5--0.6~dex) larger than Te, while \emph{s}-process enrichments of these two elements are predicted to be similar in magnitude (see Figures~\ref{fig:lmc48}--\ref{fig:smc20}). Xe abundances in PNe can be obtained from deep optical \citep[e.g.,][]{Pequignot_Baluteau_1994, Manea_Etal_2022} and/or J~band observations \citep{Sterling_2020, Dinerstein_Etal_2022_AAS}. If the \emph{i}-process is responsible for \emph{n}-capture element enrichments in LMC~73 and LMC~99, their progenitor masses may not be as large as we infer. Similarly, a comprehensive analysis of heavy element abundances in the Rb-rich OH/IR stars in the Magellanic Clouds \citep{Garcia-Hernandez_Etal_2009, Zamora_Etal_2014} would be valuable for assessing the source of enrichment.

\vspace{5 mm}

\begin{acknowledgments}

This work used the Immersion Grating Infrared Spectrometer (IGRINS), which was developed under a collaboration between the University of Texas at Austin and the Korea Astronomy and Space Science Institute (KASI), with the financial support of the US National Science Foundation under grants AST-1229522, AST-1702267 and AST-1908892, McDonald Observatory of the University of Texas at Austin, the Korean GMT Project of KASI, the Mt.\ Cuba Astronomical Foundation and Gemini Observatory.

The observations reported in this paper were obtained at the international Gemini Observatory, a program of NSF NOIRLab, which is managed by the Association of Universities for Research in Astronomy (AURA) under a cooperative agreement with the U.S. National Science Foundation on behalf of the Gemini Observatory partnership: the U.S. National Science Foundation (United States), National Research Council (Canada), Agencia Nacional de Investigaci\'{o}n y Desarrollo (Chile), Ministerio de Ciencia, Tecnolog\'{i}a e Innovaci\'{o}n (Argentina), Minist\'{e}rio da Ci\^{e}ncia, Tecnologia, Inova\c{c}\~{o}es e Comunica\c{c}\~{o}es (Brazil), and Korea Astronomy and Space Science Institute (Republic of Korea). 
We appreciate the allocation of IGRINS compensatory time to University of Texas at Austin observers that enabled pilot observations of LMC~47 and LMC~99 which served as proof of concept for the larger survey program.

We are grateful for the efforts of the Gemini-South staff to obtain these observations, some of which were taken during the COVID-19 pandemic, and thank the referee for a careful review that improved this paper. NCS acknowledges support from NSF Award AST 2307116, and HLD from AST 1715332 and 2307117. JGR acknowledges support from the Agencia Estatal de Investigaci\'on of the Ministerio de Ciencia, Innovaci\'on y Universidades (AEI-MCIU) and from the European Regional Development Fund (ERDF) under grant ``Planetary nebulae as the key to understanding binary stellar evolution'' with reference PID2022-136653NA-I00 (DOI:10.13039/501100011033). ML acknowledges the support of the LP2023-10 Lend\"ulet grant of the Hungarian Academy of Sciences and the NKFIH excellence grant TKP2021-NKTA-64.

\end{acknowledgments}

\vspace{5 mm}

\appendix \label{sec:appendix}

\section{Light Element Abundances}

ICFs for \emph{n}-capture elements require the fractional abundances of light element ions such as O$^{2+}$, S$^{2+}$, Ar$^{2+}$, and Ar$^{3+}$ \citep{Sterling_Etal_2015, Sterling_Etal_2016, Madonna_Etal_2018}. Elemental abundances are reported in the literature for targets in our sample \citep{Leisy_Dennefeld_2006, Shaw_Etal_2010}, albeit using different ionization correction methods.

To homogenize the analysis of the published optical spectra, we recomputed the total gas-phase abundances of light elements for each PN in our sample with the exception of LMC~73, using the ICF formulae of \citet{Delgado-Inglada_Etal_2014}. We adopt the ionic abundances from \citet{Shaw_Etal_2010} for SMC~13, SMC~14, and SMC~20, assuming 30\% uncertainties except for He ions (20\%). Because \citet{Leisy_Dennefeld_2006} did not report ionic abundances, we computed those with the transition probabilities and effective collision strengths used in the analysis of \citet[][see their Table~5]{Garcia-Rojas_Etal_2015}, and the electron temperatures and densities from \citet{Leisy_Dennefeld_2006}. As in our analysis of \emph{n}-capture elements, we use $T_{\rm e}$(\ion{N}{2}) to calculate ionic abundances for species with ionization potentials below 39~eV, and $T_{\rm e}$(\ion{O}{3}) for higher ionization species.

The ionic and elemental abundances for LMC~73 are taken from Mampaso et al.\ (in prep.), and will be reported in that paper. For the other targets in our sample, the recomputed light element abundances are given in Table~\ref{tab:append_lmc} for LMC PNe, and Table~\ref{tab:append_smc} for SMC PNe. These tables include ionic abundances that are used in the ICFs for \emph{n}-capture elements (\S\ref{sec:elemental}).

\section{Solar Abundances Used in Models and Empirical Analysis}

In Table~~\ref{tab:solar}, we compared the solar abundances of elements that are considered in this paper in our empirical analysis \citep{Asplund_Etal_2021}, and those used for the initial compositions (scaled by metallicity) of NuGrid \citep{Grevesse_Noels_1993}, FRUITY \citep{Lodders_2003}, and Monash \citep{Asplund_Etal_2009} models.

\begin{deluxetable*}{lccccccc}
\tablecolumns{7}
\tablewidth{0pc}
\tablenum{A1}
\tabletypesize{\scriptsize}
\tablecaption{Light Element Abundances for LMC PNe}
\tablehead{ \\[-0.6pc]
\colhead{} & \colhead{LMC 29} & \colhead{LMC 31} & \colhead{LMC 47} & \colhead{LMC 48} & \colhead{LMC 61} & \colhead{LMC 99}}
\startdata
O$^{2+}$/H$^+$ & (5.18 $\pm$ 0.65)E--05 & (1.00 $\pm$0.28)E--05 & (1.13 $\pm$ 0.27)E--04 & (1.54 $\pm$ 0.84)E--04 & (1.65 $\pm$ 0.78)E--04 & (2.52 $\pm$ 1.09)E--04 \\
S$^{2+}$/H$^+$ & (1.63 $\pm$ 0.12)E--06 & (5.28 $\pm$ 0.69)E--07 & (1.50 $\pm$ 0.22)E--06 & (4.38 $\pm$ 0.81)E--06 & (2.41 $\pm$ 0.75)E--06 & (1.52 $\pm$ 0.24)E--06 \\
Ar$^{2+}$/H$^+$ & (6.06 $\pm$ 0.53)E--07 & (2.49 $\pm$ 0.24)E--07 & (6.59 $\pm$ 0.48)E--07 & (4.72 $\pm$ 0.63)E--07 & (1.12 $\pm$ 0.15)E--06 & (3.44 $\pm$ 0.45)E--06 \\
Ar$^{3+}$/H$^+$ & (6.38 $\pm$ 0.57)E--08 & \nodata & (4.60 $\pm$ 0.64)E--07 & \nodata & (3.94 $\pm$ 1.26)E--07 & (3.55 $\pm$ 0.75)E--07 \\
He/H & 0.129 $\pm$ 0.010 & 0.047 $\pm$ 0.008\tablenotemark{a} & 0.129 $\pm$ 0.009 & 0.100 $\pm$ 0.010 & 0.117 $\pm$ 0.016 & 0.080 $\pm$ 0.006 \\
$12+\log{\mathrm{(N/H)}}$ & 8.57 $\pm$ 0.11 & 6.96 $\pm$ 0.10 & 8.51 $\pm$ 0.16 & 7.58 $\pm$ 0.22 & 7.30 $\pm$ 0.29 & 7.97 $\pm$ 0.20 \\
$12+\log{\mathrm{(O/H)}}$ & 7.95 $\pm$ 0.10 & 7.32 $\pm$ 0.06 & 8.19 $\pm$ 0.10 & 8.28 $\pm$ 0.16 & 8.62 $\pm$ 0.16 & 8.52 $\pm$ 0.15 \\
$12+\log{\mathrm{(Ne/H)}}$ & 7.22 $\pm$ 0.13 & 6.08 $\pm$ 0.08 & 7.62 $\pm$ 0.16 & 7.55 $\pm$ 0.12 & 8.31 $\pm$ 0.13 & 7.78 $\pm$ 0.24 \\
$12+\log{\mathrm{(S/H)}}$ & 6.73 $\pm$ 0.13 & 5.76 $\pm$ 0.05 & 6.54 $\pm$ 0.16 & 6.66 $\pm$ 0.07 & 6.57 $\pm$ 0.08 & 6.50 $\pm$ 0.20 \\
$12+\log{\mathrm{(Cl/H)}}$ & \nodata & \nodata & \nodata & \nodata & \nodata & 4.77 $\pm$ 0.21 \\
$12+\log{\mathrm{(Ar/H)}}$ & 6.16 $\pm$ 0.23 & 5.42 $\pm$ 0.23 & 6.07 $\pm$ 0.24 & 5.78 $\pm$ 0.28 & 6.09 $\pm$ 0.28 & 6.78 $\pm$ 0.27 \\
$[$M/H$]$\tablenotemark{b} & --0.22 $\pm$ 0.23 & --1.37 $\pm$ 0.06 & --0.50 $\pm$ 0.10 & --0.41 $\pm$ 0.16 & --0.07 $\pm$ 0.16 & --0.17 $\pm$ 0.15 \\
\enddata
\tablecomments{Light element abundances of LMC PNe recalculated from the intensities of \citet{Leisy_Dennefeld_2006} using $T_{\rm e}$ and $n_{\rm e}$ values reported by those authors. The ICF schema of \citet{Delgado-Inglada_Etal_2014} are used to determine elemental abundances, with the exception of N, for which we adopt N/O~=~N$^+$/O$^+$ \citep{Peimbert_Costero_1969, Delgado-Inglada_Etal_2015}. Ionic abundances are given for species used in \emph{n}-capture element ICF formulae (see \S\ref{sec:elemental}). For LMC~73, light element abundances will be reported in Mampaso et al.\ (in prep.).}
\tablenotetext{a}{The He abundance is underestimated in the low-excitation PN LMC~31, as it does not account for the substantial amount of neutral He in this nebula.}
\tablenotetext{b}{The metallicity of the PN, as represented by the Ar abundance relative to the solar value of \citet{Asplund_Etal_2021} for LMC~29, and by the O abundance for the other PNe (see \S\ref{sec:enrich}). The metallicity $[$O/H$]$ of LMC~73 is --0.13 $\pm$ 0.06 (Mampaso et al., in prep.).}
\label{tab:append_lmc}
\end{deluxetable*}

\begin{deluxetable*}{lccccc}
\tablecolumns{6}
\tablewidth{0pc}
\tablenum{A2}
\tabletypesize{\scriptsize}
\tablecaption{Light Element Abundances for SMC PNe}
\tablehead{ \\[-0.6pc]
\colhead{} & \colhead{SMC 3} & \colhead{SMC 13} & \colhead{SMC 14} & \colhead{SMC 20} & \colhead{SMC 25}}
\startdata
O$^{2+}$/H$^+$ & (9.75 $\pm$ 1.95)E--05 & (1.09 $\pm$ 0.33)E--04 & (1.36 $\pm$ 0.41)E--04 & (5.35 $\pm$ 1.61)E--05 & (3.30 $\pm$ 0.59)E--05 \\
S$^{2+}$/H$^+$ & (5.47 $\pm$ 1.82)E--07 & (4.86 $\pm$ 1.46)E--07 & (7.48 $\pm$ 2.24)E--07 & (2.63 $\pm$ 0.79)E--07 & (1.25 $\pm$ 0.23)E--06 \\
Ar$^{2+}$/H$^+$ & (2.14 $\pm$ 0.25)E--07 & (2.85 $\pm$ 0.86)E--07 & (3.12 $\pm$ 0.904)E--07 & (1.03 $\pm$ 0.31)E--07 & (1.42 $\pm$ 0.16)E--07 \\
Ar$^{3+}$/H$^+$ & \nodata & \nodata & (2.44 $\pm$ 0.73)E--07 & \nodata & (3.55 $\pm$ 0.73)E--07 \\
He/H & 0.086 $\pm$ 0.008 & 0.128 $\pm$ 0.019 & 0.111 $\pm$ 0.010 & 0.139 $\pm$ 0.021 & 0.122 $\pm$ 0.009 \\
$12+\log{\mathrm{(N/H)}}$ & 7.03 $\pm$ 0.17 & 7.30 $\pm$ 0.18 & 7.45 $\pm$ 0.17 & 6.95 $\pm$ 0.18 & 7.95 $\pm$ 0.14 \\
$12+\log{\mathrm{(O/H)}}$ & 8.04 $\pm$ 0.07 & 8.06 $\pm$ 0.11 & 8.26 $\pm$ 0.13 & 7.74 $\pm$ 0.11 & 7.70 $\pm$ 0.10 \\
$12+\log{\mathrm{(Ne/H)}}$ & 7.39 $\pm$ 0.07 & 7.40 $\pm$ 0.08 & 7.51 $\pm$ 0.19 & 6.95 $\pm$ 0.08 & 7.24 $\pm$ 0.15 \\
$12+\log{\mathrm{(S/H)}}$ & 5.69 $\pm$ 0.11 & 5.74 $\pm$ 0.10 & 6.25 $\pm$ 0.18 & 5.43 $\pm$ 0.11 & 6.43 $\pm$ 0.15 \\
$12+\log{\mathrm{(Ar/H)}}$ & 5.49 $\pm$ 0.23 & 5.65 $\pm$ 0.26 & 5.78 $\pm$ 0.26 & 5.22 $\pm$ 0.26 & 5.46 $\pm$ 0.24 \\
$[$O/H$]$\tablenotemark{a} & --0.65 $\pm$0.07 & --0.63 $\pm$ 0.11 & --0.43 $\pm$ 0.13 & --0.95 $\pm$ 0.11 & --0.99 $\pm$0.10 \\
\enddata
\tablecomments{Same as Table~\ref{tab:append_lmc} for SMC PNe. Intensities and physical conditions are from \citet{Shaw_Etal_2010} for SMC~13, SMC~14, and SMC~20, while those for SMC~3 and SMC~25 are from \citet{Leisy_Dennefeld_2006}.}
\tablenotetext{a}{The metallicity of the PN, as represented by the O abundance relative to the solar value of \citet[][see \S\ref{sec:enrich}]{Asplund_Etal_2021}.}
\label{tab:append_smc}
\end{deluxetable*}

\begin{deluxetable}{lcccc}
\tablecolumns{5}
\tablenum{B1}
\tablewidth{0pc}
\tabletypesize{\footnotesize}
\tablecaption{Solar Abundances of Elements Used in Analysis}
\tablehead{ \\[-0.6pc]
\colhead{Element} & \colhead{GN93} & \colhead{Lodders03} & \colhead{Asplund09} & \colhead{Asplund21}
}
\startdata
N  & 8.03 & 7.83 & 7.83 & 7.83 \\
O  & 8.93 & 8.69 & 8.69 & 8.69 \\
Ne & 8.14 & 7.87 & 7.93 & 8.06 \\
S  & 7.27 & 7.19 & 7.15 & 7.15 \\
Cl & 5.56 & 5.26 & 5.23 & 5.23 \\
Ar & 6.53 & 6.55 & 6.40 & 6.38 \\
Se & 3.41 & 3.36 & 3.34 & 3.34 \\
Br & 2.96 & 2.59 & 2.54 & 2.54 \\
Kr & 3.29 & 3.28 & 3.25 & 3.12 \\
Rb & 2.47 & 2.36 & 2.36 & 2.37 \\
Cd & 1.83 & 1.74 & 1.71 & 1.71 \\
Te & 2.30 & 2.22 & 2.18 & 2.18 \\
\enddata
\tablecomments{Solar abundances, in the format $12+\log{\mathrm{(X/H)}}$ from different references for elements used in our analysis. Meteoritic abundances are cited except for C, N, O, and noble gases. Nu models scale the initial compositions from the solar values of \citet[][GN93]{Grevesse_Noels_1993}, F models from those of \citet{Lodders_2003}, and M models from those of \citet{Asplund_Etal_2009}. Our empirical abundances are set relative to the solar abundances of \citet{Asplund_Etal_2021}.}
\label{tab:solar}
\end{deluxetable}

\bibliography{mcpn_igrins}{}
\bibliographystyle{aasjournal}

\end{document}